\documentclass[twocolumn, times]{aastex62}

\usepackage{natbib}
\usepackage{graphics}
\usepackage{multirow}
\usepackage{amsbsy}
\usepackage{amsmath}
\usepackage{mathrsfs}
\usepackage{footmisc}
\usepackage{hyperref}

\usepackage{comment}
\usepackage{booktabs}

\hypersetup{
  colorlinks = true,
  urlcolor = black,
  linkcolor=black,
  citecolor=black
}

\definecolorset{rgb}{}{}{%
    bk030, 1.0, 0.4, 0.4;%
    bk040, 1.0, 0.1, 0.1;%
    bk095, 0.8, 0.0, 0.0;%
    bk150, 0.0, 0.5, 0.0;%
    bk220, 0.4, 0.4, 1.0;%
    bk270, 0.0, 0.0, 0.6%
    }

\def\bicep{{\sc Bicep}}
\def\bk{{\sc Bicep/\it Keck}}

\def\biceptwo{{\sc Bicep2}}
\def\bicepthree{{\sc Bicep3}}

\def\planck{{\it Planck}}

\def\spider{{\sc Spider}}

\def\spt{{\sc SPT}}

\def\keck{{\it Keck}}

\def\biceparray{{\sc Bicep} Array}

\newcommand{\ukarcmin}{$\mu$K-arcmin}
\newcommand{\ukrts}{ $\mu\mathrm{K}_{\mathrm{\mbox{\tiny\sc cmb}}}\sqrt{\mathrm{s}}$ }

\def\deg{^\circ}

\newcommand{\p}{\phantom}
\newcommand{\mean}{\operatorname*{mean}}

\def\aap{Astr. \& Astroph.}

\def\apjl{Astrophys.\ J.\ Lett.}
\def\apjs{Astrophys.\ J.\ Suppl.}

\begin{document}

\title{\bicep\ / \keck\ XV: THE \bicepthree\ CMB POLARIMETER AND THE FIRST THREE YEAR DATA SET}

\correspondingauthor{Howard Hui}
\email{hhui@caltech.edu}

\author{P.~A.~R.~Ade}
\affiliation{School of Physics and Astronomy, Cardiff University, Cardiff, CF24 3AA, United Kingdom}
\author{Z.~Ahmed}
\affiliation{Kavli Institute for Particle Astrophysics and Cosmology, SLAC National Accelerator Laboratory, 2575 Sand Hill Rd, Menlo Park, California 94025, USA}
\author{M.~Amiri}
\affiliation{Department of Physics and Astronomy, University of British Columbia, Vancouver, British Columbia, V6T 1Z1, Canada}
\author{D.~Barkats}
\affiliation{Center for Astrophysics, Harvard \& Smithsonian, Cambridge, MA 02138, U.S.A}
\author{R.~Basu Thakur}
\affiliation{Department of Physics, California Institute of Technology, Pasadena, California 91125, USA}
\author{C.~A.~Bischoff}
\affiliation{Department of Physics, University of Cincinnati, Cincinnati, Ohio 45221, USA}
\author{D.~Beck}
\affiliation{Kavli Institute for Particle Astrophysics and Cosmology, SLAC National Accelerator Laboratory, 2575 Sand Hill Rd, Menlo Park, California 94025, USA}
\affiliation{Department of Physics, Stanford University, Stanford, California 94305, USA}
\author{J.~J.~Bock}
\affiliation{Department of Physics, California Institute of Technology, Pasadena, California 91125, USA}
\affiliation{Jet Propulsion Laboratory, Pasadena, California 91109, USA}
\author{H.~Boenish}
\affiliation{Center for Astrophysics, Harvard \& Smithsonian, Cambridge, MA 02138, U.S.A}
\author{E.~Bullock}
\affiliation{Minnesota Institute for Astrophysics, University of Minnesota, Minneapolis, Minnesota 55455, USA}
\author{V.~Buza}
\affiliation{Kavli Institute for Cosmological Physics, University of Chicago, Chicago, IL 60637, USA}
\author{J.~R.~Cheshire IV}
\affiliation{Minnesota Institute for Astrophysics, University of Minnesota, Minneapolis, Minnesota 55455, USA}
\author{J.~Connors}
\affiliation{Center for Astrophysics, Harvard \& Smithsonian, Cambridge, MA 02138, U.S.A}
\author{J.~Cornelison}
\affiliation{Center for Astrophysics, Harvard \& Smithsonian, Cambridge, MA 02138, U.S.A}
\author{M.~Crumrine}
\affiliation{School of Physics and Astronomy, University of Minnesota, Minneapolis, Minnesota 55455, USA}
\author{A.~Cukierman}
\affiliation{Department of Physics, Stanford University, Stanford, California 94305, USA}
\affiliation{Kavli Institute for Particle Astrophysics and Cosmology, SLAC National Accelerator Laboratory, 2575 Sand Hill Rd, Menlo Park, California 94025, USA}
\author{E.~V.~Denison}
\affiliation{National Institute of Standards and Technology, Boulder, Colorado 80305, USA}
\author{M.~Dierickx}
\affiliation{Center for Astrophysics, Harvard \& Smithsonian, Cambridge, MA 02138, U.S.A}
\author{L.~Duband}
\affiliation{Service des Basses Temp\'{e}ratures, Commissariat \`{a} l'Energie Atomique, 38054 Grenoble, France}
\author{M.~Eiben}
\affiliation{Center for Astrophysics, Harvard \& Smithsonian, Cambridge, MA 02138, U.S.A}
\author{S.~Fatigoni}
\affiliation{Department of Physics and Astronomy, University of British Columbia, Vancouver, British Columbia, V6T 1Z1, Canada}
\author{J.~P.~Filippini}
\affiliation{Department of Physics, University of Illinois at Urbana-Champaign, Urbana, Illinois 61801, USA}
\affiliation{Department of Astronomy, University of Illinois at Urbana-Champaign, Urbana, Illinois 61801, USA}
\author{S.~Fliescher}
\affiliation{School of Physics and Astronomy, University of Minnesota, Minneapolis, Minnesota 55455, USA}
\author{N.~Goeckner-Wald}
\affiliation{Department of Physics, Stanford University, Stanford, California 94305, USA}
\author{D.~C.~Goldfinger}
\affiliation{Center for Astrophysics, Harvard \& Smithsonian, Cambridge, MA 02138, U.S.A}
\author{J.~Grayson}
\affiliation{Department of Physics, Stanford University, Stanford, California 94305, USA}
\author{P.~Grimes}
\affiliation{Center for Astrophysics, Harvard \& Smithsonian, Cambridge, MA 02138, U.S.A}
\author{G.~Hall}
\affiliation{School of Physics and Astronomy, University of Minnesota, Minneapolis, Minnesota 55455, USA}
\author{G. Halal}
\affiliation{Department of Physics, Stanford University, Stanford, California 94305, USA}
\author{M.~Halpern}
\affiliation{Department of Physics and Astronomy, University of British Columbia, Vancouver, British Columbia, V6T 1Z1, Canada}
\author{E.~Hand}
\affiliation{Department of Physics, University of Cincinnati, Cincinnati, Ohio 45221, USA}
\author{S.~Harrison}
\affiliation{Center for Astrophysics, Harvard \& Smithsonian, Cambridge, MA 02138, U.S.A}
\author{S. Henderson}
\affiliation{Kavli Institute for Particle Astrophysics and Cosmology, SLAC National Accelerator Laboratory, 2575 Sand Hill Rd, Menlo Park, California 94025, USA}
\author{S.~R.~Hildebrandt}
\affiliation{Department of Physics, California Institute of Technology, Pasadena, California 91125, USA}
\affiliation{Jet Propulsion Laboratory, Pasadena, California 91109, USA}
\author{G.~C.~Hilton}
\affiliation{National Institute of Standards and Technology, Boulder, Colorado 80305, USA}
\author{J.~Hubmayr}
\affiliation{National Institute of Standards and Technology, Boulder, Colorado 80305, USA}
\author{H.~Hui}
\affiliation{Department of Physics, California Institute of Technology, Pasadena, California 91125, USA}
\author{K.~D.~Irwin}
\affiliation{Department of Physics, Stanford University, Stanford, California 94305, USA}
\affiliation{Kavli Institute for Particle Astrophysics and Cosmology, SLAC National Accelerator Laboratory, 2575 Sand Hill Rd, Menlo Park, California 94025, USA}
\affiliation{National Institute of Standards and Technology, Boulder, Colorado 80305, USA}
\author{J.~Kang}
\affiliation{Department of Physics, Stanford University, Stanford, California 94305, USA}
\affiliation{Department of Physics, California Institute of Technology, Pasadena, California 91125, USA}
\author{K.~S.~Karkare}
\affiliation{Center for Astrophysics, Harvard \& Smithsonian, Cambridge, MA 02138, U.S.A}
\affiliation{Kavli Institute for Cosmological Physics, University of Chicago, Chicago, IL 60637, USA}
\author{E.~Karpel}
\affiliation{Department of Physics, Stanford University, Stanford, California 94305, USA}
\author{S.~Kefeli}
\affiliation{Department of Physics, California Institute of Technology, Pasadena, California 91125, USA}
\author{S.~A.~Kernasovskiy}
\affiliation{Department of Physics, Stanford University, Stanford, California 94305, USA}
\author{J.~M.~Kovac}
\affiliation{Center for Astrophysics, Harvard \& Smithsonian, Cambridge, MA 02138, U.S.A}
\affiliation{Department of Physics, Harvard University, Cambridge, MA 02138, USA}
\author{C.~L.~Kuo}
\affiliation{Department of Physics, Stanford University, Stanford, California 94305, USA}
\affiliation{Kavli Institute for Particle Astrophysics and Cosmology, SLAC National Accelerator Laboratory, 2575 Sand Hill Rd, Menlo Park, California 94025, USA}
\author{K.~Lau}
\affiliation{School of Physics and Astronomy, University of Minnesota, Minneapolis, Minnesota 55455, USA}
\author{E.~M.~Leitch}
\affiliation{Kavli Institute for Cosmological Physics, University of Chicago, Chicago, IL 60637, USA}
\author{A.~Lennox}
\affiliation{Department of Physics, University of Illinois at Urbana-Champaign, Urbana, Illinois 61801, USA}
\author{K.~G.~Megerian}
\affiliation{Jet Propulsion Laboratory, Pasadena, California 91109, USA}
\author{L.~Minutolo}
\affiliation{Department of Physics, California Institute of Technology, Pasadena, California 91125, USA}
\author{L.~Moncelsi}
\affiliation{Department of Physics, California Institute of Technology, Pasadena, California 91125, USA}
\author{Y. Nakato}
\affiliation{Department of Physics, Stanford University, Stanford, California 94305, USA}
\author{T.~Namikawa}
\affiliation{Kavli Institute for Physics and Mathematics of the Universe (WPI), UTIAS, The University of Tokyo, Kashiwa, Chiba 277-8583, Japan}
\author{H.~T.~Nguyen}
\affiliation{Jet Propulsion Laboratory, Pasadena, California 91109, USA}
\author{R.~O'Brient}
\affiliation{Department of Physics, California Institute of Technology, Pasadena, California 91125, USA}
\affiliation{Jet Propulsion Laboratory, Pasadena, California 91109, USA}
\author{R.~W.~Ogburn~IV}
\affiliation{Department of Physics, Stanford University, Stanford, California 94305, USA}
\affiliation{Kavli Institute for Particle Astrophysics and Cosmology, SLAC National Accelerator Laboratory, 2575 Sand Hill Rd, Menlo Park, California 94025, USA}
\author{S.~Palladino}
\affiliation{Department of Physics, University of Cincinnati, Cincinnati, Ohio 45221, USA}
\author{T.~Prouve}
\affiliation{Service des Basses Temp\'{e}ratures, Commissariat \`{a} l'Energie Atomique, 38054 Grenoble, France}
\author{C.~Pryke}
\affiliation{School of Physics and Astronomy, University of Minnesota, Minneapolis, Minnesota 55455, USA}
\affiliation{Minnesota Institute for Astrophysics, University of Minnesota, Minneapolis, Minnesota 55455, USA}
\author{B.~Racine}
\affiliation{Center for Astrophysics, Harvard \& Smithsonian, Cambridge, MA 02138, U.S.A}
\affiliation{Aix-Marseille  Universit\'{e},  CNRS/IN2P3,  CPPM,  Marseille,  France}
\author{C.~D.~Reintsema}
\affiliation{National Institute of Standards and Technology, Boulder, Colorado 80305, USA}
\author{S.~Richter}
\affiliation{Center for Astrophysics, Harvard \& Smithsonian, Cambridge, MA 02138, U.S.A}
\author{A.~Schillaci}
\affiliation{Department of Physics, California Institute of Technology, Pasadena, California 91125, USA}
\author{R.~Schwarz}
\affiliation{School of Physics and Astronomy, University of Minnesota, Minneapolis, Minnesota 55455, USA}
\author{B.~L.~Schmitt}
\affiliation{Center for Astrophysics, Harvard \& Smithsonian, Cambridge, MA 02138, U.S.A}
\author{C.~D.~Sheehy}
\affiliation{Physics Department, Brookhaven National Laboratory, Upton, NY 11973}
\author{A.~Soliman}
\affiliation{Department of Physics, California Institute of Technology, Pasadena, California 91125, USA}
\author{T.~St.~Germaine}
\affiliation{Center for Astrophysics, Harvard \& Smithsonian, Cambridge, MA 02138, U.S.A}
\affiliation{Department of Physics, Harvard University, Cambridge, MA 02138, USA}
\author{B.~Steinbach}
\affiliation{Department of Physics, California Institute of Technology, Pasadena, California 91125, USA}
\author{R.~V.~Sudiwala}
\affiliation{School of Physics and Astronomy, Cardiff University, Cardiff, CF24 3AA, United Kingdom}
\author{G.~P.~Teply}
\affiliation{Department of Physics, California Institute of Technology, Pasadena, California 91125, USA}
\author{K.~L.~Thompson}
\affiliation{Department of Physics, Stanford University, Stanford, California 94305, USA}
\affiliation{Kavli Institute for Particle Astrophysics and Cosmology, SLAC National Accelerator Laboratory, 2575 Sand Hill Rd, Menlo Park, California 94025, USA}
\author{J.~E.~Tolan}
\affiliation{Department of Physics, Stanford University, Stanford, California 94305, USA}
\author{C.~Tucker}
\affiliation{School of Physics and Astronomy, Cardiff University, Cardiff, CF24 3AA, United Kingdom}
\author{A.~D.~Turner}
\affiliation{Jet Propulsion Laboratory, Pasadena, California 91109, USA}
\author{C.~Umilt\`{a}}
\affiliation{Department of Physics, University of Cincinnati, Cincinnati, Ohio 45221, USA}
\affiliation{Department of Physics, University of Illinois at Urbana-Champaign, Urbana, Illinois 61801, USA}
\author{C.~Verg\`{e}s}
\affiliation{Center for Astrophysics, Harvard \& Smithsonian, Cambridge, MA 02138, U.S.A}
\author{A.~G.~Vieregg}
\affiliation{Department of Physics, Enrico Fermi Institute, University of Chicago, Chicago, IL 60637, USA}
\affiliation{Kavli Institute for Cosmological Physics, University of Chicago, Chicago, IL 60637, USA}
\author{A.~Wandui}
\affiliation{Department of Physics, California Institute of Technology, Pasadena, California 91125, USA}
\author{A.~C.~Weber}
\affiliation{Jet Propulsion Laboratory, Pasadena, California 91109, USA}
\author{D.~V.~Wiebe}
\affiliation{Department of Physics and Astronomy, University of British Columbia, Vancouver, British Columbia, V6T 1Z1, Canada}
\author{J.~Willmert}
\affiliation{School of Physics and Astronomy, University of Minnesota, Minneapolis, Minnesota 55455, USA}
\author{C.~L.~Wong}
\affiliation{Center for Astrophysics, Harvard \& Smithsonian, Cambridge, MA 02138, U.S.A}
\affiliation{Department of Physics, Harvard University, Cambridge, MA 02138, USA}
\author{W.~L.~K.~Wu}
\affiliation{Kavli Institute for Particle Astrophysics and Cosmology, SLAC National Accelerator Laboratory, 2575 Sand Hill Rd, Menlo Park, California 94025, USA}
\author{H.~Yang}
\affiliation{Department of Physics, Stanford University, Stanford, California 94305, USA}
\author{K.~W.~Yoon}
\affiliation{Department of Physics, Stanford University, Stanford, California 94305, USA}
\affiliation{Kavli Institute for Particle Astrophysics and Cosmology, SLAC National Accelerator Laboratory, 2575 Sand Hill Rd, Menlo Park, California 94025, USA}
\author{E.~Young}
\affiliation{Department of Physics, Stanford University, Stanford, California 94305, USA}
\affiliation{Kavli Institute for Particle Astrophysics and Cosmology, SLAC National Accelerator Laboratory, 2575 Sand Hill Rd, Menlo Park, California 94025, USA}
\author{C.~Yu}
\affiliation{Department of Physics, Stanford University, Stanford, California 94305, USA}
\author{L.~Zeng}
\affiliation{Center for Astrophysics, Harvard \& Smithsonian, Cambridge, MA 02138, U.S.A}
\author{C.~Zhang}
\affiliation{Department of Physics, California Institute of Technology, Pasadena, California 91125, USA}
\author{S.~Zhang}
\affiliation{Department of Physics, California Institute of Technology, Pasadena, California 91125, USA}

\collaboration{(\bicep/\keck\ Collaboration)}

\begin{abstract}

We report on the design and performance of the \bicepthree\ instrument and its first three-year data set collected from 2016 to 2018.
\bicepthree\ is a 52~cm aperture, refracting telescope designed to observe the polarization of the cosmic microwave background (CMB) on degree angular scales at 95~GHz.
It started science observation at the South Pole in 2016 with 2400 antenna-coupled transition-edge sensor (TES) bolometers.
The receiver first demonstrated new technologies such as large-diameter alumina optics, Zotefoam infrared filters, and flux-activated SQUIDs, allowing $\sim 10\times$ higher optical throughput compared to the \keck\ design.
\bicepthree\ achieved instrument noise-equivalent temperatures of 9.2, 6.8 and 7.1\ukrts and reached Stokes $Q$ and $U$ map depths of 5.9, 4.4 and 4.4~\ukarcmin\ in 2016, 2017 and 2018, respectively.
The combined three-year data set achieved a polarization map depth of 2.8~\ukarcmin\ over an effective area of 585 square degrees, which is the deepest CMB polarization map made to date at 95~GHz.

\end{abstract}


\section{Introduction}
\label{sec:intro}
Inflation, a brief period of exponential expansion in the early Universe, was postulated to solve the horizon, flatness and monopole problems which arise from the $\Lambda$CDM ``standard model'' of the Universe~\citep{Brout1978, Starobinsky1980, Kazanas1980, Guth1981, Linde1982, Steinhardt1982}.
The perturbations under this paradigm are adiabatic, nearly Gaussian and close to scale-invariant, which are consistent with precise cosmic microwave background (CMB) observations~\citep{planck2018_i}.
Moreover, many models of inflation predict the existence of primordial gravitational waves (PGWs) which would leave a unique degree-scale $B$-mode polarization pattern in the CMB~\citep{Kamionkowski97,seljak1997}.
If detected, PGWs can serve as a probe of the very early Universe and high energy physics inaccessible with existing particle accelerators.

The \bicep/\keck\ experiments are a series of telescopes designed to search for this degree-scale $B$-mode polarization of CMB originating from PGWs.
These instruments are located at the Amundsen-Scott South Pole Station in Antarctica.
The $\sim$10,000 ft altitude and extreme cold make the Antarctic plateau one of the driest places on earth.
During the winter season, the 6 months of continuous darkness provides exceptionally low and stable atmospheric $1/f$ noise, which allows our telescopes to observe the sky without the need of an active instrument modulation at these large angular scales~\citep{Kuo2017}.

We first reported an excess of $B$-mode signal at 150~GHz in~\cite{BKI}. In a subsequent joint analysis with the $Planck$ collaboration, it was found that polarized emission from dust in our galaxy could account for most of the signal~\citep{bkp2015}. Dust is currently the dominant foreground contaminant to CMB polarization measurements, and is most powerful at high frequencies.
Subsequent modeling shows synchrotron may potentially be another source of foreground emission at lower frequencies~\citep{krachmalnicoff18}.
In order to probe the physics of the early Universe, we need a dedicated strategy to separate these foregrounds from the potential faint primordial signal. 

The \bicep/\keck\ instruments are small-aperture, compact, on-axis refracting telescopes, emphasizing high optical throughput and low optical loading with dedicated calibration campaigns to control instrument systematics.
Five separate instruments spanning the past two decades have been deployed to date.
\bicep1 operated from 2006 through 2008 with 98 neutron transmutation doped (NTD) germanium thermistors at 95, 150 and 220~GHz~\citep{b1_chiang, b1_Takahashi}.
\bicep2 replaced \bicep1 and observed from 2010 through 2012 with 512 planar antenna transition edge sensors at 150~GHz~\citep{BKII}.
\keck\ utilized the same optical and detector technologies as employed in \bicep2, comprising five independent receivers.
It observed at 150~GHz, and later at 95 and 220~GHz, installed in a separate telescope mount previously used for DASI~\citep{dasi2002} and QUaD~\citep{quad2008}.
It began science observations in 2012, observing until 2019~\citep{SPIE_kernasovskiy, LTD_Staniszewski}.

After \bicep2 was decommissioned at the end of 2012, \bicepthree\ was installed in the same telescope mount in November 2014 and started scientific observation in 2016 with 2400 detectors at 95~GHz.
It employed a conceptually similar design to its predecessor, but with multiple technological improvements allowing an order of magnitude increase in mapping speed compared to a single \keck\ 95~GHz receiver.
Benefiting from a modular receiver design, \keck\ was gradually adapted from an all-150~GHz receiver configuration into a high frequency `dust telescope', observing at 220 and 270~GHz, with \bicepthree\ continuing observations at 95~GHz, where foregrounds are minimal.
In late 2019, \keck\ was decommissioned and replaced with a new telescope mount~\citep{Crumrine2018} to accommodate four \bicepthree-like receivers that will form the next phase of the experiment, \biceparray.
The first receiver in \biceparray\ started observation at 30/40~GHz in 2020 to probe the low frequency polarized synchrotron signal.
\biceparray\ will cover 6 distinct bands from 30 to 270~GHz when fully deployed.
In the meantime, the \biceparray\ telescope mount carries a mixture of \keck\ and \biceparray\ receivers, while \bicepthree\ continues to observe.
Table~\ref{tab:bkfreq} shows the \bk\ experiments from 2010 to 2020 and their frequency coverage.

\begin{table*}
    \centering
    \label{tab:bkfreq}
    \caption{Frequency coverage in the \bk\ experiment from 2010 to 2020.
    Brackets in the table indicate an engineering receiver (270~GHz \keck\ in 2017 was a prototype of high-frequency focal plane, \bicepthree\ in 2015 only had a partially filled focal plane, and the 150~GHz \keck\ in 2019 was a demonstration of the $\mu$Mux readout~\citep{LTD:Cukierman2020}), and are not included in science analyses.
    \keck\ was replaced by \biceparray\ in 2020.
    In its first season, one slot was fitted with the 30/40~GHz \biceparray\ receiver,
    and three \keck\ receivers were put back into the new telescope mount.
    This paper uses the data collected by \bicepthree\ from 2016 through 2018.}
    \bgroup
    \newcommand*{\ztz}{\textcolor{bk030}{30/40GHz}}
    \newcommand*{\znf}{\textcolor{bk095}{95GHz}}
    \newcommand*{\ofz}{\textcolor{bk150}{150GHz}}
    \newcommand*{\ttz}{\textcolor{bk220}{220GHz}}
    \newcommand*{\tsz}{\textcolor{bk270}{270GHz}}
    \newcommand*{\umux}{\textcolor{bk150}{150GHz}}
    \begin{tabular}{lccccccccccc}
    \toprule
    Receiver & 2010 & 2011 & 2012 & 2013 & 2014 & 2015 & 2016 & 2017 & 2018 & 2019 & 2020 \\
    \midrule
    \biceptwo  & \ofz & \ofz & \ofz &      &      &        &      &        &      &         &      \\
    Keck Rx0   &      &      & \ofz & \ofz & \znf & \znf   & \ttz & \ttz   & \ttz & \ttz    & \ttz \\
    Keck Rx1   &      &      & \ofz & \ofz & \ofz & \ttz   & \ttz & \ttz   & \ttz & [\umux] &      \\
    Keck Rx2   &      &      & \ofz & \ofz & \znf & \znf   & \ttz & \ttz   & \ttz & \ttz    & \ttz \\
    Keck Rx3   &      &      & \ofz & \ofz & \ofz & \ttz   & \ttz & \ttz   & \ttz & \ttz    &      \\
    Keck Rx4   &      &      & \ofz & \ofz & \ofz & \ofz   & \ofz & [\tsz] & \tsz & \tsz    & \tsz \\
    \bicepthree&      &      &      &      &      & [\znf] & \znf & \znf   & \znf & \znf    & \znf \\
    BA Rx0     &      &      &      &      &      &        &      &        &      &         & \ztz \\
    \bottomrule
    \end{tabular}
    \egroup
\end{table*}

\begin{figure*}
  \centering
  \includegraphics[width=0.9\textwidth]{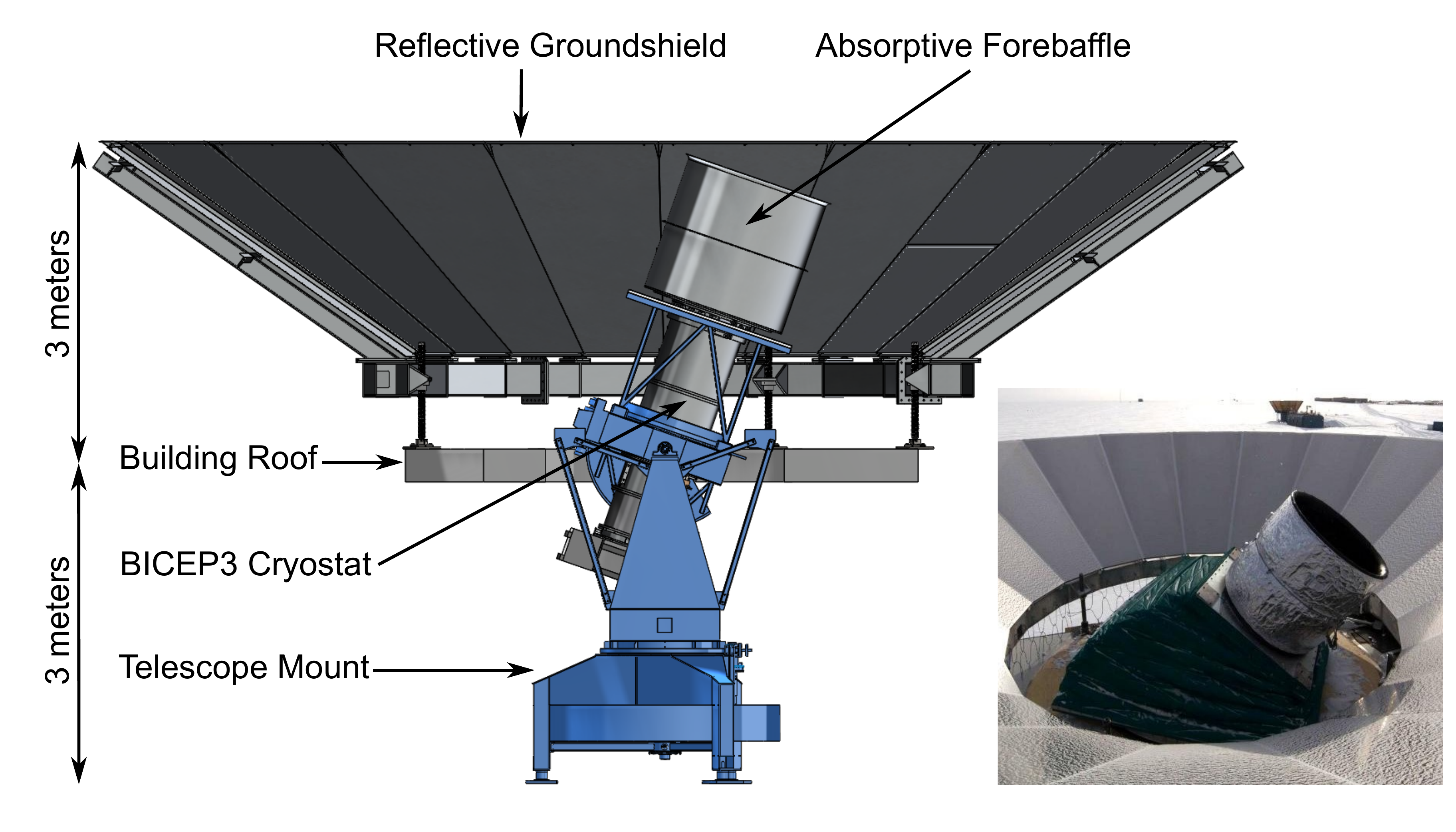}
  \caption{
The \bicepthree\ telescope in the mount, looking out through the roof of the Dark Sector Laboratory (DSL) located $\sim 1100$~m from the geographic South Pole.
The insulating environmental shield shown in the bottom right photo is hidden in the CAD layout.
The three-axis mount previously used in \bicep1 and \bicep2 allows for motion in azimuth, elevation, and boresight rotation.
A co-moving absorptive forebaffle extends skyward beyond the cryostat receiver to intercept stray light outside the designed field-of-view.
Additionally, the telescope is surrounded by a stationary reflective ground shield which redirects off-axis rays to the cold sky. 
}
  \label{fig:b3_full}
\end{figure*}

This paper provides an overview of the \bicepthree\ instrument design and performance with the three-year dataset from 2016 to 2018.
Fig.~\ref{fig:b3_full} shows the overall layout of \bicepthree\ as it is installed at the South Pole.
The following sections describe the details of each of the subcomponents: telescope mount (\S\ref{sec:telescope_mount}); optics (\S\ref{sec:optics}); cryostat (\S\ref{sec:receiver}); focal plane unit (\S\ref{sec:focal_plane}); transition-edge sensor bolometers (\S\ref{sec:detectors}); and data acquisition and control system (\S\ref{sec:daq}).

In particular, \bicepthree's 520~mm diameter aperture is $\sim2$ times the size of the \keck\ design.
This is realized by the large diameter alumina optics shown in \S\ref{optics}.
The increase in aperture size allowed us to accommodate 2400 detectors in the focal plane, compared to 288 detectors in the previous \keck\ 95~GHz receivers.
The new modular focal plane design in \S\ref{sec:focal_plane} allows rapid rework and dramatically reduces risk.
The high number of detectors also requires a mature multiplexing readout.
\bicepthree\ is the first experiment to adapt the new generation flux-activated time domain multiplexing system described in \S\ref{sec:daq}.
Most CMB experiments utilize low temperature, superconducting detectors that operate below 1~K.
Rapid development in mechanical compressor cryocoolers allowed ground-based telescopes to phase out the need of liquid Helium, but the high-pressure Helium lines in the system between the telescope and compressor induce significant wear in a continuous rotating mount.
We address this by integrating a helium rotary joint into the telescope mount system, allowing for continuous rotation while maintaining a high-pressure seal and electrical connectivity (\S\ref{sec:telescope_mount}).

The achieved performance characteristics of the receiver and detector properties of \bicepthree\ are presented in \S\ref{sec:inst_char}, the observing strategy is presented in \S\ref{sec:obs_strategy}, and in \S\ref{sec:data_set} we show the first three-year data set taken from 2016 to 2018, reporting its internal consistency validation, sensitivity, and map depth.
The cosmological analysis using \planck, WMAP, and \bicep/\keck\ observations through the 2018 observing season are presented in \cite{BKXIII}.

\section{Telescope Mount, Forebaffle and Ground Shield}
\label{sec:telescope_mount}

\subsection{Telescope mount}
\bicepthree\ is installed in the Dark Sector Laboratory building, approximately a kilometer away from the South Pole Station.
The base of the telescope mount is supported by a platform on the second floor of the building, with a 2.4~m diameter opening in the roof for telescope access to the sky (Fig.~\ref{fig:b3_full}).
The warm indoor environment of the building is extended beyond the roof level by a flexible insulating environmental shield, so that only the receiver window is exposed to the Antarctic ambient temperature.

\bicepthree\ uses a steel three-axis mount built by Vertex-RSI\footnote{Now General Dynamics Satcom Technologies, Newton, NC 28658, \texttt{http://www.gdsatcom.com/vertexrsi.php}}.
It was originally built for \bicep1 and also housed \bicep2 until 2013. 
The mount structure was modified in 2014 to accommodate the larger \bicepthree\ receiver.

The mount moves in azimuth and elevation, with the third axis rotating about the boresight of the telescope (``deck'' rotation).
The range of motion of the mount is $48^\circ$ to $110^\circ$ in elevation and $400^\circ$ in azimuth, capable of scanning at speeds of $5^\circ$/s in azimuth.
The \bicepthree\ cryostat houses a pulse tube cryocooler which limits the accessible deck angle to less than a full $360^\circ$ rotation in the \bicep\ mount.
However, the design still allows the telescope to scan with two sets of $180^\circ$ opposing deck angles, offset from each other by $45^\circ$, retaining an effective set of observation schedules in order to probe systematic errors, as shown in \S\ref{sec:obs_strategy}.

\subsection{Helium rotary joint}
The pulse tube cryogenic cooler comprises two sub-systems: a coldhead installed inside the receiver, and a helium compressor located in the building, away from the telescope mount.
This pulse tube provides cooling by expanding a high pressure helium gas volume, and requires high and low pressure helium flexible lines to be routed from the compressor, through the three mount axes (azimuth, elevation, and boresight), to the coldhead in the receiver.

\begin{figure*}
  \centering
  \includegraphics[width=0.45\textwidth]{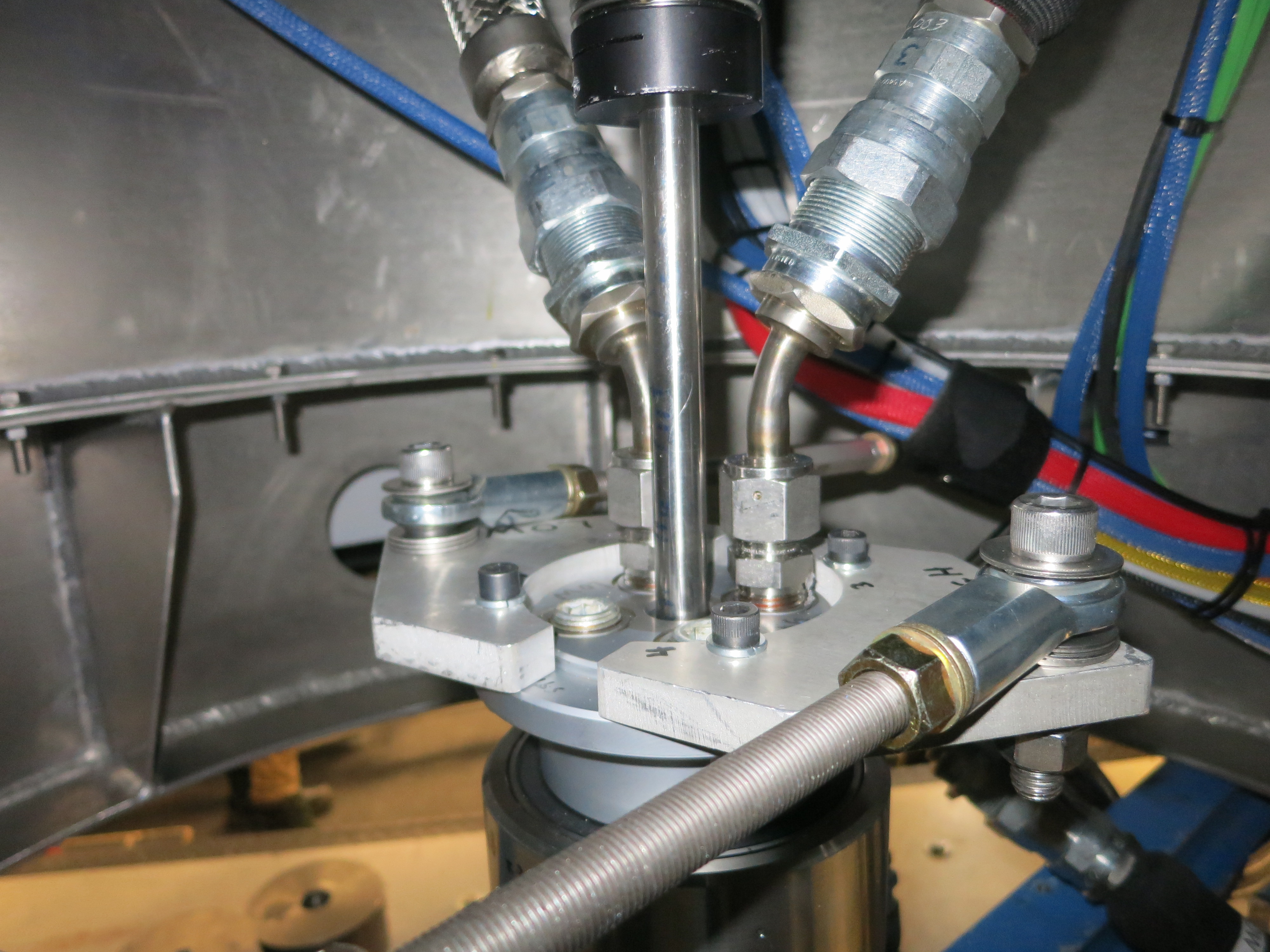}
  \includegraphics[width=0.45\textwidth]{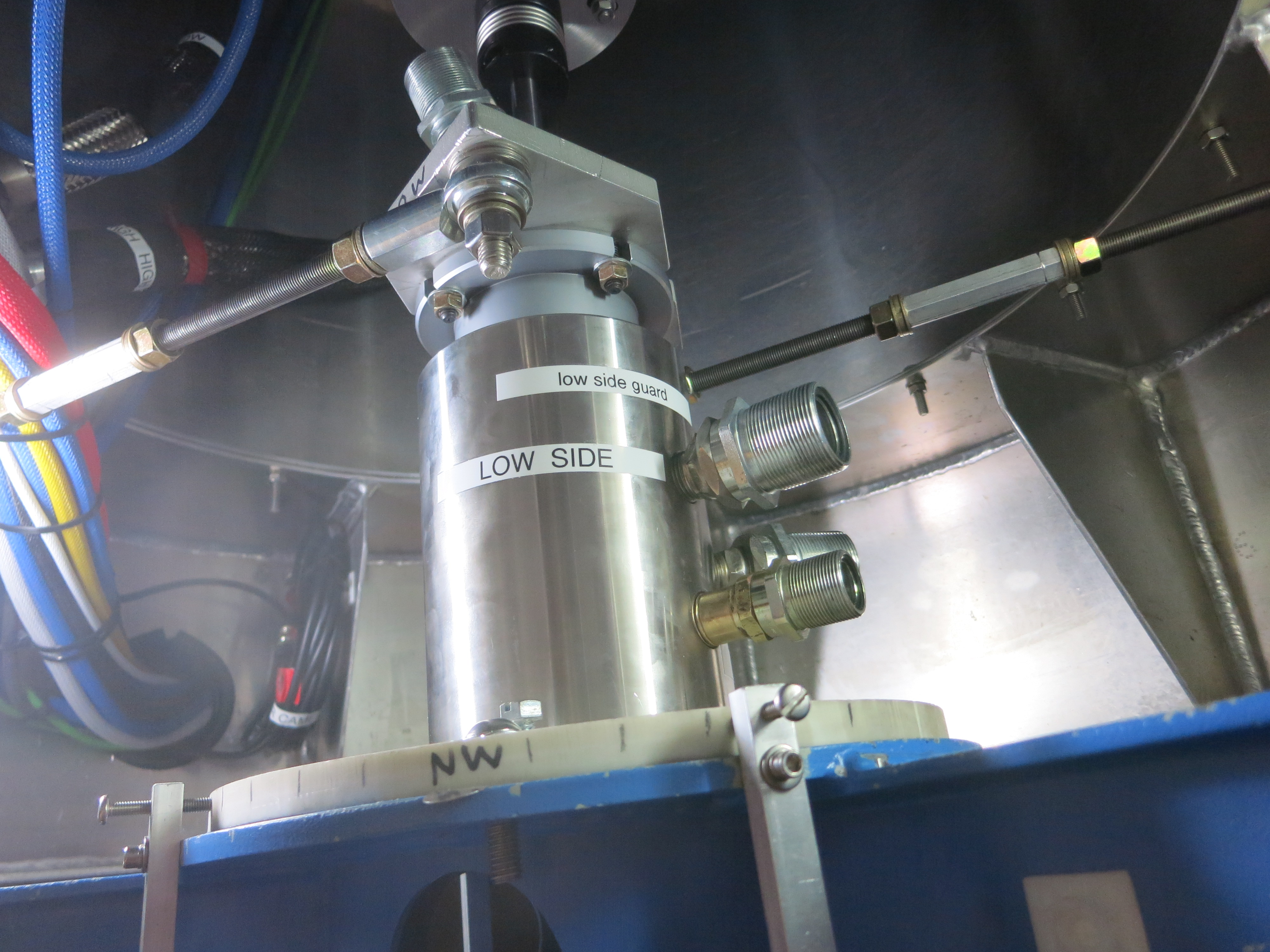}
  \caption{
Photos of the 4-channel helium Rotary Joint system. \textit{Left}: Two 30-degree bends rotate with the azimuth axis and go on to the receiver through the elevation and boresight axes. \textit{Right}: Static section with the 4 connections for the high and low pressure helium and their respective guard channels. In both photos, two ball-end rod joints act as torque arms to transmit the azimuth rotation to the rotor of the HRJ.}
  \label{fig:b3_hrj}
\end{figure*}

During an observing schedule, movements in elevation and boresight are intermittent and span a limited range of angles, unlike the azimuth axis which scans back and forth continuously in azimuth with a $130^\circ$ range.
To avoid wear on the compressor lines in the helium line wrap, \bicepthree\ uses a commercial high-pressure-gas rotary joint from DSTI\footnote{Dynamic Sealing Technologies, Inc., Andover, MN 55304, \texttt{www.dsti.com}} that enables the two pressurized helium gas lines to pass through the azimuth motion.
In this joint, shown in Fig.~\ref{fig:b3_hrj}, one set of lines remains static at the base of the mount and connects to the helium rotary joint, from which a second set of lines rotate with the azimuth axis of the mount. Therefore the azimuth cable carrier only needs to handle the much more flexible electrical cables.

During the 2015 engineering season, the original design used a basic 2-channel rotary joint (DSTI model: GP-421) to connect the compressor's high and low pressure helium channels at 290 and 90~PSI, respectively.
However, helium gas can permeate materials and gaps much more easily than other larger gas molecules, and this commercial rotary joint was not designed specifically for helium gas.
We found the overall system lost 3-5~PSI of pressure per day, originating in the dynamic seals of the rotary joint.
Such a large leak resulted in the need to refill the compressor system multiple times a week to maintain optimal pulse tube performance.
In addition to being extremely labor-intensive for the telescope operator, these repeated helium refills introduced contamination into the pulse tube, and eventually degraded the cooling performance.

To remedy the high leak rate, the rotary joint was replaced with a 4-channel model (DSTI model: GP-441) before the 2016 season.
The 4 channels were configured such that the two working high pressure helium lines would be guarded by two outer channels, serving as pressurized buffers.
Thus, the dynamic seals between the two inner high pressure lines would only `sense' the small differential pressure to the pressurized buffers ($\sim10$PSI) instead of the much larger differential to atmospheric pressure ($>100$PSI). This configuration reduced the leak rate of the active channels to between $\sim$0.1 and  0.5~PSI per day over an entire season\footnote{The guard channels still have similar leak rate as the 2-channel design, but this is acceptable since refills for them do not affect the pulse tube performance.}.
The reduced helium leak rate requires less frequent refills, and enables optimal pulse tube performance throughout a full season.
The HRJ dynamic seals receive a complete replacement once per year.

\subsection{Ground shield and absorptive baffle}
\label{sec:baffle}
A warm, absorptive forebaffle as shown in Fig.~\ref{fig:b3_full} extends skyward beyond the cryostat receiver to intercept stray light outside the designed field-of-view. 
The forebaffle is mounted directly to the receiver and therefore co-moving with the axes of motion of the telescope. 
The forebaffle is constructed from a large aluminum cylinder, 1.3~m in diameter and height, with a rolled top edge lined by microwave-absorptive Eccosorb HR-10 foam.
Heater tape keeps the forebaffle a few degrees above the Antarctic ambient temperature to help avoid snow accumulation, and a layer of closed-cell polyethylene foam (Volara) protects the Eccosorb from accumulating moisture.
Based on radiative loading on the detectors observed once the forebaffle is installed, the forebaffle intercepts $\sim10\%$ of the total optical power.
The source of this wide-angle response is likely a combination of scattering and multiple reflections.

Additionally, the telescope is surrounded by a stationary reflective ground shield.
It is fixed to the roof of the building to act as a second barrier against stray light and signal contamination from nearby ground sources and reduces the large radiative gradient between the sky and the ground. 
The ground shield is 10~m in diameter and 3~m in height, constructed with aluminum honeycomb panels and steel beams.
The combination of the baffle and the ground shield is designed such that off-axis rays from the telescope must diffract at least twice before intercepting the ground.

\subsection{Star camera}
An optical camera is used to determine mount pointing parameters (\S\ref{sec:star_pointing}).
It is attached to the side of the receiver vacuum jacket, and looks up through a hole in the bottom of the forebaffle.
An optical baffle reduces stray light when using the star camera during daylight and twilight conditions, but is removed for CMB observations.
The telescope is a Newtonian reflector, with a 10~cm aluminum-coated\footnote{Edmund Optics, Inc., Barrington, NJ, USA} objective and a 44~cm focal length.
A 700~nm low-pass edge filter removes much of the Rayleigh-scattered sunlight during daylight and twilight.
The camera is a CCD with video readout\footnote{Astrovid StellaCam Ex, Adirondack Video Astronomy, 72 Harrison Avenue, Hudson Falls, New York, USA; the CCD is a Sony ICX248AL B/W, Sony Group Corporation, 1-7-1 Konan Minato-ku, Tokyo, 108-0075 Japan}, and the video-to-digital conversion is done with a video capture card\footnote{Sensoray, 7313 SW Tech Center Dr., Tigard, OR, USA} in one of the control computers.
The field of view is approximately $0.8^\circ\times0.6^\circ$.
The CCD is on a linear stage to allow focusing via a remote controller used by the operator.

\section{Optics}
\label{sec:optics}

\subsection{Optical design}
\label{optics}

\bicepthree\ utilizes the same concept as previous \bk\ receivers, using a compact, on-axis, two-refractor optical design that provides a wide field of view and a telecentric focal plane.
It has a 4~K aperture of 520~mm and beam width given by a Gaussian radius~$\sigma \sim 8.9'$.
The lenses and filters operate at cryogenic temperatures inside of the cryostat receiver to minimize excess in-band photon loading.
The HDPE plastic cryostat window is at ambient temperature.
Thermal filters mounted behind the window cool radiatively.
Table~\ref{tab:opdes} shows \bicepthree's optical design parameters compared to previous \bicep/\keck\ receivers. 

\begin{table}
  \centering
  \caption{Optical design parameters for \bicep2 at 150~GHz and \bicepthree\ at 95~GHz.}
  \label{tab:opdes}
  \begin{tabular}{c c c}
    \toprule
    & \bicep2/ & \bicep3 \\
    & \keck\ & \\
    \midrule
    Aperture dia.   & 264~mm & 520~mm \\
    Field of view   & 15$^{\circ}$ & 27.4$^{\circ}$ \\
    Beam width~$\sigma$      & $12'$ & $8.9'$ \\
    Focal ratio		& $f/2.2$ & $f/1.6$ \\
    \bottomrule
    \vspace{1 mm}
  \end{tabular}
\end{table}

\begin{figure}
  \centering
  \includegraphics[width=0.45\textwidth]{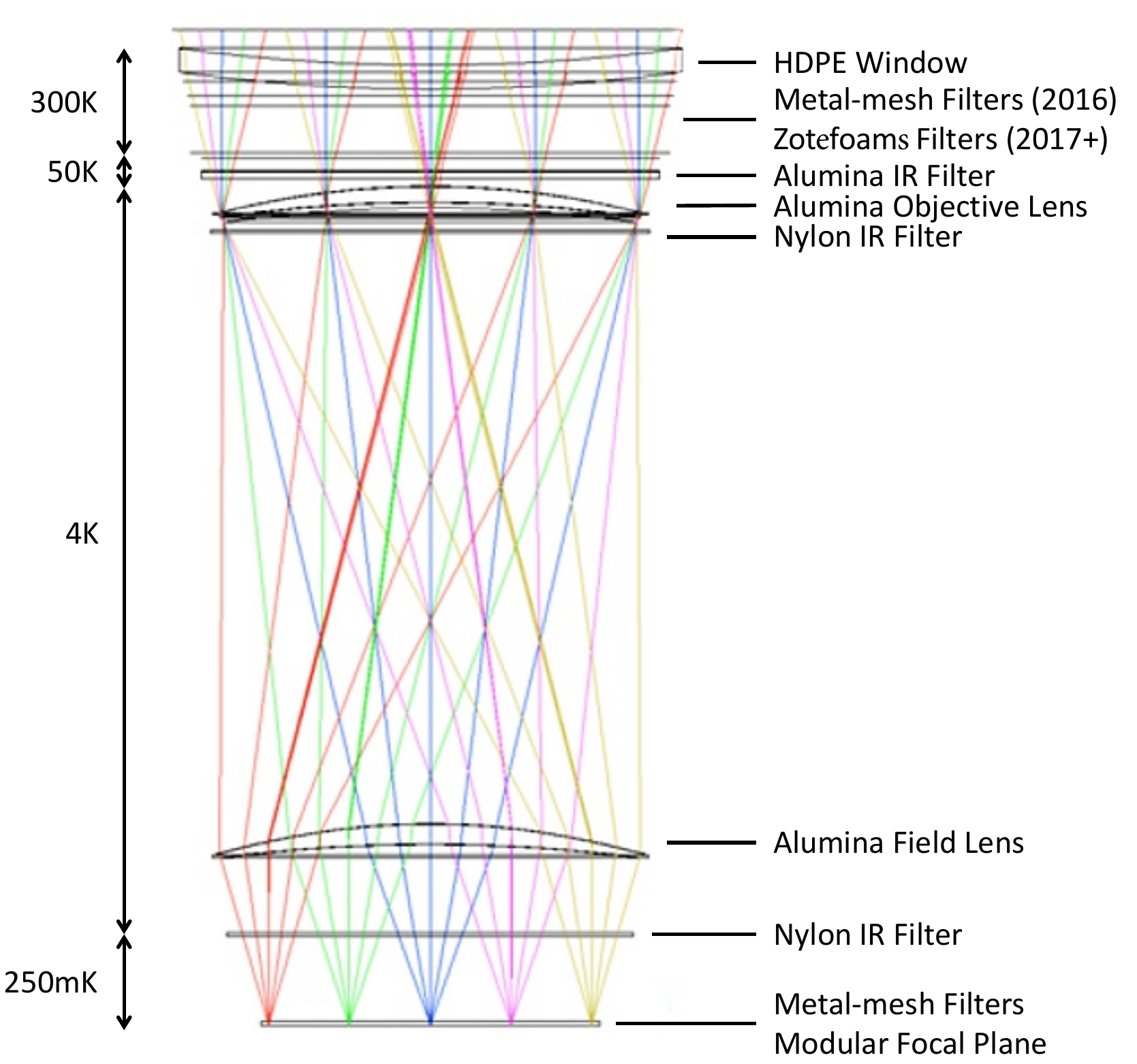}
  \caption{
Ray diagram including the elements of the optical chain.
The 300~K metal-mesh filters were replaced by a stack of 10 Zotefoam filters in 2017, which improved both the IR loading on the cryostat, and the in-band power incident on the detectors.
}
  \label{fig:opticaldiagram}
\end{figure}

The ray diagram and full optical chain are shown in Fig.~\ref{fig:opticaldiagram}. 
The radially symmetric optical design allows for well-matched beams for two idealized orthogonally polarized detectors at the focal plane.

\subsection{Vacuum window and membrane}

The first optical element in the receiver is the vacuum window.
\bicep2/\keck\ used laminated Zotefoam\footnote{Plastazote HD30 from Zotefoams, Inc., Walton, KY 41094, USA, \texttt{www.zotefoams.com}}~(Zotefoam HD30), but \bicepthree\ instead uses a 31.75~mm thick, 73~cm diameter high-density polyethylene (HDPE) window due to the larger aperture setting more stringent requirements on its mechanical strength.
The surfaces of the HDPE window are coated with a $\lambda/4$ anti-reflective (AR) layer made of Teadit 24RGD\footnote{TEADIT North America, Pasadena, TX, USA}~(expanded PTFE sheet).
The AR coating adheres to the window with a thin layer of low-density polyethylene (LDPE) plastic, melted in a vacuum oven press.

In front of the window is a 22.9~$\mu$m thick biaxially oriented polypropylene (BOPP) membrane to protect the window from snow and create an enclosed space below, which is slightly pressurized with room temperature nitrogen gas to evaporate snow that falls onto the membrane surface.

\subsection{Large-diameter 300~K filters}
\label{sec:foam_filters}

Inside the receiver and directly behind the vacuum window is a set of infrared filters to reduce the thermal loading in the receiver.
These are a stack of 10 thin filters mounted on a set of aluminum rings mechanically connected to the room-temperature vacuum jacket.
The filters reflect or absorb infrared radiation in stages, and radiatively equilibrate at progressively lower temperatures to reduce the thermal infrared power into the cryostat.

The original design used a set of metal-mesh filters, composed of 3.5~\micron\ Mylar or 6~\micron\ polypropylene/polyethylene (PP/PE) film, pre-aluminized to a 40~nm deposition thickness and laser ablated to form a grid of metal squares \citep{mesh2014}.
However, we found that the performance of the metal-mesh filter depended on the etching process of the metal on the thin film, and minor defects in fabrication introduced excess in-band scattering from the filters.
The in-band scattering was slightly polarized, leading to additional mm-wave power on the detectors, and associated photon noise.
Furthermore, simulations using high-frequency structure simulator (HFSS\footnote{Ansys, \texttt{www.ansys.com}}) software indicated of order 0.5\% specular reflection per layer even without defects.

\begin{figure}
  \centering
  \includegraphics[width=0.45\textwidth]{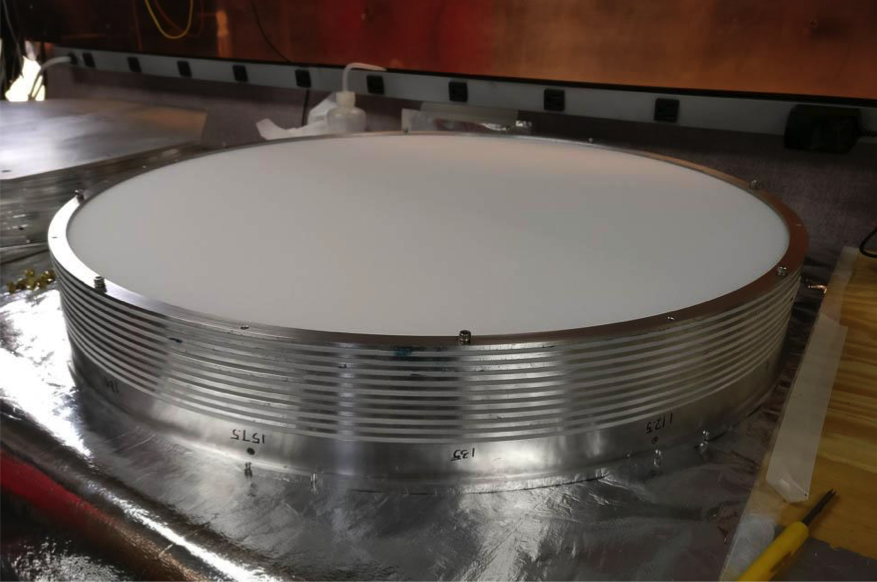}
  \caption{
Stack of 10 layers room temperature IR filters installed in \bicepthree, immediately behind the vacuum window.
This photo shows the current configuration, with each layer composed of 3.17~mm thick Zotefoam, glued onto a stack of aluminum frames with 3.17~mm spacing.
The original design was a stack of metal-mesh filters, which was replaced in 2017.
}
  \label{fig:hd30}
\end{figure}

All the metal-mesh filters except for one placed behind the 50~K alumina filter were replaced in 2017 with a set of ten, 3.17~mm thick Zotefoam layers~(Fig.~\ref{fig:hd30}).
These filters are nitrogen-expanded polyethylene foam layers that scatter infrared radiation (IR) isotropically and therefore act as floating radiative layers, while maintaining $>$99~\% transmission in-band.
Using room-temperature transmission measurements, we estimate 8~\% improvement in-band transmission compared to the metal-mesh filters.
Table~\ref{tab:mesh2016} details the individual filters used in \bicepthree.

\begin{table}
  \centering
  \caption{Room temperature IR filters installed in \bicepthree.
  The main stack of 10 filters behind the window are listed in order beginning with the closest filter to the window.
  The metal-mesh filters were replaced by Zotefoam in 2017.}
  \label{tab:mesh2016}
  \begin{tabular}{c| c c| c}
    \toprule
    & 2016 & Square/pitch & 2017+ \\
    Location & Substrate & [$\mu$m] & Substrate\\
    \midrule
    Behind window               & 3.5~$\mu$m Mylar & 50/80 & HD-30 foam\\
    ($\sim290$~K)               & 3.5~$\mu$m Mylar & 40/55 & HD-30 foam\\
                                & 3.5~$\mu$m Mylar & 50/80 & HD-30 foam\\
                                & 3.5~$\mu$m Mylar & 40/55 & HD-30 foam\\
                                & 3.5~$\mu$m Mylar & 90/150 & HD-30 foam\\
                                &   6~$\mu$m PP/PE & 40/55 & HD-30 foam\\
                                & 3.5~$\mu$m Mylar & 50/80 & HD-30 foam\\
                                & 3.5~$\mu$m Mylar & 40/55 & HD-30 foam\\
                                & 3.5~$\mu$m Mylar & 50/80 & HD-30 foam\\
                                & 3.5~$\mu$m Mylar & 90/150 & HD-30 foam\\
    \midrule
    Behind 50~K                 & 3.5~$\mu$m Mylar & 90/150 & 3.5~$\mu$m Mylar\\
    Alumina filter & & & \\
    \bottomrule
  \end{tabular}
\end{table}

\subsection{Alumina thermal filter and optics}

Motivated by the larger aperture diameter and faster $f/1.6$ speed in \bicepthree, we developed large-diameter alumina filters and their anti-reflection coating.
Alumina lenses are much thinner and less-aggressively shaped than their HDPE equivalents owing to the significantly higher index of refraction at $n=3.1$.
The alumina optics are 21~mm and 27~mm thick at the center for the field and objective lens respectively, compared to $>67$~mm for a comparable HDPE design.
Both the lenses and 50~K filter are made from 99.6\% pure alumina sourced from CoorsTek\footnote{CoorsTek, Golden, CO 80401, USA, \texttt{www.coorstek.com}}.

The reduction in thickness and high thermal conductivity of alumina ($0.5\ \text{Wm}^{-1} \text{K}^{-1}$ at 4~K) enables the optical elements to cool to base temperatures more rapidly and limits any thermal gradient across the lenses to less than 1~K from center to edge.
Lab measurements of similar alumina materials indicate low in-band absorption at room temperature which decreases with temperature~\citep{PennAlford97, Inoue2014}.
Our own measurements at room temperature indicated significant differences between various formulae, and the CoorsTek AD-996 Si used for the filter and lenses was the best we tested.
After deployment, we also confirmed substantially decreased loss at 77~K.
A single 10~mm thick alumina disk serves as an absorptive thermal filter, mounted on the 50~K cryogenic stage.
The high mid-IR absorption and high thermal conductivity make alumina a choice material for this application.

The AR coating used for the alumina optics is a mixture of Stycast 1090 and 2850FT with a homogeneous refractive index of $n=1.74$.
The epoxy is poured and rough-molded to 1~mm thickness on the alumina surface, then either machined (lenses) or abrasively ground (flat filter) to the final 0.452~mm thickness.
The thickness of the coating is controlled to less than 25~\micron\ tolerance by referencing pre-coating surface measurements of the alumina.

\begin{figure}
\centering
  \includegraphics[width=0.45\textwidth]{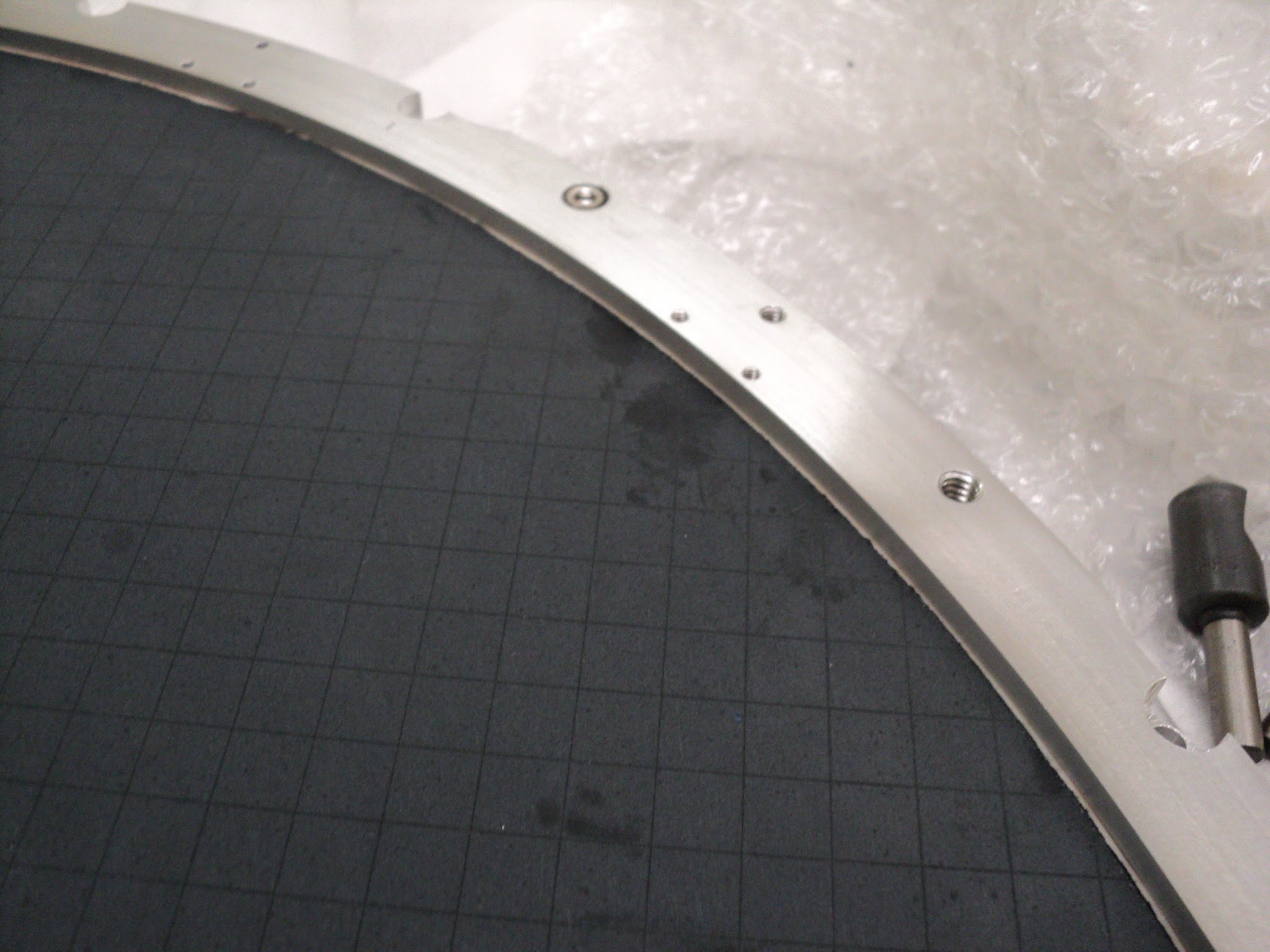}
  \caption{
AR coated alumina filter in \bicepthree.
The alumina filter is coated with a mix of Stycast 1090 and 2850FT.
The epoxy is machined to the correct thickness and laser diced to 1~cm squares to mitigate differential thermal contraction between alumina and the epoxy.
}
  \label{fig:alumina_ar}
\end{figure}

Historically, alumina optics were limited to small sizes unless accommodation for differential contraction between the alumina and the epoxy was made.
\cite{Inoue2014, rosen2013} put slices through their coatings to allow cryogenic operation.
We adopted a laser cutting technique using Laserod\footnote{Laserod Technologies LLC, Torrance, CA 90501, USA}, the same commercial laser machining company that etched the IR blocking metal-mesh film filters described above.
The laser cuts in the AR epoxy are $\sim 30\mu$m wide, tuned to reach the alumina surface, and spaced every 10~mm in a square grid pattern (Fig.~\ref{fig:alumina_ar}).

\subsection{Nylon IR blocking filters}
\label{sec:nylon_filters}

Following the same machining and coating approach in \bicep2/\keck~\citep{BKII}, two Nylon IR blocking filters are placed in the receiver.
One is behind the aperture stop; the other is behind the 
field lens, above the focal plane (FPU) assembly, both at 4~K.
Nylon strongly absorbs far-infrared radiation \citep{1986Halpern} and thus reduces radiated power from 50~K from reaching the 280~mK focal plane.

\subsection{Metal mesh low-pass edge filters}
\label{sec:ade_filters}

A set of metal mesh low-pass edge filters \citep{ade2006} with a cutoff at 4~cm$^{-1}$ were used to control any out-of-band response in the detectors.
They are made from multiple polypropylene substrate layers, each coated with copper grids in different sizes, and hot-pressed together to form a resonant filter.

Prior to the 2017 season, these filters were cut into 76~mm~$\times$~76~mm squares and independently mounted onto each detector module (\S\ref{sec:focal_plane}) at 280~mK.
We found anomalous detector spectral responses in the 2016 FTS measurements described in \S\ref{sec:FTS}.
Upon examining the filters at the end of the 2016 season, we found the layers had delaminated.
It was determined that the cause of delamination was likely insufficient oven temperature during fabrication.
Furthermore, the cutting of individual, smaller filters introduced extra stress on the edge contributing to the delamination.

New filters were fabricated using a higher oven temperature in the fusing process.
The filter design was modified to a larger $\sim$23~cm$\times$15~cm size covering 5 detector modules.
This change reduced mechanical stress at the filter edges caused by the dicing process.
Extra spring loaded washers and widened mounting slots allowed the filter to slide more freely during thermal contraction.
These modifications were done at the end of the 2016 season and subsequent FTS measurements showed no evidence of filter delamination.

\subsection{Optical loading reduction}
\label{sec:loading}

The dominant noise source in \bicepthree\ is the photon noise of the in-band signal power.
For better sensitivity, it is important to minimize the internal non-sky instrument load.

We calculated the total internal loading by measuring the detector responsivity with a flat aluminum sheet mounted just beyond the cryostat window (procedure describe in \S\ref{sec:loadcurve}).
The detector beams were reflected and therefore received power only from within the cryostat.
The co-moving forebaffle loading is measured by differencing of the detector response with and without the forebaffle during clear weather while looking at zenith.

The optical power coupled to the detectors due to each individual optical element is calculated by using the transmission properties of the material and finding the source temperature  distribution.
The calculation incorporates the cumulative optical efficiency from the detector up to the source, and includes the emissivity of the source itself.
For the 2016 design, a simple scattering model is used for the room temperature metal-mesh filters, in which each filter isotropically scatters a small fraction of the radiation to wide angles and warms surfaces around them.
Table~\ref{tab:preddetload} shows the modeled in-band loading estimate for each individual elements and the total measured loading.
The agreement between them validates the model assumption.

A significant contributor to the cryostat internal loading was the scattering of the metal-mesh IR-reflective filters before their replacement in 2017 with the Zotefoam filters.
The reduction of scattered radiation coupling to the filters and telescope forebaffle results in a decrease of the total instrument loading by 30\%.

Non-sky loading also comes from the room temperature HDPE cryostat window, which now dominates the internal power.
We are developing thinner materials that can potentially replace the window in future seasons \citep{spie:denis}.

\begin{table} 
  \caption{
Per-detector in-band optical load.
The total loadings listed in \textbf{bold} are direct measurements from detector load curves, which are in good agreement with the individual modeled optical elements.
A stack of 300~K metal-mesh filters used in the 2016 season were replaced by
HD-30 foam filters for the 2017 season.
}
  \label{tab:preddetload}
  \centering
  \begin{tabular}{l c c} 
    \toprule 
    Source   &    Load [pW] & $T_\mathrm{RJ}$ [K]\\
    \toprule 
    4K lenses \& elements       &   0.15    & \p{0}1.0  \\
    50K alumina filter          &   0.12    & \p{0}0.9  \\
    Metal-mesh filters (2016)   &   0.63    & \p{0}5.2  \\
    HD-30 foam filters (2017+)  &   0.10    & \p{0}0.8  \\
    Window                      &   0.69    & \p{0}5.9  \\
    \midrule
    Total cryostat internal (2016)  &   \textbf{1.60} &   \textbf{13.0} \\
    Total cryostat internal (2017+) &   \textbf{1.10} &   \textbf{\p{0}8.6}\\
    \midrule
    Forebaffle (2016)               &   \textbf{0.31}   &   \textbf{\p{0}2.7}   \\
    Forebaffle (2017+)               &   \textbf{0.14}   &   \textbf{\p{0}1.1}   \\
    Atmosphere                 &  1.10 & \p{0}9.9 \\
    CMB                        &  0.12     & \p{0}1.1 \\
    \midrule
    Total (2016)                &   \textbf{3.13} & \textbf{27}   \\
    Total (2017+)                &   \textbf{2.46} & \textbf{21}   \\
    \bottomrule
    \vspace{1mm}
  \end{tabular}
\end{table}

\section{Cryostat receiver}
\label{sec:receiver}

\subsection{Overview}
\begin{figure*}[t]
  \centering
  \includegraphics[width=0.85\textwidth]{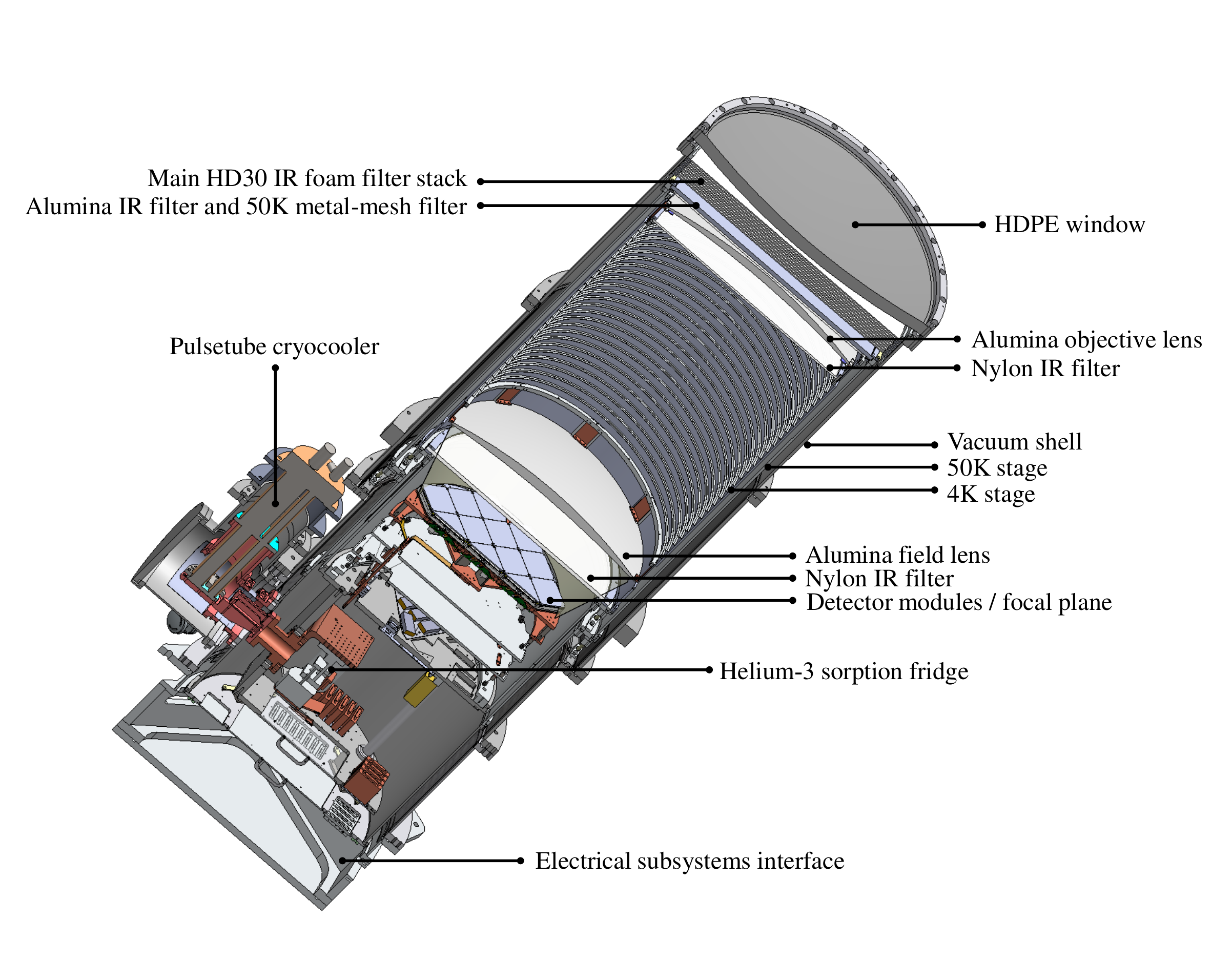}
  \caption{
Cutaway view of the \bicepthree\ cryogenic receiver.
The thermal architecture is separated into a two-stage pulse tube cryocooler
(50~K, 4~K stages) and a three-stage helium sorption fridge
(2~K, 350~mK, 250~mK stages).
All thermal stages are mechanically supported by sets of carbon fiber and G-10 fiberglass support.
The focal plane, with 20 detector modules and 2400 detectors, is located at the 250~mK stage, surrounded by multiple layer of RF and magnetic shielding.
}
  \label{fig:b3cutaway}
\end{figure*}

The cryostat receiver is a compact, cylindrical design that allows for a large optical path while maintaining sub-Kelvin focal plane temperatures (Fig.~\ref{fig:b3cutaway}).
The fully populated receiver weighs about 540~kg without the attached electronics subsystems.
The outermost aluminum vacuum jacket is 2.4~m tall along the optical axis and 73~cm in diameter, excluding the pulse tube cryocooler extension.
It maintains high vacuum for thermal isolation and is capped at one end by the HDPE plastic window, as described in \S\ref{optics}.

The wide-field refractor design allows for ground-based characterization in the optical far field.
The optical design further allows the use of a co-moving, absorptive forebaffle (\S\ref{sec:baffle}) that terminates stray light and wide angle response from the receiver.
Cooling most of the optical elements, including the internal baffling between the lenses, to less than 4~K reduces the thermal photon noise seen by the detectors, maximizing the sensitivity of the instrument.

\subsection{Cryogenic and thermal architecture}

Nested within the room-temperature vacuum jacket are the 50~K and 4~K stages, each comprised of cylindrical aluminum radiation shields and cooled by the 1\textsuperscript{st} and 2\textsuperscript{nd} stages of the PT-415 pulse tube cryocooler\footnote{Cryomech Inc., Syracuse, NY 13211, USA (\texttt{www.cryomech.com})}, which provides continuous cooling to 35~K at the `50~K stage' under typical 26~W load and 3.3~K at `4~K stage' under 0.5~W load.
The stages are mechanically supported off each other and the vacuum jacket by low thermal-conductivity, G-10 fiberglass.
Multi-layer insulation (MLI) wrapped around radiation shields minimizes radiative heat transfer between the 300-50-4~K stages.

\begin{figure*}
  \centering
  \includegraphics[width=0.9\textwidth]{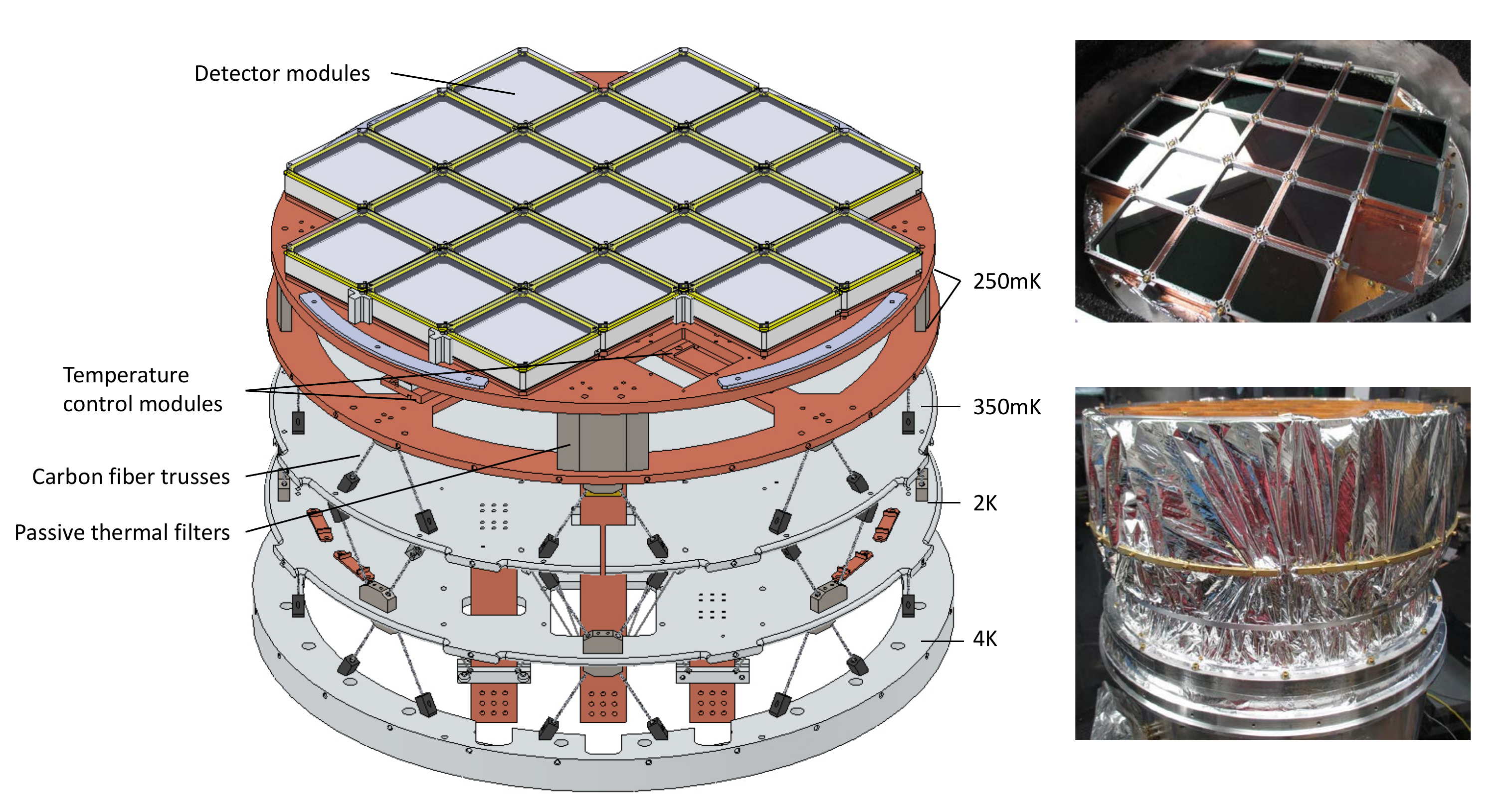}
  \caption{
\textit{Left}: Exploded view of the \bicepthree\ sub-Kelvin stages.
Each temperature stage is mechanically supported by sets of carbon fiber trusses.
Sets of stainless steel supports connect the two 250~mK copper plates,
passively low-pass filtering thermal fluctuations,
and two active temperature control modules maintain thermal stability over observation cycles.
\textit{Right, top}: The assembled focal plane with 20 detector modules
installed into the 250~mK stage without metal-mesh edge filters.
The empty module slot in the lower right is absent due to the capacity of the readout electronics.
\textit{Right, bottom}: A thin aluminized Mylar shroud extends from the top of the
focal plane assembly to the bottom of the 4~K plate to close the 4~K Faraday cage.
}
  \label{fig:b3_focal_plane}
\end{figure*}

A non-continuous, three-stage (\textsuperscript{4}He/\textsuperscript{3}He/\textsuperscript{3}He) helium sorption fridge from Chase Research Cryogenic\footnote{Chase Research Cryogenics Ltd., Sheffield, S10 5DL, UK (\texttt{www.chasecryogenics.com})} is heat sunk to the 4~K stage and cools the sub-Kelvin focal plane and supporting structures.
The focal plane and ultra-cold (UC 250~mK) stage is a planar copper assembly mounted in a vertical stack on two buffer stages, the inter-cooler \textsuperscript{3}He (IC 350~mK) and \textsuperscript{4}He (2K) stages, each supported and isolated by carbon fiber trusses (Fig.~\ref{fig:b3_focal_plane}).
The UC stage cools a 9~mm thick, 46~cm diameter focal plane plate that supports the detector modules and a thinner secondary plate.
These plates are made from gold-plated, oxygen-free high thermal conductivity (OFHC) copper.
The secondary plate and the focal plane are separated by seven 5~cm tall stainless steel blocks that serve as passive low-pass thermal filters to dampen thermal fluctuations to the focal plane. 
The focal plane and the UC stage are actively temperature controlled in a feedback loop to 274~mK and 269~mK, respectively, using Neutron transmutation doped (NTD) Germanium thermometers and a resistive heater.
Thermal fluctuations on the focal plane during CMB observation are controlled to $<0.1$~mK.

\subsection{Thermal performance}
\label{sec:thermal}

The sum of all incident thermal power on the 50~K and 4~K stages determines the temperature profile of the elements along each stage and the base operating temperature of the pulse tube.

The room-temperature HDPE plastic window emits $\sim$110~W of power into the receiver while the pulse tube cryocooler is rated for less than 40~W on the 50~K stage.
We employed two different types of thermal filters at 300~K mounted just behind the cryostat window to reject the majority of the IR load:
(1) a stack of thin film, IR-reflective, capacitive metal-mesh filters in 2016; and
(2) a stack of Zotefoam filters starting in 2017. 
The reason for switching the design is discussed in \S\ref{sec:foam_filters}.

An alumina filter is heatsunk to the 50~K stage, to provide absorptive IR filtering due to Alumina's mid-infrared absorption and high thermal conductivity.
This replaced the filters used in previous telescopes.
Two additional nylon filters are placed in the 4~K stage of the receiver to reduce thermal loading on the sub-Kelvin focal plane by absorbing infrared radiation. 
Table~\ref{tab:IRload} shows the final temperature and power deposited onto each cryogenic stage.
Switching from metal mesh filters to Zotefoam filters in 2017 reduced the thermal loading and improved the cryogenic hold time of the sub-Kelvin fridge from $\sim50$ to $>80$~hours, with 6~hours of recycling time.
This permits the continuous three-day observation schedule shown in \S\ref{sec:schedule}.

\begin{table}
  \centering
  \caption{Measured final temperature and thermal loading on each temperature stage in \bicepthree.}
  \label{tab:IRload}
  \begin{tabular}{l c c}
    \toprule 
     & \multicolumn{1}{c}{2016} & \multicolumn{1}{c}{2017+} \\
    Stages & Temp/Load & Temp/Load \\
    \toprule
    50~K tube top    & 58~K & 53~K \\
    50~K tube bottom & 52~K & 49~K \\
    50~K tube loading & 19~W & 13~W \\
    \midrule
    4~K tube top    & 4.96~K & 4.68~K \\
    4~K tube bottom & 4.58~K & 4.33~K \\
    4~K tube loading & 0.18~W & 0.15~W \\
    \midrule
	350~mK stage      & 354~mK & 352~mK \\
    250~mK stage      & 245~mK & 244~mK \\
    Focal Plane & 268~mK & 268~mK\\
    \bottomrule
    \vspace{1mm}
  \end{tabular}

\end{table}

\subsection{Cryogenic thermal monitoring and control}

For general thermometry down to 4~K, we use silicon diode thermometers (Lakeshore\footnote{Lake Shore Cryotronics, Westerville, OH 43082, USA (\texttt{www.lakeshore.com})} DT-670), with thin-film resistance temperature detectors (Lakeshore Cernox RTDs) on the sub-K stages.
NTD Germanium thermometers are integrated in the secondary UC stage, copper focal plane, and each detector module for more sensitive measurements of the temperatures.
The NTDs on the secondary UC stage and focal plane are packaged with a heater to provide active temperature control on their respective temperature stages.

Thermal operations are controlled by a custom-built system similar to the one used in \bicep2/\keck.
It contains electronic cards used to bias and read out thermometers, control heaters and provide temperature control servos, and is mounted directly to the cryostat vacuum jacket and interfaces with MicroD (MDM) connectors\footnote{Glenair Inc., Glendale, CA 91201, USA (\texttt{www.glenair.com})} at the cryostat.
Signals from the system are routed to the rack-mounted BLASTbus ADC system \citep{thesis:wiebe} next to the telescope which generates the AC bias used for the resistive thermometers and the NTDs, and demodulates the thermometer signals, which are then digitized at $\sim$100~Hz.

\subsection{Radio frequency shielding}

Several levels of radio frequency (RF) shielding are designed into the 4~K stage and sub-Kelvin structures to minimize RF coupling to the detectors. 
All cabling inside the cryostat uses twisted pairs, except for the short lengths of flex ribbon cable connecting the detector modules to the focal plane readout circuit board.
These ribbon cables are shielded by the detector module, copper focal plane module cutout, and the ground plane of the circuit board that accepts the cable.
The 4~K non-optics volume is designed as a Faraday enclosure, with all seams taped with conductive aluminum tape and cabling passing through inductive-capacitive PI-filtered connectors\footnote{Cristek Inc., Anaheim, CA 92807, USA (\texttt{www.cristek.com})}.
The cage is continued to enclose the stack of sub-Kelvin stages by wrapping and sealing a single layer of aluminized mylar between the 4~K stage and the edge of the focal plane.
The niobium enclosure of each detector module and detector tile ground plane close the sky side of the Faraday cage.
Upon exiting the cryostat, all of the detector signal lines immediately interface with a capacitive filtered connection on the readout electronics box that is directly mounted on the cryostat.

During the 2015 engineering season, we found an azimuth-synchronous signal strongly affecting the detectors, largely common-mode across a large fraction of detectors within each readout system.
These interference showed variation 1000 times larger than the $50\mu$K CMB temperature variations,
causing `SQUID jumps' because of the strong signals.
Our detector readout scheme works through feedback to maintain linearity in the SQUID amplification curve (\S\ref{subsec:readout}), but large current variations can disrupt the feedback and cause the readout to jump to a different part of the SQUID curve.
We discovered that this interference signal was caused by radio-frequency emission from the South Pole station land mobile radio (LMR) system at 450MHz, coupling into the cryostat and detectors through the cryostat window.
\bicepthree\ is inherently more susceptible to this 450~MHz signal than \keck\ due to its larger aperture, which has a cutoff frequency at 340MHz at the optics cylinder.

Prior to the 2016 season, we applied silver loaded paste between the detector modules and copper focal plane, so that reliable electrical conductivity was maintained from the modules to the focal plane.
In 2015, only 9 out of 20 detector modules were filled, leaving large gaps at the top of the focal plane.
In 2016, having the full population of 20 detector modules provided a better RF shielded enclosure.
After implementing the improved internal cryostat shielding, RF susceptibility in the range of 400--500MHz was reduced by 10~dB.
In addition, the LMR antenna was changed to a directional sector antenna with reduced power output towards the telescope.
Attenuators were also installed to reduce the overall broadcast power, which was tested to be much more powerful than necessary to maintain radio communication across the base.
In total, the LMR source power seen at the telescope was reduced by 35~dB.
Azimuth scanning tests conducted after these changes have shown none of the visible structure seen in 2015.

\subsection{Magnetic shielding}

Earth's magnetic field ($\sim50\mu$T) is the most dominant variable magnetic environment.
While this azimuth-fixed signal is largely filtered out during analysis, instrumental magnetic shielding is crucial to minimize coupling to the TES detectors and the SQUID amplifiers.
We incorporate two methods of shielding in the cryostat.
First we use a cylindrical, high-permeability Amumetal-4K (A4K)\footnote{Amuneal Manufacturing Corp., (\texttt{www.amuneal.com/})} structure, with open ends to avoid interference with the optics and allow data cabling through the bottom.
It is a split shield on the inner surface of the vacuum jacket of the cryostat.
The two halves overlap midway, near the focal plane level given the constraints of the cryostat and optics.
Additionally, a shorter, superconducting Niobium cylinder is mounted on the 4~K stage, surrounding the focal plane.
Lab tests showed a shielding factor of $\sim$~30 for the magnetic field amplitude along the cylindrical axis of the cryostat.

The detector module, which includes layers of niobium, aluminum and high-$\mu$ A4K (\S\ref{modular}), provides further shielding of the first-stage SQUID amplifier chips on the sub-Kelvin focal plane.
The series SQUID array (SSA) on the 4~K stage are packaged in niobium boxes and additionally wrapped with 10 layers of high-$\mu$ Metglas 2714A\footnote{Metglas Inc., (\texttt{www.metglas.com})}.
Overall, \bicepthree\ shows an induced response $\sim$~7~$\mu\text{K}_\text{CMB}/\mu$T, or 350~$\mu\text{K}_\text{CMB}/B_{Earth}$ by directly measuring the SQUID amplifier response to a Helmholtz coil.

\begin{figure}
  \centering
  \includegraphics{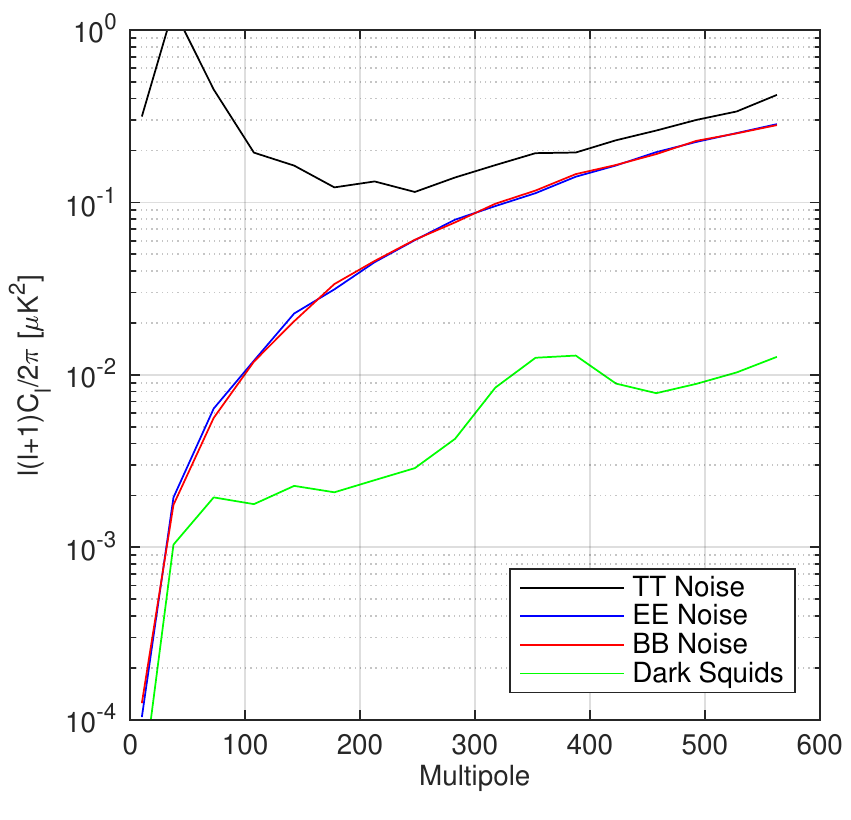}
  \caption{
Temperature (black) and polarization (red and blue) noise of \bicepthree.
The dark squids (green) is a readout channel that is disconnected from the detector, but sensitive to external magnetic field. This comparison demonstrates the magnetic pickup in science data is subdominant.
}
  \label{fig:b3_DS}
\end{figure}

Another method to estimate impact on external magnetic field to the data is using the dark SQUID channels in the readout system (\S\ref{subsec:readout}).
These channels are connected to the SQUID amplifiers, but not the detectors, hence ideally only respond to external magnetic fields as the telescope scans.
Lab measurements show the properties and calibrations of these dark SQUID channels are similar to the other SQUID channels in the same SQUID chip.
Then using the neighboring optical channel calibrations and telescope pointing, we constructed `maps' of these dark SQUIDs.
Comparing these maps and the associate spectrum with single optical channel maps, we found the dark SQUID responses are factor of 2 to 20 smaller than the Q/U noise of the same bins (Fig.~\ref{fig:b3_DS}).

\section{Focal Plane}
\label{sec:focal_plane}

\begin{figure*}
  \centering
  \includegraphics[width=0.4\textwidth]{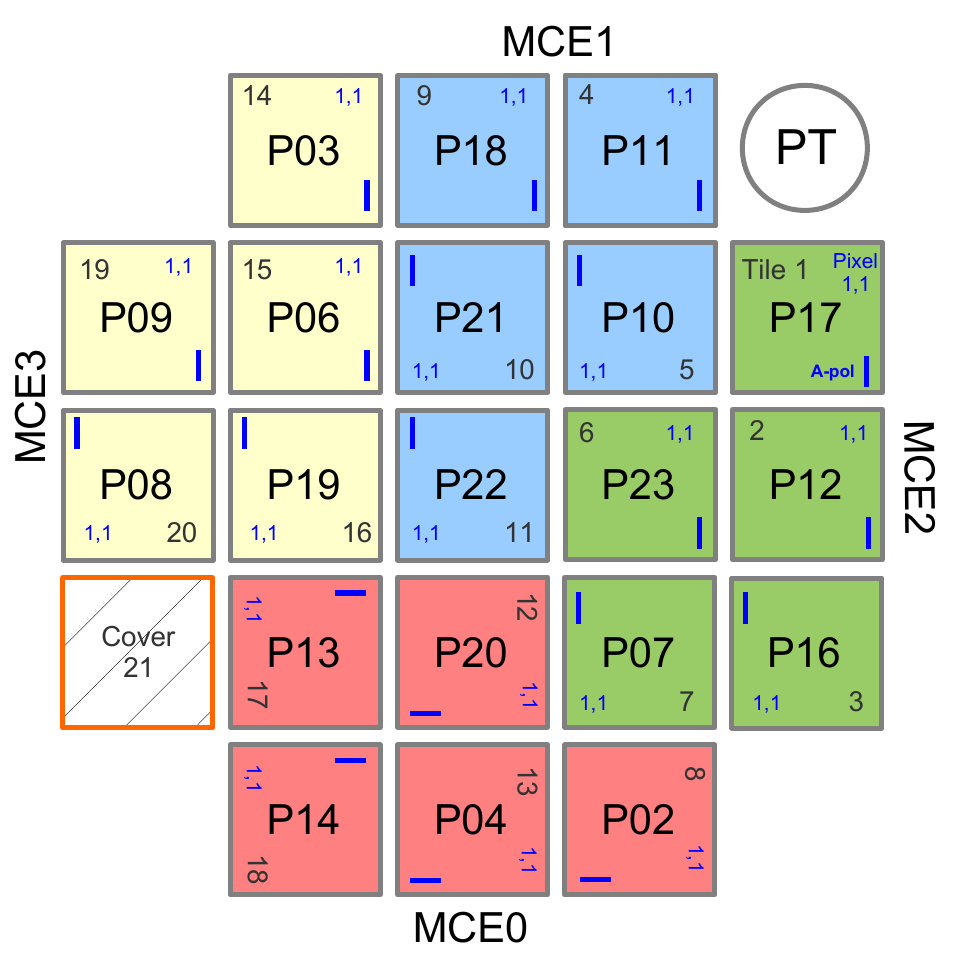}
  \includegraphics[width=0.4\textwidth]{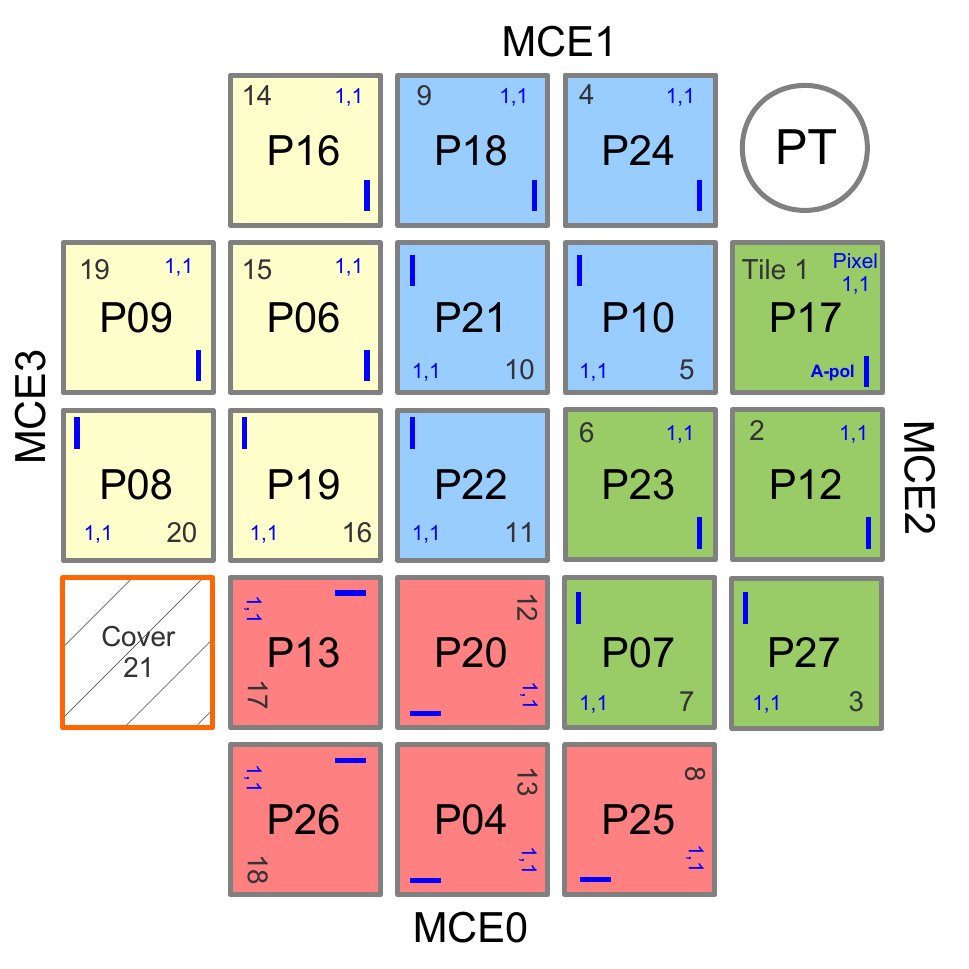}
  \caption{
\textit{Left}: \bicepthree\ focal plane layout in 2016.
\textit{Right}: 4 modules were replaced prior to the 2017 season.
In each detector module, the serial number \textit{P}xx is labeled in the center.
The orientation of the module is indicated by the pixel (1,1) and the polarization A direction in blue, along with the tile number 1-20.
The background color of the module indicates the 4 readout MCE units.
The location of the pulse-tube cooler is shown by the PT marking in the upper right.
Slot 21, shown in an orange outline, does not contain a detector module, since the readout electronics were designed to support 20 modules. 
It was covered with a thin copper sheet, which created extra reflection for Tile 1 as discussed in \S\ref{subsec:FSL}.
}
  \label{fig:fpu_layout}
\end{figure*}

\bicepthree\ has 2400 detectors, a factor of 9 greater than a single \keck\ receiver
at the same frequency.
The detectors are fabricated on 20 silicon wafers, each consisting of 60 dual-polarized detector pairs.
Each wafer is packaged into a focal plane module with its cold readout
electronics and installed onto the copper focal plane to form the \bicepthree\ focal plane (Fig.~\ref{fig:fpu_layout}).
The focal plane base plate provides the necessary thermal stability, magnetic shielding, and mechanical alignment to operate the detectors.
The modular design allows individual detector tiles to be replaced with minimum impact to the rest of the receiver.

\subsection{Modular packaging}
\label{modular}

\begin{figure}
  \centering
  \includegraphics[width=0.45\textwidth]{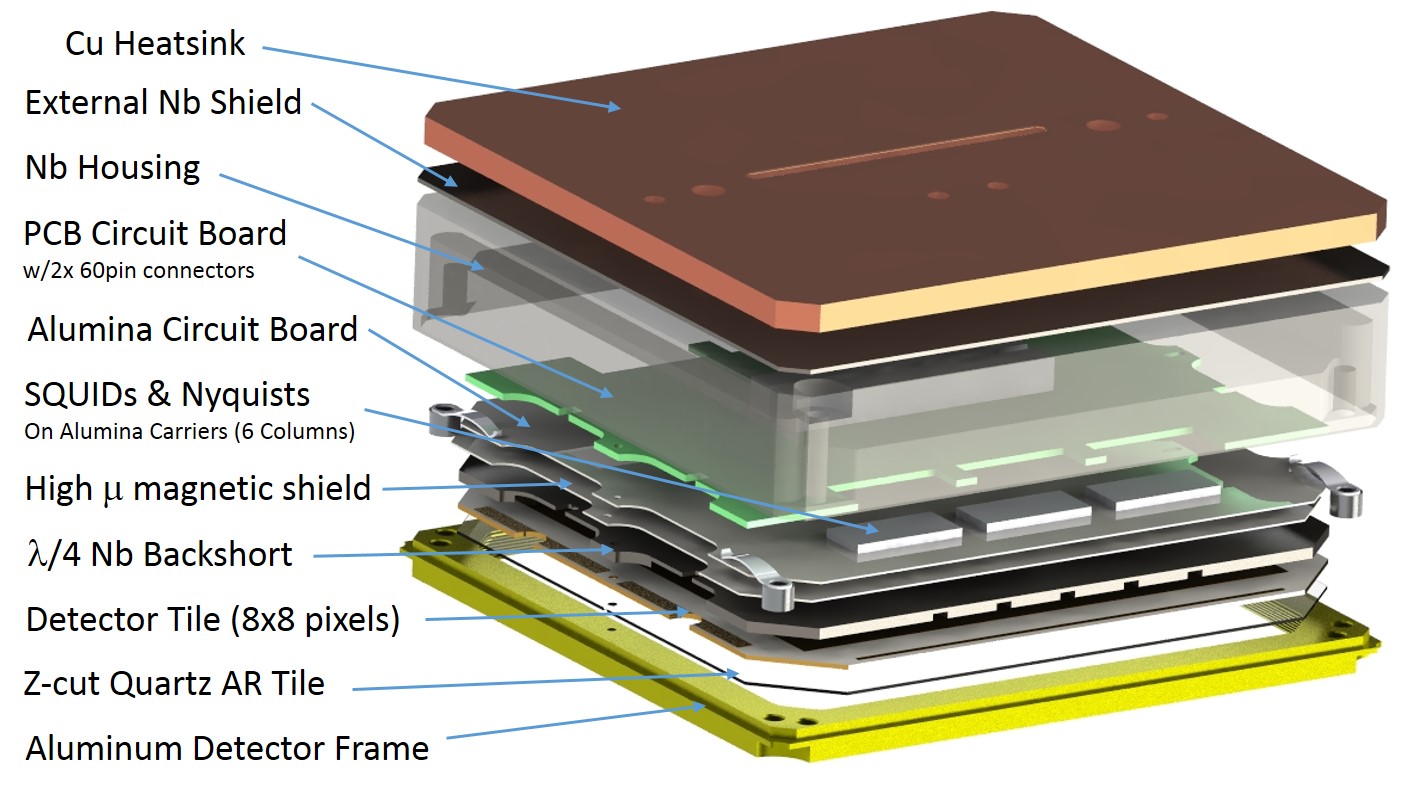}
  \caption{
Exploded view of the \bicepthree\ detector module.
Sky-side is facing downward in this diagram.
The multiplexing SQUIDs and circuit boards are mounted directly behind the detector wafer, separated by a $\lambda /4$ Nb backshort and A4K magnetic shield.
The backside is enclosed by a Nb cover and plate for magnetic shielding.
}
  \label{fig:detmodule}
\end{figure}

Each detector module is $79\times79\times22$~mm in size (Fig.~\ref{fig:detmodule}).
Two 60-pin, 0.5~mm pitch flex ribbon cables connect between each module and the focal plane circuit board, via a pair of zero-insertion force (ZIF) surface-mount connectors.
The module mounts on the focal plane on all four corners.

The detector module consists of a quartz anti-reflection coating, detector tile, niobium (Nb) $\lambda/4$ backshort, A4K magnetic shield, and alumina and PCB readout circuit boards.
These sub-components are stacked together on the aluminum detector frame, secured at the corners with commercially available copper clips\footnote{Ted Pella Inc., (\texttt{www.tedpella.com})}, and aligned with a 2~mm diameter copper pin-slot pairs located on opposing edges of the wafer.
The module is enclosed with a niobium housing, an external niobium magnetic shield covering the ribbon cables, and a copper heatsink.

\begin{figure}
  \centering
  \includegraphics[width=0.4\textwidth]{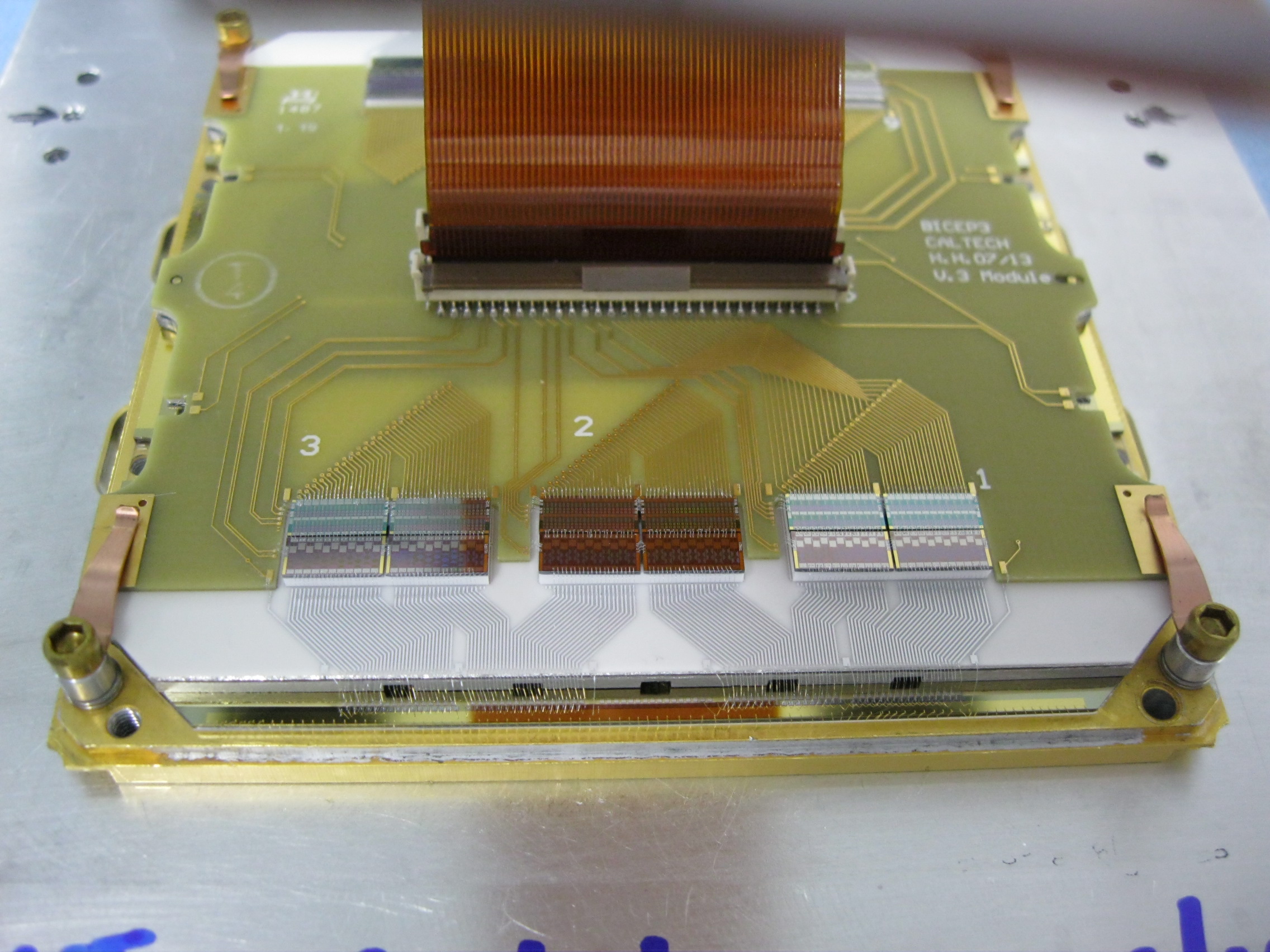}
  \caption{
Backside of the detector module.
Aluminum wirebonds connect the detectors to SQUIDs chips via an alumina circuit board.
Two 60-pin Kapton/Cu flex-circuit ribbon cables connect to the ZIF connectors and travel out of the niobium casing through a thin slot to matching connectors on the focal plane board. 
}
  \label{fig:module_mux}
\end{figure}

The readout circuit boards are composed of an alumina and a FR4 printed circuit board.
The 0.25~mm thick alumina circuit board has 0.13~mm wide, 0.23~mm pitch aluminum traces, creating a superconducting path between the detectors and the SQUID chips.
Twelve readout SQUID and Nyquist chips are mounted onto individual alumina carriers which are glued onto the alumina circuit board.
On top of the alumina board is a two-layer FR4 circuit board with standard copper traces.
These components are electrically connected to each other with aluminum wire bonds, and two 60-pins surface mount connectors are soldered on top of the FR4 circuit board for the flex cables shown in Fig.~\ref{fig:module_mux}.

\subsection{Thermal sinking and magnetic shielding}

The detector tiles are thermally sunk to the aluminum frame on all four sides with $\sim$~500, 0.5~mm pitch gold wirebonds.
Additionally, these wirebonds form the top RF shield of the system by connecting the aluminum frame and detector Nb ground plane.

The detector module is mounted to a copper heat-sink at the back of the niobium housing.
It is supported by 3 thermally isolated alumina spacers in the corners, making the center-most point the only point of thermal contact between the copper and niobium.
This single contact point cools the module housing from the middle, ensuring that the niobium superconducting transition begins from the center and continues radially outwards, avoiding trapped magnetic flux during cool-down.

\begin{figure}
  \centering
  \includegraphics[width=0.45\textwidth]{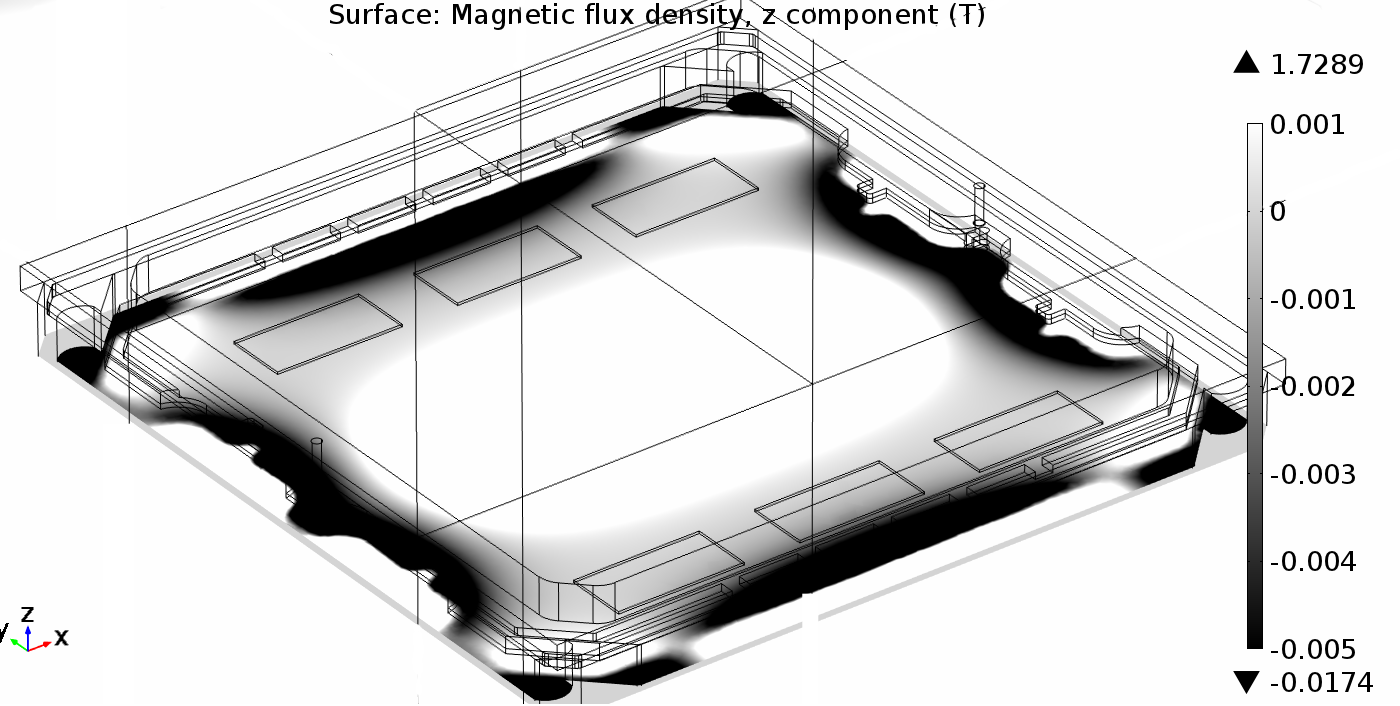}
  \caption{
COMSOL Multiphysics simulation of the detector module, the scale is saturated to highlight the location of the SQUIDs (MUX).
Simulation shows the external magnetic field is reduced to $\sim$0.1\% at the SQUIDs.
}
  \label{fig:magmodule}
\end{figure}

The backside of the module is completely enclosed with a Nb housing, with only a $35$~mm~$\times 1.3$~mm slit to allow the flex cables to exit the module.
A 0.5~mm-thick Nb sheet with an offset slit is placed at the back of the housing, creating a near continuous superconducting magnetic shield. 
Additionally, a sheet of high-permeability Metglas 2714A is placed 1.27~mm away from the SQUID chips inside the module to create the lowest magnetic field environment at the location of the SQUID chips (Fig.~\ref{fig:magmodule}).

\subsection{Corrugated frame}
\label{sec:corrugation}

The interaction between the detector module metal frame and the edge-adjacent planar slot antenna causes differential pointing within that detector pair.
Although most of the systematic errors caused by the differential pointing are mitigated during analysis, we corrugated the frame to minimize residual beam systematics of these pixels.

\begin{figure}
  \centering
  \includegraphics{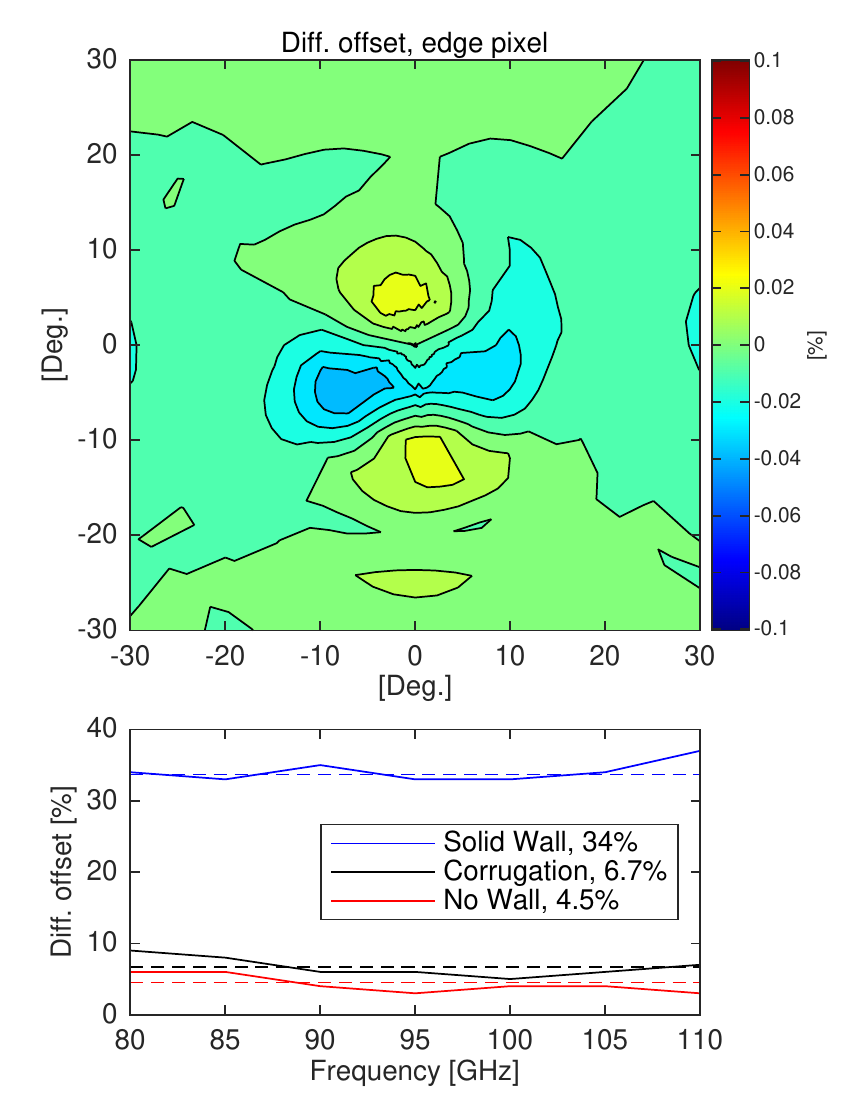}
  \caption{
\textbf{Top:} Simulated peak normalized differential beam map for an edge pixel closest to the corrugation frame over 25\% bandwidth at 95~GHz.
\textbf{Bottom:} Peak-to-peak differential beam amplitude over design frequency band.
Solid lines are the simulated value, and dash lines are the average value over the 25\% bandwidth.
Red lines show the case without metal frame, black lines show the current corrugated frame design, and blue lines show the pixel next to a solid metal wall.
This modeling is confirmed by our measurements shown in the middle panel of Fig.~\ref{fig:nfbm_mismatch}.
}
  \label{fig:corrugation_fig}
\end{figure}

The corrugated frame has $\lambda/4$ depth and pitch, and is placed $\lambda/2$ away from the closest antenna, where $\lambda$ is the design band center~\citep{Soliman_SPIE_2018}.
Fig.~\ref{fig:corrugation_fig} presents a differential beam map model using CST Studio Suite\footnote{Dassault Systemes (\texttt{www.3ds.com})}, showing a reduction in residual beam mismatch from $\sim$~34\% with a flat frame to $\sim6.7\%$ with the corrugated frame, evaluated over the 25\% spectral bandwidth.

\section{Detectors}
\label{sec:detectors}

\bicepthree\ inherits the planar phased-array antenna and transition-edge sensor (TES) bolometer detector technology from \bicep2/\keck\ \citep{dets2015}. 
The planar antenna design does not require feed horns or similar coupling optics to free space.
The 95~GHz band is defined by lumped-element filters along the microstrip feed line to the bolometers.
Each silicon detector tile contains an 8$\times$8 array of pixels and each pixel is made of two co-located, orthogonally-polarized sub-antenna networks and two TES bolometers.

Holding the edge taper on the pupil fixed, \bicepthree's faster focal ratio enables a denser detector packing than the 6$\times$6 array of pixels in each 95~GHz \keck\ detector tile.

The TES detectors are voltage-biased, providing electro-thermal stability and linearity, where electrical power compensates for variation in optical power.
Each TES is made up of a titanium and aluminum film in series; the titanium transition maximizes sensitivity ($T_c\approx0.5$~K) during normal science observation, while the higher $T_c\approx1.2$~K aluminum TES provides higher saturation power to observe high-temperature calibration sources, though at reduced sensitivity.

Together, the two independent signals from the co-located orthogonal antenna and bolometer pairs on each pixel can be summed to get total incident power or differenced to measure polarization in the vertical-horizontal direction.

\subsection{Tapered antenna networks}

Optical radiation couples to the detectors through a planar phased antenna array, combined in phase with a summing network that controls the amplitude and phase in each sub-slot.
The illumination pattern is controlled through the microstrip feed network that sums signals from the sub-antennas to deliver power to the TES bolometer.
Previous designs, used in \bicep2/\keck\ and \spider, drive each of the sub-antennas in phase with equal field strength, synthesizing a top-hat illumination and thus a sinc pattern in the far field.
Such a pattern has side-lobes with peak levels at -13~dB below the main lobe.
In these instruments, side-lobes are terminated onto the 4~K aperture stop with limited impact on the sensitivity.

Programmatically, some optical designs would benefit from lower side-lobe levels and \bicepthree\ was used to advance this capability.
The side-lobe levels of antenna arrays can be controlled by tapering the illumination such that the central sub-slots have higher coupling than those at the edge.
The array factor with non-uniform illumination can be generalized as

\begin{eqnarray}
 A(\theta) &=&\sum_{m=-(N-1)/2}^{(N-1)/2} E_m e^{-j mk\sin(\theta)s}  \nonumber \\
 &\simeq&\int_S E(x',y')e^{j(k_x x'+ k_y y')} \,dx'dy' 
 \label{taper_dist}
\end{eqnarray}
where the last line approximates the sum as an integral across the antenna aperture, and
$k_x$ and $k_y$ are the components of the tangential free-space wavevector~$k\sin(\theta)$.
This expresses the far-field antenna pattern as the Fourier transform of the illumination pattern.
For \bicepthree, the feed network was designed to generate a Gaussian illumination with an electric field waist radius of 6.3~mm, compared to the physical aperture size of 7.5 $\times$ 7.5~mm.
This reduces the side-lobe levels to -16~dB and the integrated spillover to 13\% compared to the 17\%  that would have been achieved with a uniform feed.
The result is an illumination that is close to uniform, as the instrument's aperture stop requires, but also allows our team to develop flexibility for other instruments.
We also developed and tested two designs with stronger tapering, which would be advantageous in optical systems with a warm pupil stop.

\section{Data Acquisition system}
\label{sec:daq}

\bicepthree\ uses a SQUID time-division multiplexed (TDM) system for the bolometer readout, which includes the SQUID multiplexing chips, the room temperature Multi-Channel Electronic (MCE) system and the overall Generic Control Program (GCP).

\subsection{Time-division multiplexing SQUID readout}
\label{subsec:readout}

\bicepthree\ uses a low-noise SQUID TDM readout system, which amplifies the small current flowing through the TES while adding noise sub-dominant to the detector itself, and transforms the small $\sim 60$~m$\Omega$ impedance of the TES in two amplification stages.
The TDM architecture is similar to previous experiments~\citep{dekorte2003} but uses a new generation MUX11 model that takes advantage of superconducting-to-resistive, flux-activated switches (FS) in the multiplexing sequence \citep{irwin2011}.
The SQUID amplifier chips were developed and fabricated by National Institute of Standards and Technology (NIST), and the operating firmware and characterization were first demonstrated in \bicepthree\ \citep{spie:Hui_2016}.
Since then the technology has been deployed in Advanced ACTPol~\citep{Henderson_2016}, CLASS~\citep{Dahal_2020} and \biceparray~\citep{spie:Hui_2018}.

In the readout architecture each independent detector is inductively coupled to a single SQUID array (SQ1) by an input coil, and the amplifier is operated in a flux-lock loop to linearize the output and increase the dynamic range of the periodic SQUID response.
As the flux from the input coil changes in response to the TES current, a compensating flux is applied by the feedback coil to cancel it.
This flux feedback serves as the signal output of the TES.
The second stage SQUID array (SSA) at 4~K provides additional amplification that impedance matches to the room temperature MCE, providing $\sim1\Omega$ dynamic resistance and a $\sim100\Omega$ output impedance.

\begin{figure}[t]
  \centering
  \includegraphics[width=0.45\textwidth]{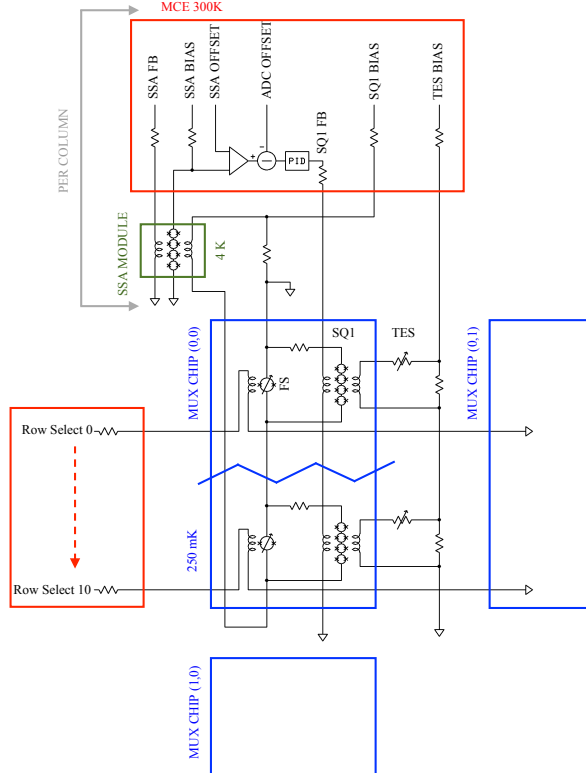}
  \caption{
MUX11 SQUID readout and feedback schematic modified from Fig.~C.1 in \cite{thesis:grayson}, explicitly showing two RS of one MUX column.
Red outlines denote MCE-sourced signals at 300~K, green the SSA modules at 4~K, and blue the MUX chips at 250~mK.
This diagram does not show the Nyquist chip outline or the Nyquist inductor.
A single bias line is used to bias both the SQ1 and the flux switches (FS).
}
  \label{fig:squids}
\end{figure}

The multiplexing circuit is arranged according to `columns' and `rows' of detectors.
\bicepthree\ uses total of 4 MCE, each MCE readout 660 channels, which are mapped into 30 MUX columns and 22 MUX rows (600 optical channels, 20 dark detectors, and 40 dark SQUID channels.)
Multiplexing is done across rows, such that 22 channels of each column are read out in TDM sequence.
The SQ1 on each column lie in series, and the entire line is resistively shunted by the FS.
The FS are a four-Josephson-junction design that behaves like a SQUID with a critical current about twice the critical current of the SQ1 when no flux is applied to them.
The SQ1 and FS are shared by the same SQ1 bias line; a zero flux applied to an FS leaves it superconducting, while half of a flux quantum applied sends the FS normal with a resistance much greater than SQ1.
Each FS is coupled to an input inductor, which can apply a flux and is driven by a row-selected input line (RS) which runs across columns.
At a given moment in time during the multiplexing sequence, only one RS line applies a half flux quantum to the switches of its row, while the remaining 21 RS lines remain zeroed, resulting the shared SQ1 bias current shorts through those 21 FS. This bypasses those SQ1s and only flows through the single SQ1 whose FS is highly resistive.
The circuit diagram of the TDM system is shown in Fig.~\ref{fig:squids}.

Additionally, the TES bias circuits includes elements of a `Nyquist' chip (NYQ), which consists of a shunt resistor $R_{sh}\sim3$~m$\Omega$ parallel to the TES, and an inductor $L_{\text{NYQ}}$ in series with the TES.
The inductor is included to create a RL filter with the TES resistance, with $L_{\text{NYQ}}\sim2\mu$H, $R_{\text{TES}}\sim50$~m$\Omega$ resulting a roll off at $\sim4$~kHz to avoid aliasing of higher frequency noise.

\subsection{Warm multiplexing hardware}
Control of the MUX system and feedback-based readout of the TES data are done via the room temperature MCE system \citep{mce2008}.

The MCE samples the raw SSA output at 50~MHz, givens a 90~samples each row in the MUX sequence per switch.
This gives a 25.3~kHz visitation rate.
The data are filtered and downsampled in the MCE before being output to the control system.
The MCE uses a fourth-order digital Butterworth filter before downsampling by a factor of 168, then the the final electronic stage applies a second filtering using an acausal,
zero-phase-delay finite impulse response (FIR) low-pass filter, down-sampled by another factor of 5,
giving an archived sample rate of 30.1~Hz.
The full multiplexing parameters used in \bicepthree\ are shown in Table~\ref{tab:muxparam}.

\begin{table}
  \centering
  \caption{Summary of multiplexing readout parameters used by \bicepthree.}
  \begin{tabular}{l l l}
    \toprule
    Raw ADC sample rate          & 50~MHz \\
    Row dwell                    & 90 samples \\
    Row switching rate           & 556~kHz \\
    Number of rows               & 22 \\
    Same-row revisit rate        & 25.3~kHz \\
    Output data rate per channel & 150~Hz \\
    Archived data rate           & 30.1~Hz \\
    \bottomrule
    \vspace{1mm}
  \end{tabular}
  \label{tab:muxparam}
\end{table}

\subsection{Control system}
All of the telescope systems are controlled and read out by a set of six Linux computers running GCP, inherited and modified from \bicep/\keck\ and other CMB experiments \citep{story2012}. 
These control computers interface with all telescope subsystems, including mount movement control, detectors and SQUIDs via the MCEs, and thermometry. 
All data timestreams are packaged into archived files on disk and streamed back with telescope operation logs to North America via daily satellite uplink. 
Observation and fridge-cycle scheduling are scripted within GCP and executed automatically with periodic monitoring by the operator.

\section{Instrument characterization}
\label{sec:inst_char}

\subsection{Detector bands}
\label{sec:FTS}

\bicepthree\ detectors are designed for a frequency band centered at 95~GHz with $\sim 25\%$ fractional bandwidth.
The band is chosen to avoid the broad oxygen absorption band around 60~GHz as well as the oxygen spectral line at 118.8~GHz.

The spectral response of each detector is measured \texttt{in situ} with a custom-built Martin-Puplett Fourier Transform Spectrometer (FTS) mounted above the cryostat window. 
The apparatus and measurement procedure are described in \cite{karkare2014}.
The band center $\langle \nu \rangle$ in frequency $\nu$ is defined as
\begin{eqnarray}
\langle \nu \rangle = \int \nu S\left(\nu\right) d\nu
\end{eqnarray}
where $S\left(\nu\right)$ is the spectral response, and its bandwidth $\Delta \nu$
is defined as 
\begin{eqnarray}
\Delta \nu = \frac{\left(\int S\left(\nu\right) d\nu\right)^2}{\int S^2\left(\nu\right) d\nu}
\end{eqnarray}

The FTS beam is smaller than the pupil, but illuminates several detectors at once in angular extent. This leads to frequency-dependent beam truncation in the measurement, with a spectral shape that depends on the beam size at the aperture, the size of the FTS entrance port, and the nominal detector frequency. 
A correction $\nu^a$ is applied to the final spectra $S(\nu)$, where $a$ is calculated using models from a intensity measurements over the pupil (\S\ref{subsec:nfbm}).
We measured a median band center at $96.1 \pm 1.5$~GHz with median band width at $26.8 \pm 1.3$~GHz, corresponding to a fractional spectral bandwidth of 27~\%.

\begin{figure}
  \centering
  \includegraphics{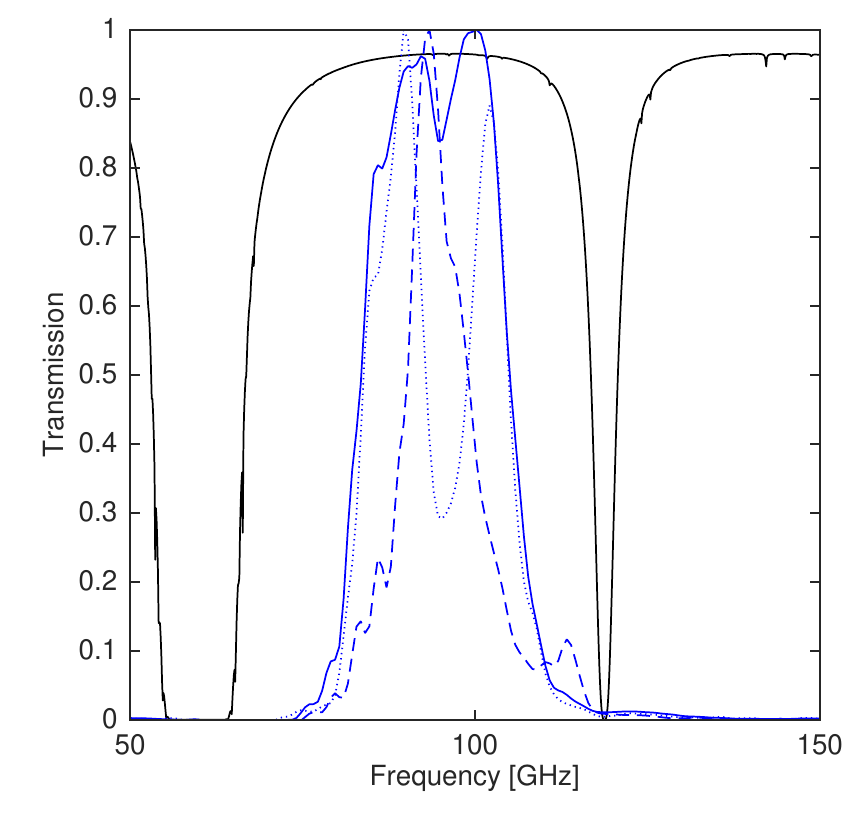}
  \caption{
    The peak-normalized, average spectral response of \bicepthree\ detectors
    (solid blue) shown against the atmospheric transmission at the South Pole (black).
    Also plotted are two extreme example cases of the type of bandpass variation caused
    by delamination of the low-pass edge filters in 2016:
    a spike-like spectrum (dashed blue, $e=-0.51$) and
    a dip-like spectrum (dotted blue, $e=0.20$).
    All low-pass edge filters were replaced for the 2017 season,
    and subsequent measurements are similar to the average spectrum for all detectors.
}
  \label{fig:b3_fts}
\end{figure}

\begin{figure}
  \centering
  \includegraphics{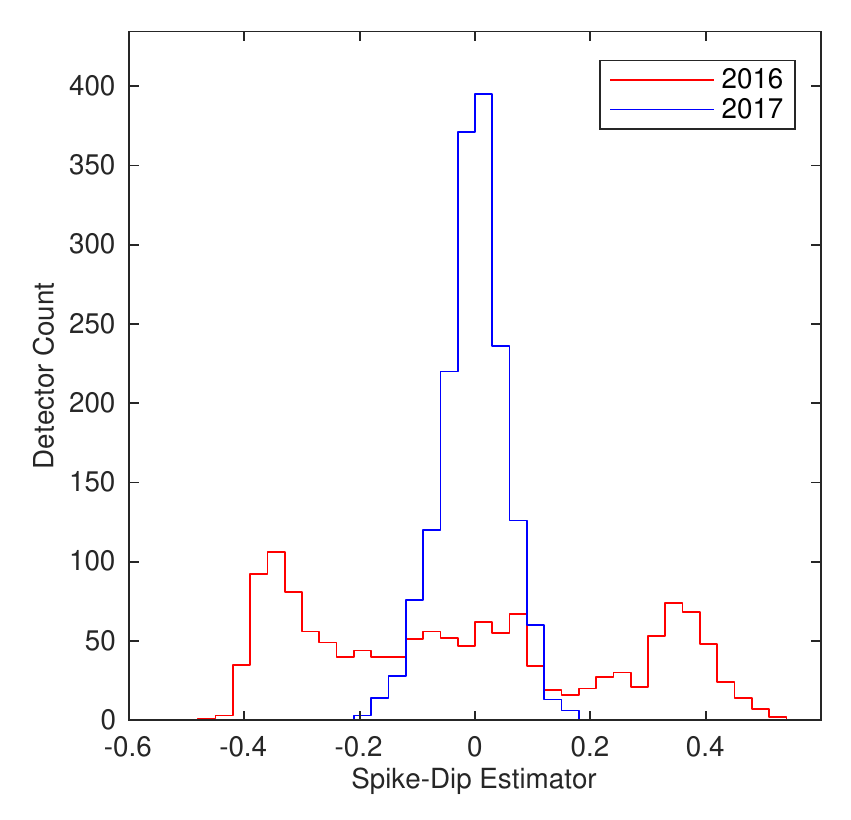}
  \caption{
  The spectral (spike-dip) estimator $e$ with the 2016 and 2017 focal planes, calculated with Eq.~\ref{eq:spec_estimator}.
  The peaks in $e=\pm\sim0.4$ indicate that many detectors in the 2016 focal plane exhibited spike and dip-like features. 
}
  \label{fig:b3_spikedip_tilemap}
\end{figure}

The delamination of the low-pass edge filters described in \S\ref{sec:ade_filters}
resulted in non-uniform spectral features in the detectors during the 2016 season
(Fig.~\ref{fig:b3_fts}) as well as a decrease in optical efficiency.
We broadly observed two types of spectral variations: one where the response
was suppressed at the outer edges of the nominal band which created a ``spike'' shaped
profile, and one where the response was suppressed at the center of the band and
created a ``dip''.
To further characterize this feature, we define an estimator $e$:
\begin{eqnarray}
 \label{eq:spec_estimator}
 e &=& \mean_{\substack{84 <\nu < 90\text{GHz} \\ 98 < \nu < 106\text{GHz}}}\left[ B'(\nu) \right]- \mean_{\substack{90 <\nu < 98\text{GHz}}} \left[ B'(\nu) \right]
\end{eqnarray}
where $B'(\nu) = G(\nu;\sigma) * B(\nu)$.
$B(\nu)$ is the peak-normalized detector spectral response from FTS measurements,
$G(\nu; \sigma)$ is a Gaussian with $\sigma = 1$~GHz,
and $B'(\nu)$ is the detector spectrum convolved with the Gaussian smoothing kernel.
The estimator $e$ takes on values $-1\le e \le 1$, with $e<0$ being spike-like and
$e>0$ being dip-like.
The distribution of estimator values is shown in Fig.~\ref{fig:b3_spikedip_tilemap}.
All the low-pass edge filters were replaced at the end of the 2016 season, and no evidence of delamination has been found since their replacement.
In order to determine the impact of these spectral features on the 2016 CMB data, we developed an additional jackknife test discussed further in \S\ref{sec:jackknifes}.

High-frequency blue leaks originating from direct-island coupling to the TES bolometer are measured using a chopped liquid nitrogen source and a stack of thick grill high-pass filters \citep{Timusk1981}.
These filters are machined metal plates with hex-packed circular holes corresponding to waveguide cutoff frequencies at 120, 170 and 247~GHz.
Measurements showed response to a Raleigh-Jeans source of approximately 0.76~\%, 0.61~\% and 0.55~\% above the 120~GHz, 170~GHz and 247~GHz edge, respectively.

\subsection{Optical efficiency}
\label{sec:loadcurve}

We measured changes in optical power through detector load curves, by applying a high detector bias voltage to first drive the detector normal, then stepping down the bias voltage until the detector is superconducting.
From these load curves, we can measure changes in optical power, assuming the total optical and electrical power is constant.
The power difference is compared against the expected optical loading from an aperture-filling source at a known temperature, giving the end-to-end optical efficiency of the full system.

The change in optical loading is obtained by taking load curves while observing a source at ambient temperature ($\sim 266$~K) and a liquid nitrogen (LN2) source at 74.2~K (at South Pole atmospheric pressure). 
For an aperture-filling, Rayleigh-Jeans source, the optical power $Q_{opt}$ deposited on a single-moded polarization-sensitive detector is
\begin{equation}
    Q_{opt} = \frac{\eta}{2} \int \lambda^2 S\left(\nu\right)B\left(\nu, T\right) d\nu
    \label{eq:opt_eff}
\end{equation}
where $\eta$ is the optical efficiency, $B\left(\nu\right)$ is the Planck
blackbody spectrum at temperature $T$, and $S\left(\nu\right)$ is the detector
spectral response.
In the Rayleigh-Jeans limit ($h\nu \ll kT$), Eq.~\ref{eq:opt_eff} simplifies to
\begin{equation}
    Q_{opt} = kT\eta \int S\left(\nu\right) d\nu = kT\eta \Delta \nu
\end{equation}
where $\Delta \nu$ is the bandwidth and $\eta$ is the optical efficiency of the system.
Observations of the sky and an aluminum mirror redirecting light into the cryostat provide estimates of the atmosphere and the internal cryostat photon load. 

\begin{figure}
  \centering
  \includegraphics{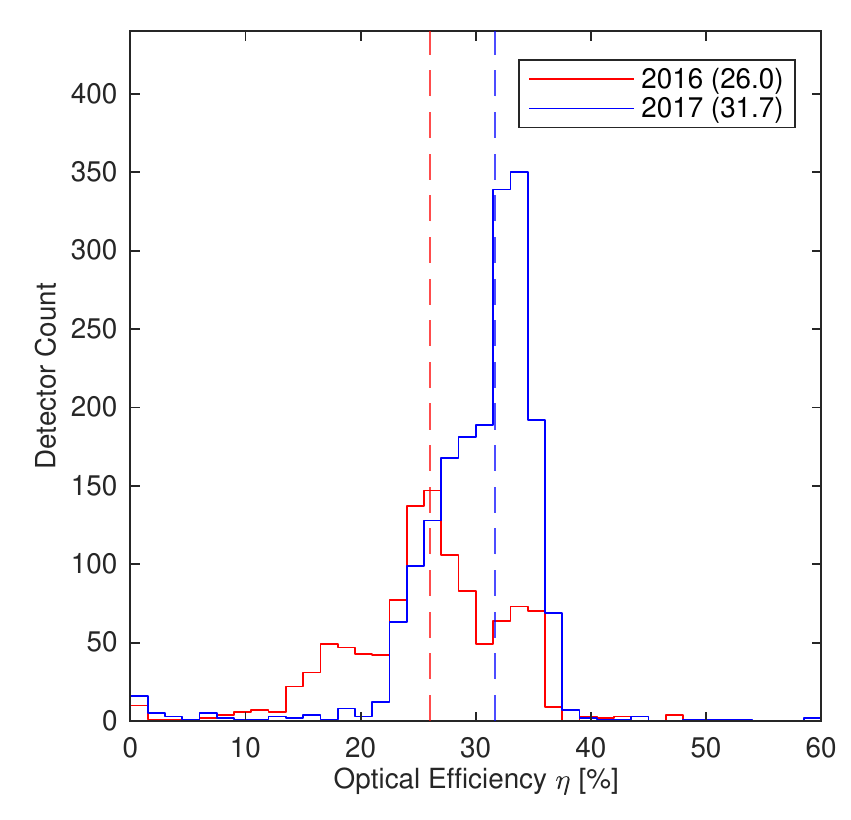}
  \caption{
  Optical efficiencies of \bicepthree\ detectors.
  The median efficiency increased from $26\%$ in 2016 (red) to
  $32\%$ in 2017 (blue).
  Because the optical efficiency measurement is performed with the detectors biased on the aluminum transition, the detector yield shown here is lower than the yield for CMB observations, when the detectors are biased on the titanium transition.
  }
  \label{fig:b3_dpdt}
\end{figure}

The \bicepthree\ per-detector optical efficiencies as measured using this method are shown in Fig.~\ref{fig:b3_dpdt}.
The median end-to-end optical efficiency improved from $26\%$ in the 2016 season to $32\%$ in 2017.
This is mostly due to the change in 300~K thermal filters described in \S\ref{sec:foam_filters} and the replacement of the delaminated metal-mesh edge filters described in \S\ref{sec:ade_filters}.

\subsection{Measured detector properties}
We designed the thermal conductance of the detector to avoid saturation during science observations while minimizing phonon noise.
The design saturation power of the \bicepthree\ detectors is 5~pW, which has a safety factor of 2-2.5 from the expected optical load during nominal observing conditions, giving a target thermal conductance $G_c = 40$~pW/K for the titanium TES bolometer with transition temperature $T_c = 500$~mK and a bath temperatures $T_o = 280$~mK, where we assume a thermal index $\beta\sim2$ from bare silicon nitride supports.

The detectors are screened prior to deployment to ensure that the properties are near target values.
The thermal conductance $G_c$, thermal conductance index $\beta$, and transition temperature $T_c$ are given by
\begin{eqnarray}
 P_{\text{sat}}=G_c T_o\frac{\left(T_c/T_o\right)^{\beta+1}-1}{\beta+1}
\end{eqnarray}
where $P_{\text{sat}}$ is the saturation power of the detector.
These parameters, shown in Table~\ref{tab:det_param}, are measured by taking load curves at multiple bath temperature $T_o$ in a ``dark'' configuration, where the cryostat is optically sealed to prevent light coupling to the detectors.

\begin{figure}
  \centering
  \includegraphics{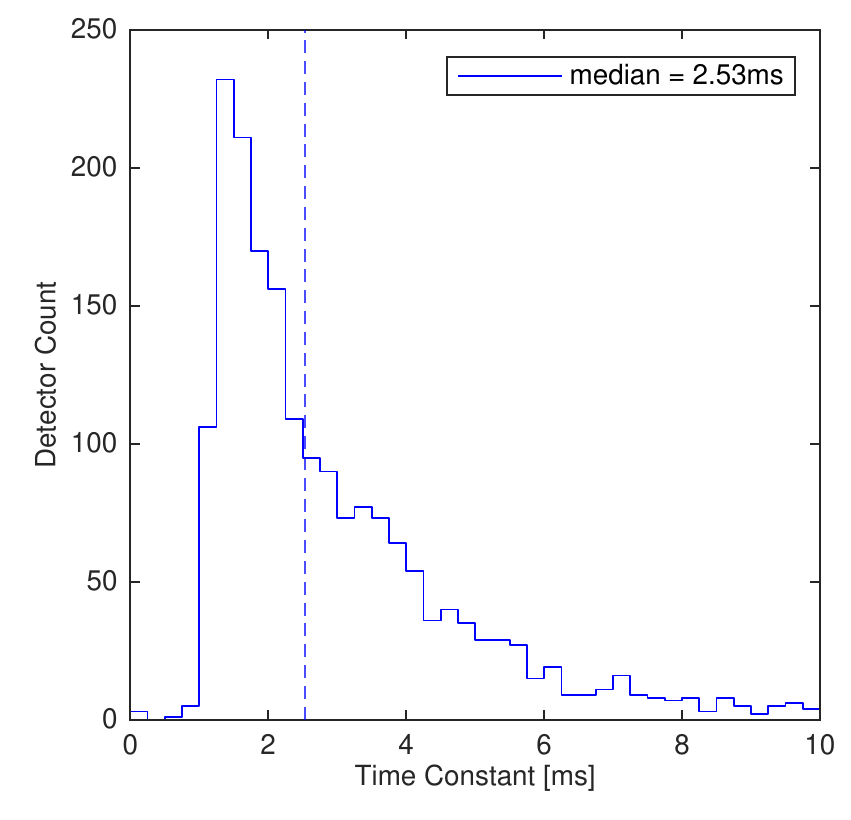}
  \caption{
  Detector time constants in 2021, measured at the nominal TES bias voltage used for CMB observations.
  }
  \label{fig:det_tau}
\end{figure}

The effective thermal time constant $\tau$ of an ideal voltage-biased TES bolometer is given by
\begin{eqnarray}
 \tau=\frac{C/G}{1+\mathscr{L}\left(V\right)}
\end{eqnarray}
where $C$ is the heat capacity of the bolometer island and $\mathscr{L}$ is the effective loop gain of the electrothermal feedback at detector bias voltage $V$.
The detector time constants $\tau$ in \bicepthree\ are calculated by measuring the detector response to square-wave modulations in the bias voltage (Fig.~\ref{fig:det_tau}).

\begin{table*}
  \centering
  \caption{Average detector parameters for each \bicepthree\ module.
  Four of the modules, listed in parentheses, were replaced before the 2017 season.
  The normal resistance, saturation power, thermal conductance and transistion temperature are measured in a dark TES configuration before deployment for this subset of modules.
  The optical efficiency and spectral response were measured \texttt{in situ} at the South Pole.
  Fig.~\ref{fig:fpu_layout} shows the placement of the detector modules over the focal plane.
  }
  \label{tab:det_param}
  \begin{tabular}{l r c c c c c} 
    \toprule 
    Module & Norm. Rest. & $\text{P}_{\text{sat}}@300$mK & Thermal Conductance & Tran. Temp. & Band Center & Band Width\\
    \midrule
    (P02) & 81~m$\Omega$ & 3.71~pW & 27.2~pW/K & 503~mK & 91~GHz & 19~GHz\\
    (P03) & 83~m$\Omega$ & 3.01~pW & 39.1~pW/K & 514~mK & 94~GHz & 22~GHz\\
    P04 & 92~m$\Omega$ & 4.76~pW & 41.2~pW/K & 492~mK & 94~GHz & 23~GHz\\
    P06 & 65~m$\Omega$ & 4.25~pW & 32.5~pW/K & 501~mK & 93~GHz & 24~GHz\\
    P07 & 61~m$\Omega$ & 4.56~pW & 32.1~pW/K & 507~mK & 95~GHz & 25~GHz\\
    P08 & 63~m$\Omega$ & 4.33~pW & 32.2~pW/K & 505~mK & 92~GHz & 22~GHz\\
    P09 & 63~m$\Omega$ & 5.61~pW & 46.5~pW/K & 479~mK & 93~GHz & 23~GHz\\
    P10 & 49~m$\Omega$ & 5.03~pW & 35.4~pW/K & 513~mK & 93~GHz & 24~GHz\\
    (P11) & 138~m$\Omega$ & 6.72~pW & 72.4~pW/K & 478~mK & 92~GHz & 21~GHz\\
    P12 & 78~m$\Omega$ & 5.57~pW & 41.1~pW/K & 494~mK & 92~GHz & 22~GHz\\
    P13 & 78~m$\Omega$ & 5.97~pW & 46.1~pW/K & 487~mK & 93~GHz & 26~GHz\\
    (P14) & 153~m$\Omega$ & -- & -- & -- & 92~GHz & 22~GHz\\
    P16 & 91~m$\Omega$ & 4.71~pW & 31.6~pW/K & 474~mK & 95~GHz & 18~GHz\\
    P17 & 105~m$\Omega$ & 4.22~pW & 56.2~pW/K & 452~mK & 93~GHz & 20~GHz\\
    P18 & 86~m$\Omega$ & 4.41~pW & 36.4~pW/K & 458~mK & 95~GHz & 23~GHz\\
    P19 & 73~m$\Omega$ & 4.03~pW & 31.3~pW/K & 438~mK & 93~GHz & 24~GHz\\
    P20 & 79~m$\Omega$ & 4.41~pW & 32.4~pW/K & 460~mK & 93~GHz & 22~GHz\\
    P21 & 74~m$\Omega$ & 4.16~pW & 32.3~pW/K & 474~mK & 95~GHz & 26~GHz\\
    P22 & 72~m$\Omega$ & 5.71~pW & 46.4~pW/K & 485~mK & 93~GHz & 21~GHz\\
    P23 & 70~m$\Omega$ & 5.14~pW & 42.3~pW/K & 484~mK & 93~GHz & 24~GHz\\
    \midrule
    P24 & 59~m$\Omega$ & 3.05~pW & 26.4~pW/K & 483~mK & 93~GHz & 24~GHz\\
    P25 & 64~m$\Omega$ & 4.20~pW & 32.7~pW/K & 461~mK & 93~GHz & 24~GHz\\
    P26 & 49~m$\Omega$ & 3.34~pW & 24.9~pW/K & 479~mK & 93~GHz & 24~GHz\\
    P27 & 62~m$\Omega$ & 3.14~pW & 27.2~pW/K & 474~mK & 93~GHz & 24~GHz\\
    \bottomrule
  \end{tabular}
\end{table*}

\subsection{Detector bias}
\begin{figure}
  \centering
  \includegraphics{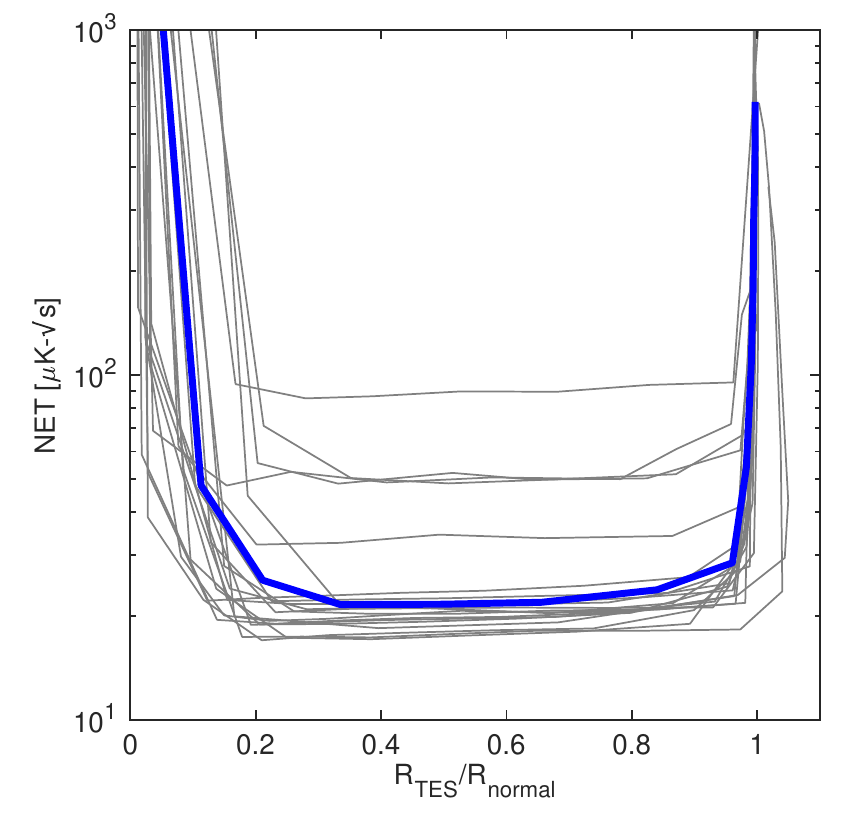}
  \caption{
  Noise equivalent temperature (NET) in units of CMB temperature as a function of detector resistance.
  The gray lines are the NET for each detector in one sample readout column, and the blue line is the average response for that column.
  The optimal bias point for the readout column is determined when average NET is at its minimum.
  The ``noise stares'' were taken under conditions of low atmospheric loading with an assumed sky temperature.
  Variation in sky temperature will affect the absolute NET values shown in this figure, but do not impact the selection of optimal bias point.
  }
  \label{fig:net_bias}
\end{figure}

The detectors operate in strong electrothermal feedback to linearize the response and to speed up the time constant.
The usable bias range is limited by thermal instability at low bias, and detector saturation at high bias as shown in Fig.~\ref{fig:net_bias}.
We believe thermal instability arises from finite thermal conductivity internal to the island which becomes problematic at the higher backgrounds at higher observing frequencies \citep{Sonka2017}.

The sensitivity of the detectors as a function of bias voltage is measured before each observing season to select the optimal bias.
This is done by taking three-minute ``noise stares'' with the telescope at the nominal elevation for CMB observation,
and comparing the noise levels obtained with the optical response as inferred from elevation nods (described in further detail in \S\ref{sec:abscal}).
Due to the multiplexing design, all detectors in one readout column share a common bias voltage.
This limitation only modestly reduces system sensitivity as the per-detector NET is sufficiently insensitive to bias to allow a wide range of operating bias points.

\subsection{Crosstalk}

\begin{figure}
  \centering
  \includegraphics{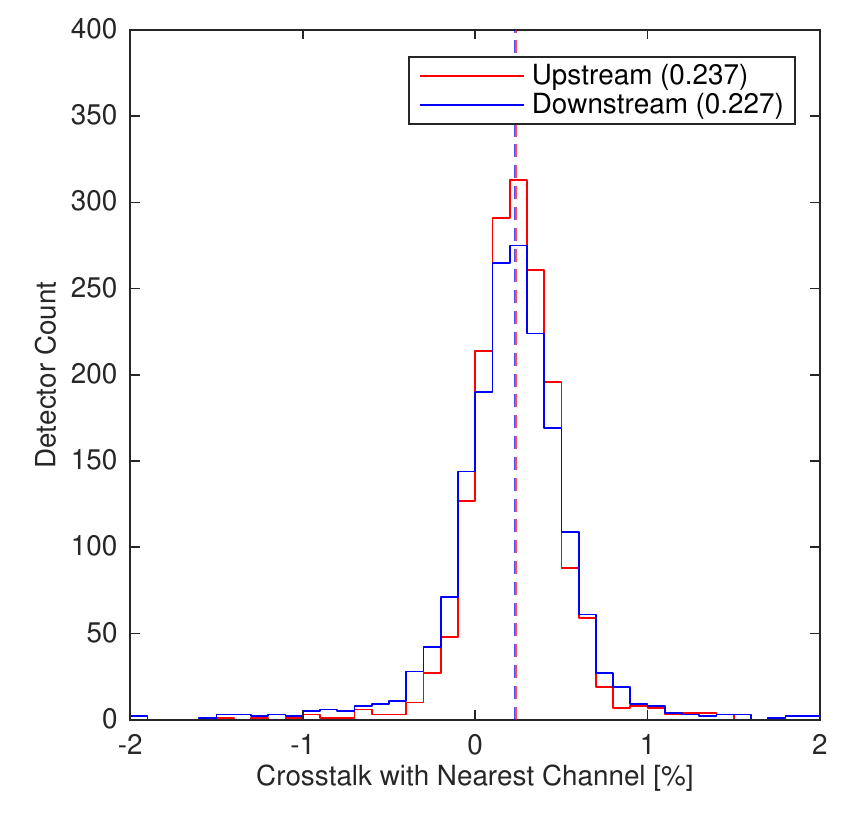}
  \caption{
Measured nearest-neighbor crosstalk in each detector using a cosmic ray analysis with median value shown in parentheses.
The red histogram shows upstream crosstalk (response seen in the detector visited before the target detector in the time domain) and the blue histogram shows downstream crosstalk (response seen in the detector sampled after the target detector).
  }
  \label{fig:b3_xtalk}
\end{figure}

Crosstalk can occur between neighboring detectors within a readout column in the TDM system.
One way to quantify the level of crosstalk through the readout chain is using cosmic rays.
When a cosmic ray hits a detector, it generates a transient signal which may also trigger a faint signal in neighboring detectors in a readout column.
We set up a custom analysis which searches unfiltered data to locate spikes from cosmic rays, stacks multiple events to increase the S/N, and compares the response in neighboring channels.
Through this analysis, we find the crosstalk level in \bicepthree\ is consistent with previous experiments at $\sim0.3\%$.
The fact that upstream is very similar to downstream crosstalk (see Fig.~\ref{fig:b3_xtalk}) argues that inductive crosstalk dominates over crosstalk from settling time crosstalk.
The CMB temperature-to-polarization leakage due to crosstalk is quantified in the beam simulations shown in Appendix F of \cite{BKXIII}.

\subsection{Timestream noise}
\label{time_stream}

While the science audio band in \bicepthree\ is $\sim$0.1-1~Hz, the TDM readout system can alias higher frequency noise at multiples of the multiplexer’s Nyquist frequency into the science band.
So we must model this high frequency noise, particularly in our readout electronics, and check that aliased noise does not compete with photon and phonon noise.
 
\begin{figure*}
  \centering
  \includegraphics{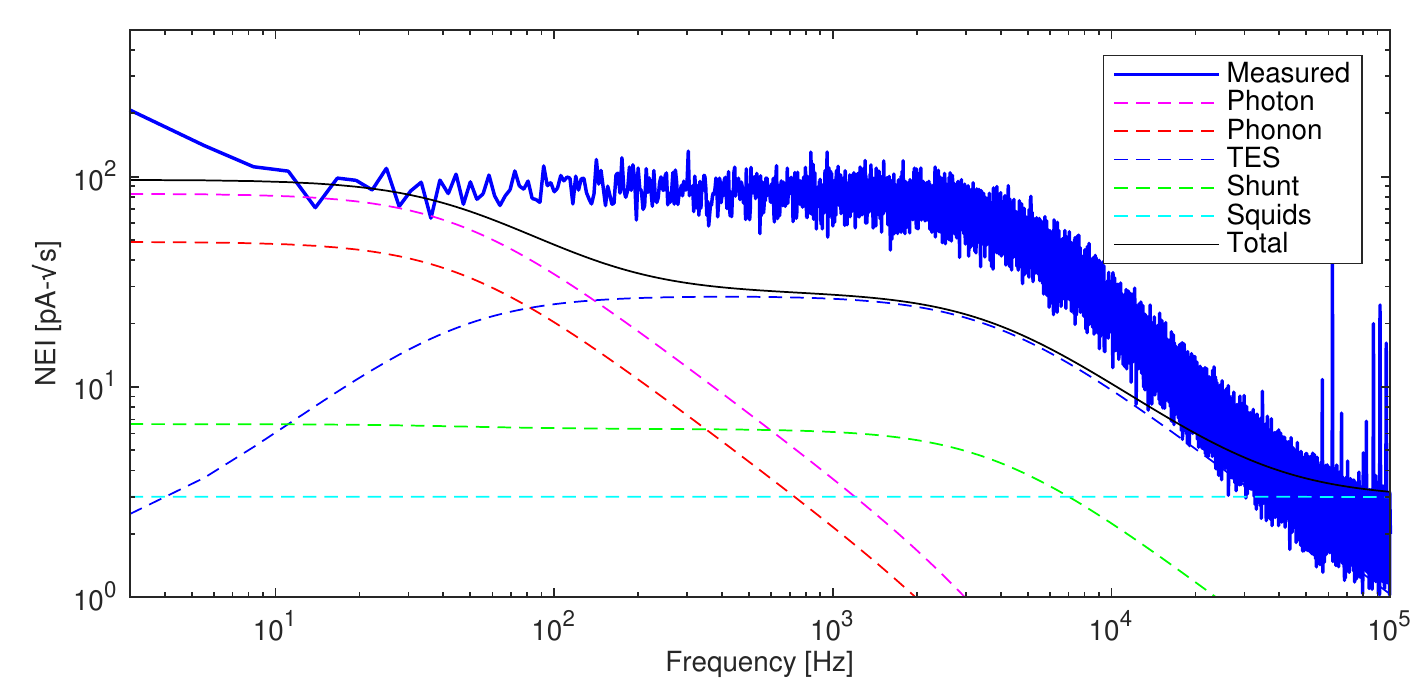}
  \caption{
  Measured and modeled noise for a nominal single undifferenced detector in \bicepthree.
  We plot noise equivalent current (NEI) in the SQUID and readout electronics.
  The noise in Eq.~\ref{eq:NEP_photon} and \ref{eq:NEP_phonon} have NEP=NEI/$S$, where the responsivity  $S=dI/dP=1/V*\mathscr{L}/(\mathscr{L}+1)*1/(1+j\omega\tau)$, $\mathscr{L} \geq10$ and the effective time constant $\tau$ is measured to be $\tau\sim 2$ms (Fig.~\ref{fig:det_tau}).  The 1/f knee at 8 Hz in the measured spectra is from atmospheric fluctuations, which is suppressed by an order of magnitude down to 0.1 Hz after pair-difference polarization pairs (see Fig.~\ref{fig:net_year_time}).  
The $1/f$ knee at 8~Hz in the measured spectra from atmospheric fluctuations, which are suppressed by an order of magnitude to 0.1~Hz in pair-difference polarization pairs.
}
  \label{fig:noise_model}
\end{figure*}

Fig.~\ref{fig:noise_model} shows the noise spectrum of a single detector under nominal observing conditions, as well as a model of the component contribution.
The photon noise is
\begin{align}
	\label{eq:NEP_photon}
    \mathrm{NEP}^2_{\text{photon}} &= 2h\nu Q_{\mathrm{load}} +
        \frac{2Q^2_{\mathrm{load}}}{\nu \frac{\Delta \nu}{\nu}},
\end{align}
where $\nu$ is the frequency, $\frac{\Delta \nu}{\nu}$ is the fractional bandwidth, and $Q_{load}$ is the sum of astrophysical, atmospheric, and internal cryostat power loading.
Table~\ref{tab:preddetload} summarizes all optical sources that contribute to $Q_{load}$, dominated by the atmosphere.
The computed photon noise dominates over the other internal noise mechanisms in the detector and electronics \citep{irwin2005}.
The next most significant noise contribution comes from phonon noise,
\begin{align}
	\label{eq:NEP_phonon}
    \mathrm{NEP}^2_{\mathrm{phonon}} &= 4k T^2_c G_c F\left(T_c, T_{\mathrm{bath}}\right),
\end{align} 
where $G_c$ is the thermal conductance, and $F(T_c,T_{\mathrm{bath}})$ accounts for the distributed thermal conductance and is estimated to be $\sim0.5$.
The Johnson noise, suppressed by the TES thermal feedback loop gain $\mathscr{L}$, and the SQUID amplifier noise are subdominant at low frequencies.
 
As seen in Fig.~\ref{fig:noise_model}, the measured total noise exceeds the calculated total at frequencies $\geq100$~Hz.
This excess noise lies above that predicted by simple noise models, and tends to be proportional to the slope of the superconducting transitions as described in \cite{Gildemeister2001}.
However, by setting the multiplexing rate to 25~kHz, we minimize aliasing most of this excess noise into the science audio band, leading to only a small contribution to the overall sensitivity (discussed in 10.4).
Fortunately, \bicepthree's parameter choices with relatively low $G$ and relatively high multiplexing rate avoid the level of aliased excess noise observed in comparable experiments \citep{dets2015}.

\subsection{Near-field Beam Mapping}
\label{subsec:nfbm}

We measured the near-field angular response above the \bicepthree\ window during the austral summer at the South Pole in 2016 and 2017.
These maps were obtained 53.5~cm above the primary lens (pupil), which represent a truncated map of the antenna response of each focal plane detector in its far field.
These maps allow us to probe for various pathologies endemic to both the focal plane and optical elements before an extensive mapping campaign of the telescope far-field beams.
The near-field beam maps are measured by observing a chopped $\sim500$~K thermal source that is mounted on linear translation stages which allow for X/Y motion just above the aperture plane.
The mapping apparatus is mounted directly onto the window such that the hot source is placed as close to the aperture stop as possible without incurring damage to the polyethylene vacuum window.
The source is scanned across the plane of the aperture in a $50\times50$ grid.
At each step in this grid, the source remains stationary for $\sim10$ seconds before proceeding to the next step. 

\begin{figure}
  \centering
  \includegraphics[width=.45\textwidth]{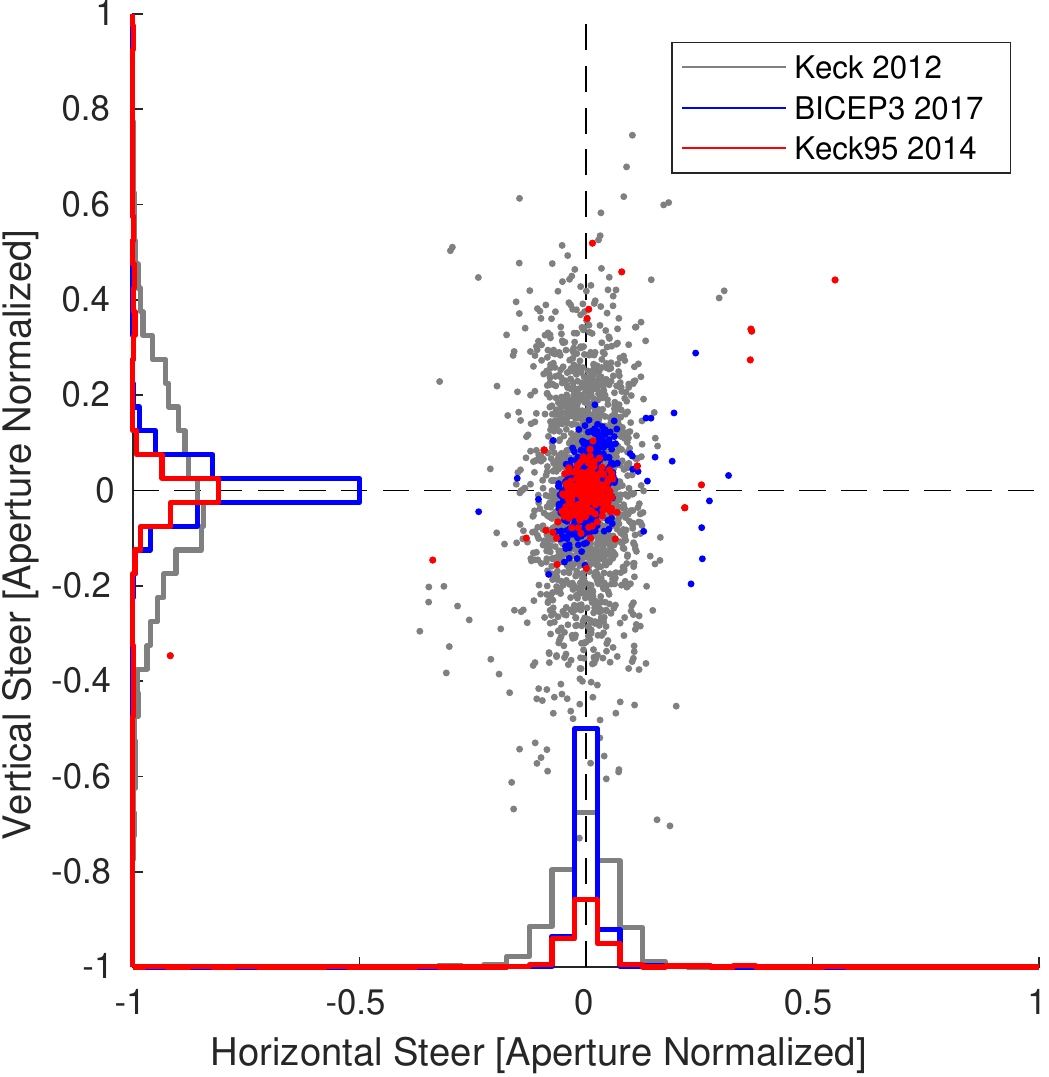}
  \caption{Near-field beam centers of \bicepthree\ detectors for the 2017 observing season (blue) compared to those of all \keck\ 150 GHz focal planes in 2012 (gray) and all \keck\ 95 GHz focal planes in 2014 (red).
The new ``etch-back'' procedure in detector fabrication was implemented after 2012.}
  \label{fig:nfbm_steer}
\end{figure}

Some \biceptwo\ and early \keck\ detectors demonstrated off-center near-field beam centers with a large truncation at the aperture stop, which both introduced distortions in the far-field beams and reduced the optical efficiency \citep{thesis:wong}.
This effect was traced to niobium contamination from the liftoff process during detector fabrication that introduced a phase shift across the planar antennas.
Changing the fabrication to an ``etch-back'' process drastically reduced beam steer in
\keck\ focal planes thereafter, and the same etching process was
used in fabrication of \bicepthree\ modules~\citep{spie:buder}.
Fig.~\ref{fig:nfbm_steer} shows beam steer measured across all detectors of \bicepthree\ compared to all detectors of the 95~GHz focal planes on \keck\ (both using the new etching process) and to \keck\ 150~GHz detectors (using the old process).
While some beam steer still exists, the improved fabrication method led to a significant reduction in beam truncation at the aperture stop.

\begin{figure*}
  \centering
  \includegraphics{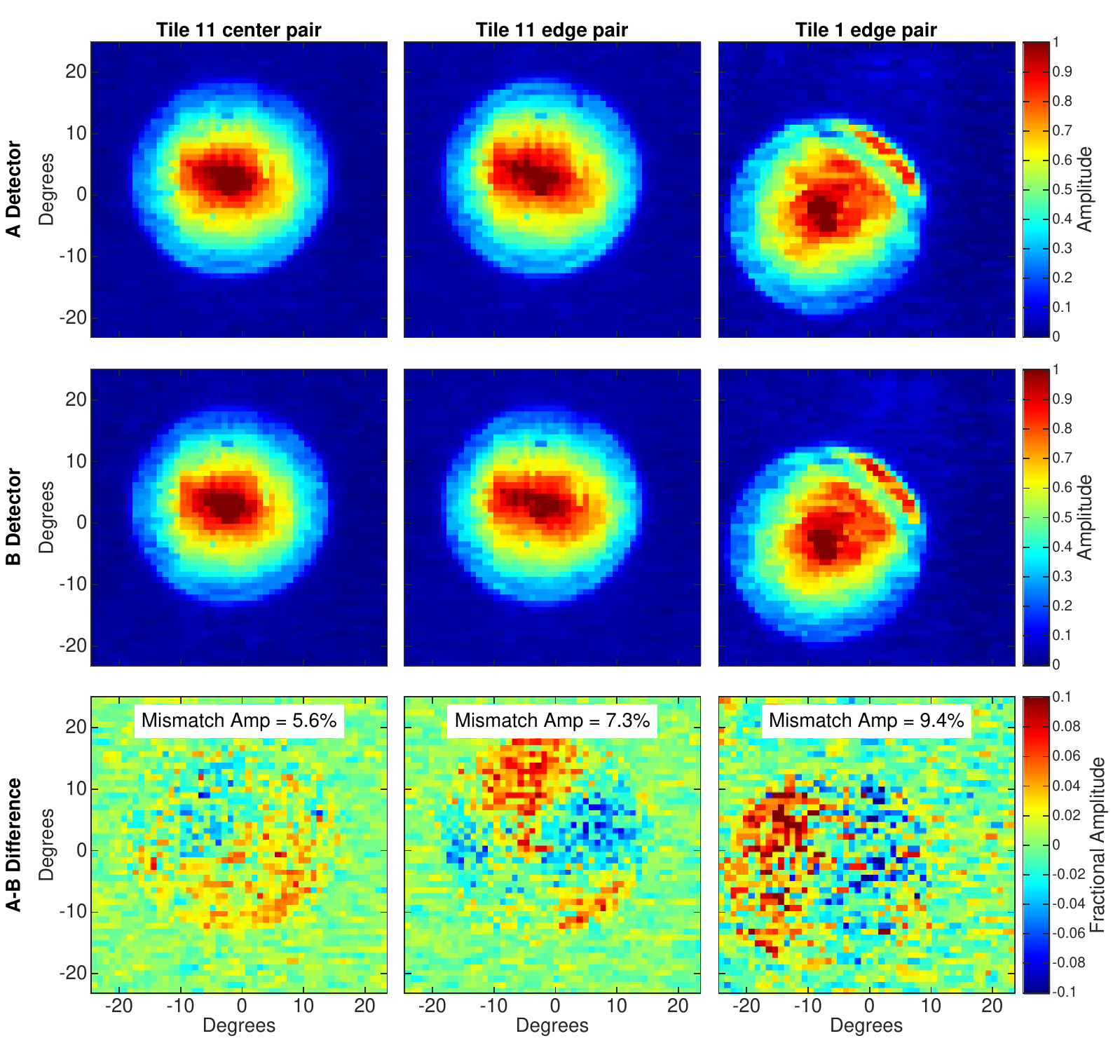}
  \caption{Amplitude-normalized near-field beam maps for detector pairs individually (top and middle rows) and their difference beams (third row) for a pair at the center of both the focal plane and its respective tile (left column); a pair near the center of the focal plane but at the edge of its tile (middle column); and a pair that is both at the edge of its tile and the whole focal plane.
  The pair centered both in the tile and focal plane demonstrates the minimum typical A/B mismatch that can be expected by detectors under ideal conditions.
  The pairs central on the FPU but at the edge of the tile confirms both the level and shape of the simulations in Fig.~ \ref{fig:corrugation_fig} where differential pointing is slightly exacerbated by the proximity of a detector pair to the corrugation frame.
  The pair at the edge of both the FPU and its tile demonstrates how a small subset of beams near the edge are steered into the aperture stop, which we attribute to beam truncation at the camera lens. }
  \label{fig:nfbm_mismatch}
\end{figure*}

These measurements also characterize the mismatch in near-field beam centers between detectors within a given pair, which can arise either from a preferential steering of one detector in a pair, or from interactions between a detector at the tile edge and the surrounding corrugated frame (see \S\ref{sec:corrugation}).
As shown in Fig.~\ref{fig:nfbm_mismatch}, a detectable increase in mismatch can be seen in detectors close to the corrugation frame which is at a level consistent with the simulations shown in Fig.~\ref{fig:corrugation_fig}.
Combining this near-field beam map dataset with metrics derived from the far-field beam maps described in \S\ref{subsec:ffbm}, we found no significant correlation between near-field beam steer and far-field beam shape in detectors which contribute to the final CMB data set.

\subsection{Far-field Beam Mapping}
\label{subsec:ffbm}

Prior to the start of each observing season, \bicepthree\ undergoes an extensive far-field beam mapping (FFBM) campaign to characterize the shape of each beam in the telescope far-field. 
The compact aperture allows measurement of the far field ($\sim$170 m for \bicepthree) by placing a chopped source on a nearby ground-base location, the adjacent Martin A. Pomerantz Observatory (MAPO) building (where \keck\ was, and \biceparray\ is, stationed), 200 m away from \bicepthree. 
A 1.7$\times$2.5~m flat aluminum mirror is erected at a 45$^{\circ}$ angle above \bicepthree, allowing the telescope to observe the source that is otherwise obstructed by the ground shield.
The source is mounted on a 40~ft vertical mast above MAPO, consisting of a 24~inch aperture that is chopped between ambient blackbody ($\sim 250$ K) and the sky at zenith ($\sim 10$ K). 
Details of the setup and results of the pre-2016 season measurement are found in \cite{spie:kirit}. 

The raw beam map timestreams are demodulated at the chop rate to isolate the signal from the chopped source, and are binned into component maps with 0.1$^{\circ}$ square pixels.  Each beam is then fit to a 2D elliptical Gaussian:
\begin{equation}
    B(\mathbf{x}) = \frac{1}{\mathrm{A}}e^{-\frac{1}{2}(\mathbf{x}-\mathbf{\mu})^{T}\Sigma^{-1}(\mathbf{x}-\mathbf{\mu})}
\end{equation}
where $\mathbf{x}$ is the two-dimensional coordinate of the beam center, $\mathbf{\mu}$ is the origin, $\mathrm{A}$ is the normalization, and $\Sigma$ is the covariance matrix, defined as
\begin{eqnarray}
 \Sigma = \quad 
 \begin{pmatrix} 
  \sigma^2 (1 + p) & c\sigma^2 \\
  c\sigma^2 & \sigma^2(1-p) 
 \end{pmatrix}
\end{eqnarray}
where $\sigma$ is the beamwidth, and $p$ and $c$ are plus and cross ellipticy, respectively. The fit values for \bicepthree\ are shown in Table~\ref{tab:FFBMparam} where the individual measurement uncertainty is the spread in parameter values over all component maps for a given detector, and is generally smaller than the detector-to-detector scatter.
The median Gaussian beamwidth for \bicepthree\ is 0.161$^{\circ}$, equivalent to a FWHM of 0.379$^{\circ}$.

\begin{table}
  \centering
  \caption{\bicepthree\ median far-field beam parameters. These values are taken from 2017 FFBM data, and are presented as median over all detectors $\pm$ scatter over all detectors $\pm$ individual measurement uncertainty.
  }
  \label{tab:FFBMparam}
  \begin{tabular}{l r} 
    \toprule 
    Parameter & Median $\pm$ Scatter $\pm$ Unc.\\
    \toprule
    Beamwidth $\sigma$ (degrees)    & $0.161 \pm 0.003 \pm 0.001$ \\
    Ellipticity plus $p$ & $0.008 \pm 0.002 \pm 0.002$ \\
    Ellipticity cross $c$ & $-0.010 \pm 0.020 \pm 0.019$ \\
    Diff. beamwidth $d\sigma$ (degrees) & $0.000 \pm 0.001 \pm 0.001$ \\
    Diff. ellipticity plus $dp$ & $-0.003 \pm 0.011 \pm 0.002$\\
    Diff. ellipticity cross $dc$ & $-0.003 \pm 0.004 \pm 0.002$ \\
    Diff. pointing $dx$ (arcmin) & $-0.060 \pm 0.120 \pm 0.050$ \\ 
    Diff. pointing $dy$ (arcmin) & $0.000 \pm 0.130 \pm 0.050$ \\
    \bottomrule
    \vspace{1mm}
  \end{tabular}
\end{table}

\begin{figure}
  \centering
  \includegraphics{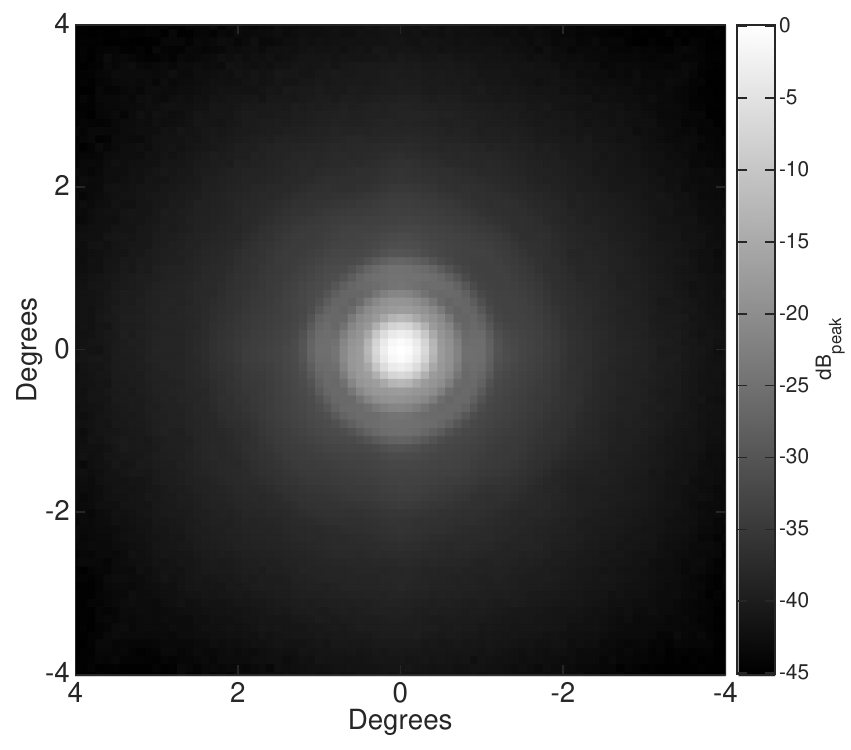}
  \caption{
    The \bicepthree\ average beam, made by coadding composite beam maps
    from all optically-active detectors.
  }
  \label{fig:ffbeam}
\end{figure}

The component beam maps for each detector are then averaged together to create high-fidelity, per-detector composite beam maps.
These composite beam maps are then coadded over all detectors to form the receiver-averaged beam, shown in Fig.~\ref{fig:ffbeam}.
The receiver-averaged beams are then Fourier transformed and azimuthally averaged into the beam window function $B(\ell)$ which is inverted to recover the sky power spectrum.
The per-detector composite beams are also used to quantify the temperature-to-polarization leakage in a given data set, which is done in Appendix F of \cite{BKXIII}.

\subsection{Polarization Response}
\label{subsec:RPS}
\begin{figure}
  \centering
  \includegraphics{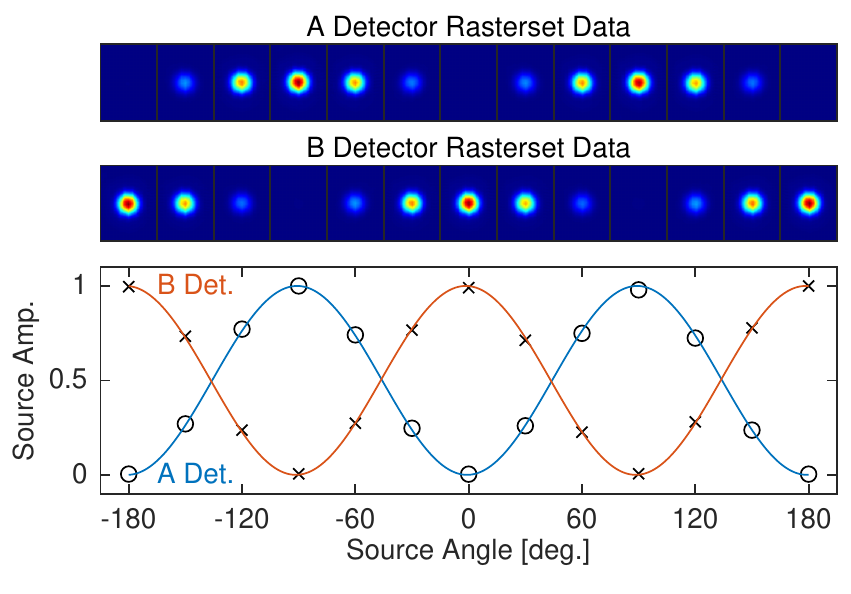}
  \caption{
    Beam maps of individual RPS rasters for an A/B-polarization detector
    (\textbf{top/middle}) and the corresponding normalized modulation curve
    (\textbf{bottom}), where the blue and orange lines are the best-fits to the
    A- and B-polarization detectors, respectively.
  }
  \label{fig:modcurve}
\end{figure}

Unlike previous generations of \bicep\ receivers, \bicepthree\ did not use an aperture-filling rotating polarized source as shown in \cite{b1_Takahashi}.
Instead the polarization response is acquired through observations of a rotating polarized quasi-thermal noise source (RPS).
The source is placed on the same mast used in far-field beam measurements and observed via the large flat mirror described in \S\ref{subsec:ffbm}.
For a single observation, maps are created by rastering across the RPS in azimuth and stepping in elevation while keeping the polarization axis of the RPS at a fixed angle.
13 beam maps are created for RPS angles spanning $360^\circ$ in $30^\circ$ increments.
Fig.~\ref{fig:modcurve} shows the resulting modulation in amplitude of the beams as a function of source polarization angle that produces a sinusoidal curve from which we derive detector polarization properties.
The details and results from the RPS observations used here are described in \cite{spie:jac}.

The relative polarization angles between detector pairs are measured to a precision of $<0.04^\circ$, with a measured variation among pairs within each tile of $0.13^\circ$ rms, and variation of the median angles across tiles of $0.32^\circ$ rms.
The small size of these variations allows us to use ideal, rather than measured per-detector polarization angles when creating \bicepthree's CMB polarization maps.
Global polarization rotation is not as well constrained due to systematics arising in the geometry of the calibration itself.
Instead, we estimate and subtract the global rotation angle from an EB/TB-minimization procedure to mitigate false B-mode signals \citep{2014selfcal}.

The relative calibration of the polarization maps to the CMB temperature map shown in \S\ref{sec:abscal} depends on the level of cross-polar response of each detector, which is dominated by the crosstalk between the two detectors within a pair.
The median cross-polar response is measured to be $0.7\pm0.2\%$, consistent with the crosstalk within the measurement uncertainty.
While any difference between the measured and assumed cross-polar response contributes to additional uncertainty on the absolute gain calibration of the $E$- and $B$-mode polarization maps, it does not introduce any additional bias in the $B$-mode signal.

\subsection{Far-sidelobe Mapping}
\label{subsec:FSL}

All \bk\ receivers use two levels of warm baffling, to ensure that any ray must diffract twice to couple to the ground as described in \S\ref{sec:baffle}.
Any excess power in the far-sidelobe (FSL; roughly defined as the part of the beam outside the region captured by the co-moving forebaffle) should be coupled to an ambient-temperature absorber or redirected via the ground shield to cold sky.
However, this power still increases the loading of the detectors, and if polarized, could lead to leakage that may be difficult to constrain.
We therefore take measurements using a high-powered noise source to map the far-sidelobe region response.
This measurement used the same noise source described in \S\ref{subsec:RPS}, which has variable attenuation that gives $\sim$70dB of dynamic range needed to map out all regions of the beam.
The source is mounted on a mast on the same building as the \bicepthree\ instrument. 

\begin{figure}
  \centering
  \includegraphics[width=0.45\textwidth]{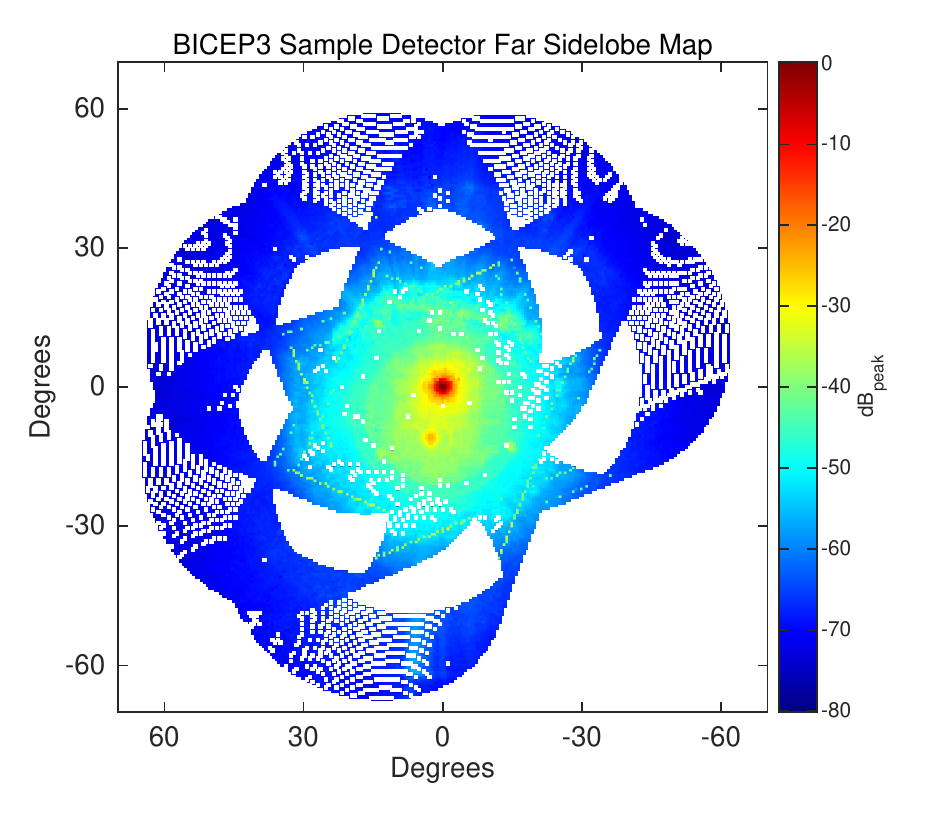}
  \caption{
    Sample far-sidelobe map of a \bicepthree\ detector, made by stitching together measurements from three power settings and coadding maps made at both source polarization orientations.
    The forebaffle was on for this measurement, representing the true beam response on sky as during CMB observations.
    The main beam and extent of the \bicepthree\ aperture are clearly seen.
    The feature just below the main beam in the map is the ``ghost'' beam described in the text.
  }
  \label{fig:fsl_beam}
\end{figure}

A typical far-sidelobe schedule takes 380$\deg$ scans in azimuth, with an elevation range of 34$\deg$ in 0.5$\deg$ steps, all repeated over multiple boresight rotation angles.
The measurement is often repeated with both the co-moving forebaffle on and off, as an external check of the amount of power intercepting the forebaffle.
A waveguide twist can be placed before the source output horn that couples to free space, in order to take measurements in two orthogonal source polarizations.
Three different power settings are used to map out the entire beam at each boresight rotation angle, where the power settings are changed by adjusting the attenuation in the source.
A ``low'' power setting maps the main beam, ``medium'' maps the mid-sidelobe, and ``high'' maps the far-sidelobe.
The maps made with each setting are stitched together to create a single map for each detector.
The maps at each polarization (made with and without the waveguide twist installed) can be coadded together to create an effective unpolarized FSL map.
An example of this for a single \bicepthree\ detector is shown in Fig.~\ref{fig:fsl_beam}.

\begin{figure}
  \centering
  \includegraphics[width=0.5\textwidth]{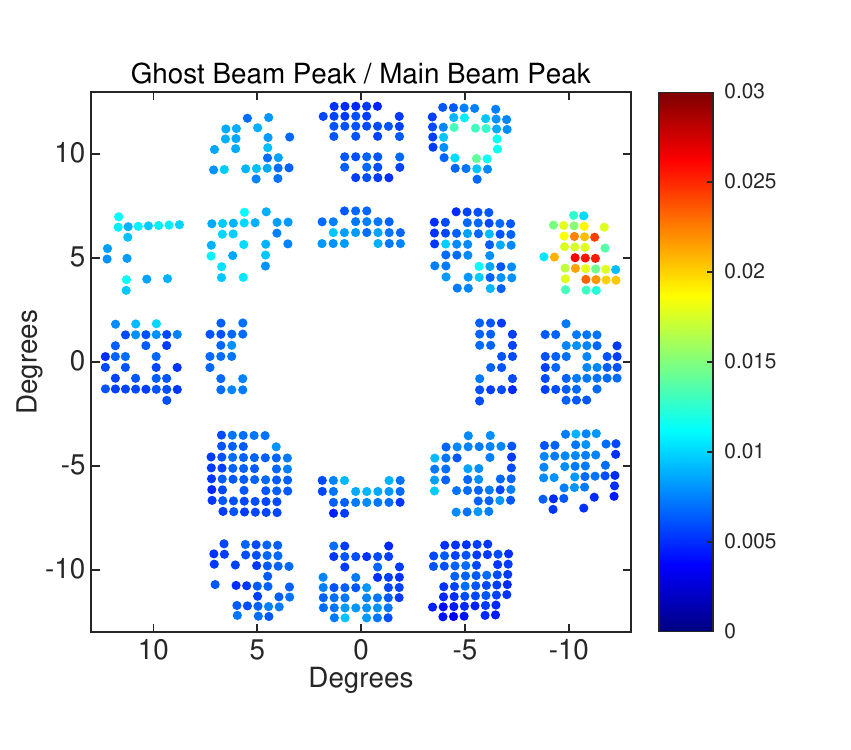}
  \caption{
Per-pair values of the ratio of ghost beam peak power to the main beam peak power.
Each point is the average between both detectors in a pair.
For detectors near the center of the focal plane, ghost beams cannot be confidently separated from the main beam and are therefore omitted from this plot.
Tile 1 (top-right corner) has higher-amplitude ghost beams due to the increased reflection from the blanked port on the opposite side of the focal plane.
}
  \label{fig:b3_buddybeam}
\end{figure}

The FSL maps also reveal a small-amplitude, well-formed ``ghost beam'' located on the opposite side of the boresight from each detector's main beam.
In the time-reverse sense, the beam partially reflects off one of the flat 50~K filters and travels back through the 4~K optics, refocuses and reflects again off the focal plane to the emerging on sky.
This feature has been seen in previous \bk\ receivers~\citep{BKIII} and for most detectors, the integrated power of this ghost beam is $ < 1$\% of the integrated main beam power.
However, tile~1 shows integrated ghost beam power that is $2-3\times$ larger than that of the other detectors (Fig.~\ref{fig:b3_buddybeam}).
This anomalous ghost beam power is likely due to the non-symmetrical focal plane layout (Fig.~\ref{fig:fpu_layout}).
\bicepthree\ is equipped with 20 detector modules --- slot~21, which is directly opposite to tile~1 on the focal plane, has no module and is covered by a reflective copper plate.
Data from all detectors in tile~1 were thus removed from the final analysis in order to pass internal consistency checks.

\section{Observing strategy}
\label{sec:obs_strategy}

\subsection{Observing field}

\bicepthree\ observes the same sky patch as \bicep2/\keck, covering $-60^{\circ}<\text{RA}<60^{\circ}$ and $-70^{\circ}<\delta<-40^{\circ}$.
However, its effective sky area is $\sim600 ~\text{deg}^{2}$, larger than the $\sim400 ~\text{deg}^2$ in \bicep2/\keck\ due to the larger instantaneous field of view of \bicepthree.
To avoid regions of high dust contamination in this extended field, we slightly shifted the field center to $\text{RA} = 0 \text{hr}$, $\text{dec} = -55^{\circ}$, compared to $\text{RA} = 0 \text{hr}$, $\text{dec} = -57.5^{\circ}$ for \bicep2/\keck.
About 10\% of the observing time is used to map a part of the Galactic plane, centered at $\text{RA} = 15:42 \text{hr}$, $\text{dec} = -53^{\circ}$. 

\begin{figure}
  \centering
  \includegraphics{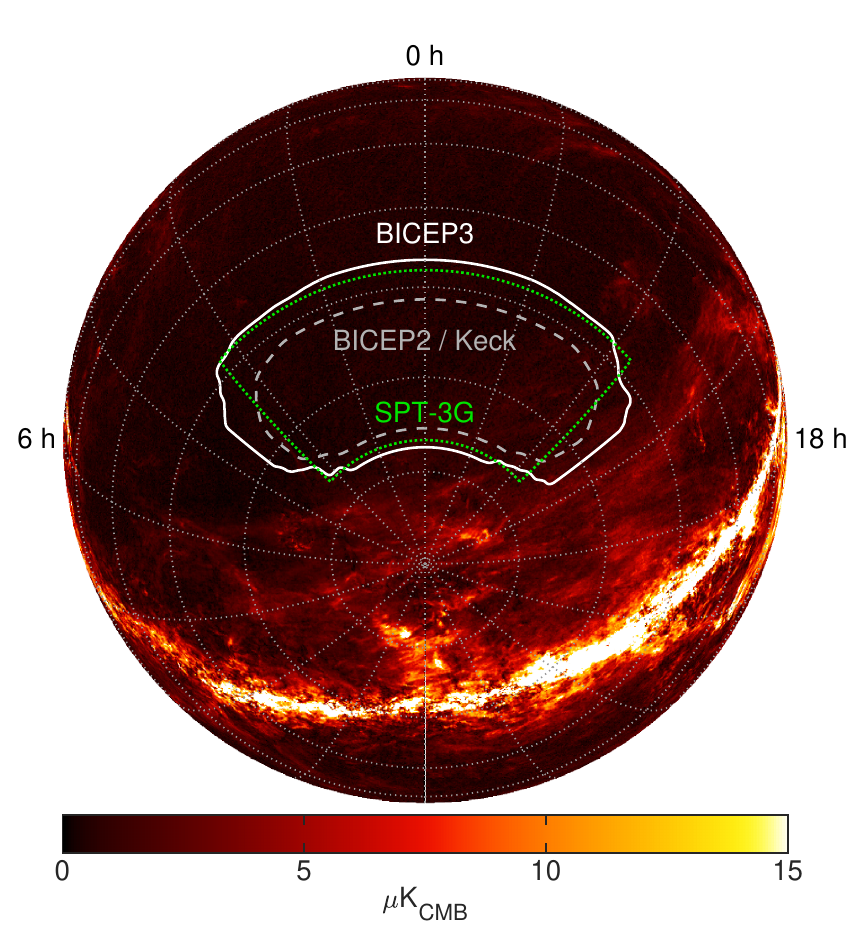}
  \caption{
    The \bicepthree\ CMB observing field (solid white) on the southern
    celestial sphere, together with the smaller \bicep2/\keck\ field (dashed white) and the
    SPT-3G $1500~\text{deg}^{2}$ survey (dotted green)~\citep{sobrin2021}.
    The background image shows the polarized intensity $P = \sqrt{Q^2 + U^2}$
    of the \planck\ component-separated (SMICA) dust map \citep{planck2018_iv},
    rescaled in amplitude from $353\,\mathrm{GHz}$ to $95\,\mathrm{GHz}$
    assuming a graybody spectrum with temperature
    $T_\mathrm{d} = 19.6\,\mathrm{K}$ and spectral index $\beta_\mathrm{d} = 1.5$.
  }
  \label{fig:b3_obsfield}
\end{figure}

This observing field is known to have very low polarized foregrounds, and is covered by other experiments (Fig.~\ref{fig:b3_obsfield}), providing the possibility for joint analyses.
For example, we demonstrated in \cite{wu2021} a method for separating the lensing $B$-mode signal from the potential PGW signature in collaboration with \spt.

\subsection{Scan pattern and schedule}
\label{sec:schedule}

\bicepthree\ observes its target CMB sky patch continuously through the austral winter season. 
At the South Pole, the telescope azimuth and elevation axes conveniently map to Right Ascension (RA) and Declination in equatorial coordinates, respectively.
The sky patch then rotates in azimuth but does not move in elevation, allowing us to track it continuously.

\begin{figure*}[]
  \centering
  \includegraphics{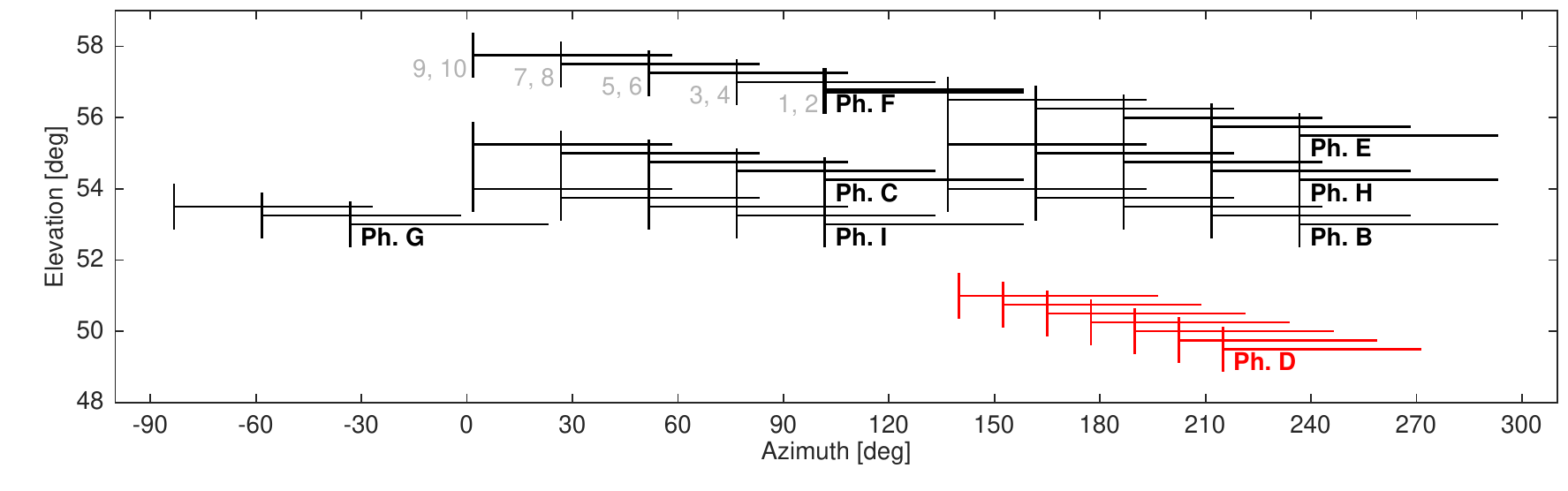}
  \caption{
Observing pattern of a typical three-day schedule in ground-based coordinates.
The first scanset of phase F is shown in bold.
Horizontal lines indicate the field scans and the vertical lines indicate the bracketing elevation nods.
The telescope scans at a fixed elevation during each scanset.
For the CMB field scans, we observe two scansets before changing elevation.
Phase D is on the galactic plane.
}
  \label{fig:obsmount}
\end{figure*}

The fundamental observing block is a constant-elevation `scanset', consisting of 50 back-and-forth scans in azimuth, at $2.8^{\circ}$/s spanning $64.4^{\circ}$ over 50~mins.
Because the sky drifts by $12.5^{\circ}$ during each scanset, the azimuth center is shifted every other scanset by $25^{\circ}$ to track the change in RA of the target sky patch (Fig.~\ref{fig:obsmount}).
This azimuth-fixed scan pattern allows us to remove ground-fixed pickup and terrestrial magnetic contamination with a simple ground-subtraction template.
The elevation is stepped every other scanset by $0.25^{\circ}$ to fill in coverage between the spatially separated detector beams.

The overall schedule contains cryogenic service and CMB and Galactic plane scansets.
These scansets are grouped into `observing phases', and each phase contains between 6 to 10 scansets along with the accompanying calibrations.
During a three-day schedule, the telescope completes one cryogenic cycle, six 10-hour phases on the CMB field, one 6-hr phase on the CMB field, and one 6-hr phase on the Galactic plane (Table~\ref{tab:phases}).

The telescope is rotated about its optical axis to a different boresight angle for each schedule. 
A total of four boresight angles at $23^{\circ}$, $68^{\circ}$, $203^{\circ}$ and $268^{\circ}$ are used.
The $45^{\circ}$ pairs are required to measure both the Stokes $Q$ and $U$ parameters.
The whole set of four angles is clocked to optimize the coverage symmetry and homogeneity over the target CMB sky patch. 

\begin{table}
  \centering
  \caption{Observation schedule for \bicepthree. Start times are listed by Local Sidereal Time (LST).
  The 2016 season used a two-day schedule without the bolded phases G, H and I.}
  \label{tab:phases}
  \begin{tabular}{c c c c} 
    \toprule 
    Phase & LST & Field & No. of Scansets\\
    \midrule
    A & Day 0 23:00 & Fridge re-cycling & \\
    B & Day 1 05:00 & CMB & 10\\
    C & Day 1 14:00 & CMB & 10\\
    D & Day 1 23:00 & Galactic &  7\\
    E & Day 2 05:00 & CMB & 10\\
    F & Day 2 14:00 & CMB & 10\\
    \textbf{G} & \textbf{Day 2 23:00} & \textbf{CMB} &  \textbf{6}\\
    \textbf{H} & \textbf{Day 3 05:00} & \textbf{CMB} & \textbf{10}\\
    \textbf{I} & \textbf{Day 3 14:00} & \textbf{CMB} & \textbf{10}\\
    \bottomrule
    \vspace{1 mm}
  \end{tabular}
\end{table}

\subsection{Detector calibrations}
\label{sec:abscal}
The detectors in \bicepthree\ are calibrated in two steps.
First, a relative gain calibration is applied to ensure the timestream data from each detector 
pair accurately subtracts the large common-mode unpolarized signals from the atmosphere, telescope and CMB.
Each scanset is bracketed by an elevation nod (elnod) where the telescope is stepped upward by 0.6$^{\circ}$ then downward by 1.2$^{\circ}$, and finally upward again by 0.6$^{\circ}$ to return to the starting position over the course of one minute.
This motion causes all of the detectors to measure varying levels of atmospheric emission according to the relative opacity $\kappa$ of the atmosphere is described by
\begin{eqnarray}
 \kappa &=&\frac{1}{\sin\left(\text{el}\right)} 
 \label{el_nod}
\end{eqnarray}
down to elevation el~$=30^\circ$.
Each leading and trailing elnod gives a mean gain of detector native feedback units (FBU) per airmass for each channel for the scanset, allowing us to relatively calibrate all the detectors.
A larger `sky dip' spanning 50$^{\circ}$ to 90$^{\circ}$ is performed before each phase as an additional calibration data point for confirming the atmospheric profile.

In addition to the relative gain calibration, the final data set requires an absolute gain calibration.
We apply a single scale factor to convert from detector FBU to final CMB temperature units.
This scale factor $g_{\text{abs}}$ is determined by computing the ratio of the cross spectra of the \bicepthree\ map with external, calibrated maps from \planck.
We first calculate
\begin{equation}
    g_b = \frac{\left<\tilde{m}_{\text{cal}}\times\tilde{m}_{\text{ref}}\right>_b}{\left<\tilde{m}_{\text{real}}\times\tilde{m}_{\text{ref}}\right>_b},
\end{equation}
where $\tilde{m}_{\text{real}}$ is the uncalibrated \bicepthree\ temperature map, and $\tilde{m}_{\text{cal}}$ and $\tilde{m}_{\text{ref}}$ are the \planck\ 95 and 145~GHz maps respectively.
Two separate external maps are used to reduce noise.
The \planck\ maps are smoothed by \bicepthree's beam and reobserved using the same filtering applied to the \bicepthree\ maps.
The ratio of these two spectra is a set of bandpower calibration factors $g_b$.
The final scale factor $g_{\text{abs}}$ uses the mean of the first five bandpowers $g_b$ of the \bk\ bins.

\subsection{Star pointing}
\label{sec:star_pointing}

`Star pointing' optical pointing measurements are made with the star camera every few weeks.
It has become a routine observation to track the small movements of the telescope mount built on top of snow, though the frequency and quality of star pointings varies depending on the weather.
The goal of star pointing observations is to verify the parameters used in the telescope pointing model are stable, including three tilt and two zero-offset parameters along the azimuth and elevation directions, as well as two parameters associated with the offset of the star camera from the mount's boresight rotation axis.

Two separate lists of stars for summer (12 stars) and winter (29 stars) accommodate the change in visibility.
The entire set of stars is observed at three distinct deck angles, yielding three sets of independent data points.
The sequence involves centering each star, in turn, on the star camera's boresight crosshairs (in practice, one predefined pixel in the CCD), and recording the mount encoders and the current time.
This is done for each star at three different boresight rotation angles.
The pointing model parameters can be found by fitting the telescope pointing against the star positions.
In this process, outliers are dropped and only those schedules with more than 24 data points (in the winter) are kept for data fitting.
The root mean square of the fit residuals typically reach $\sim 20$ arcsec.

During the analysis of the 2016 data, we noticed the temporal split jackknife (see discussion in \S\ref{sec:jackknifes}) showed exceptionally low statistical probabilities.
It was eventually traced to shift in the elevation offset parameter derived from star pointings. 
There was a clear $0.02^{\circ}$ shift in this parameter roughly halfway through the season, leading to the jackknife test failures.
We associated this shift with a mechanical slip in the star camera mount that was not aligned with the boresight of the telescope, and we added a flexure term in the pointing model to capture this effect and correct the jackknife failures.
In the subsequent investigation, we correlated per-schedule CMB maps against reobserved \planck\ CMB maps.
This provided a high temporal resolution boresight tracker to monitor pointing shifts to sub arc-min levels, and corroborated our assumption about the pointing shift.
This analysis has been incorporated to track anomalous changes in tilting and offset parameters in \bicep\ and \keck\ telescopes.

\section{Three-year data set}
\label{sec:data_set}

\subsection{Data selection and cuts}
\bicepthree\ was installed in the Dark Sector Laboratory (DSL) at the South Pole on December of 2014 with a partially filled focal plane.
It was populated with a full complement of 20 detector tiles in December of 2015 and began scientifically meaningful observations in early 2016.
\cite{spie:jimmy} and \cite{spie:Hui_2016} show preliminary data quality of \bicepthree\ during the 2016 season.
After the 2016 season, the telescope was removed from the mount and modified during the austral summer.
These modifications improved the detector yield, the observation efficiency, and the noise performance, ultimately improving the sensitivity by $\sim30\%$ compared to the previous season \citep{spie:jhk}.

The detector wafers in \bicepthree\ were fabricated at the JPL Microdevices Laboratory.
90\% of the wafers achieved $\geq 82\%$ in room temperature impedance screening.
The majority of the detector yield losses are associated with wirebond failures in the readout chain.
There are more than 22,000~wirebonds in \bicepthree, connecting between detector wafers, SQUIDs amplifier chips, and various circuit boards.
Failures in critical wirebonds can disable an entire readout column of 22 detectors or an entire row of 30 detectors.

\begin{figure*}[t]
  \centering
  \includegraphics[width=0.9\textwidth]{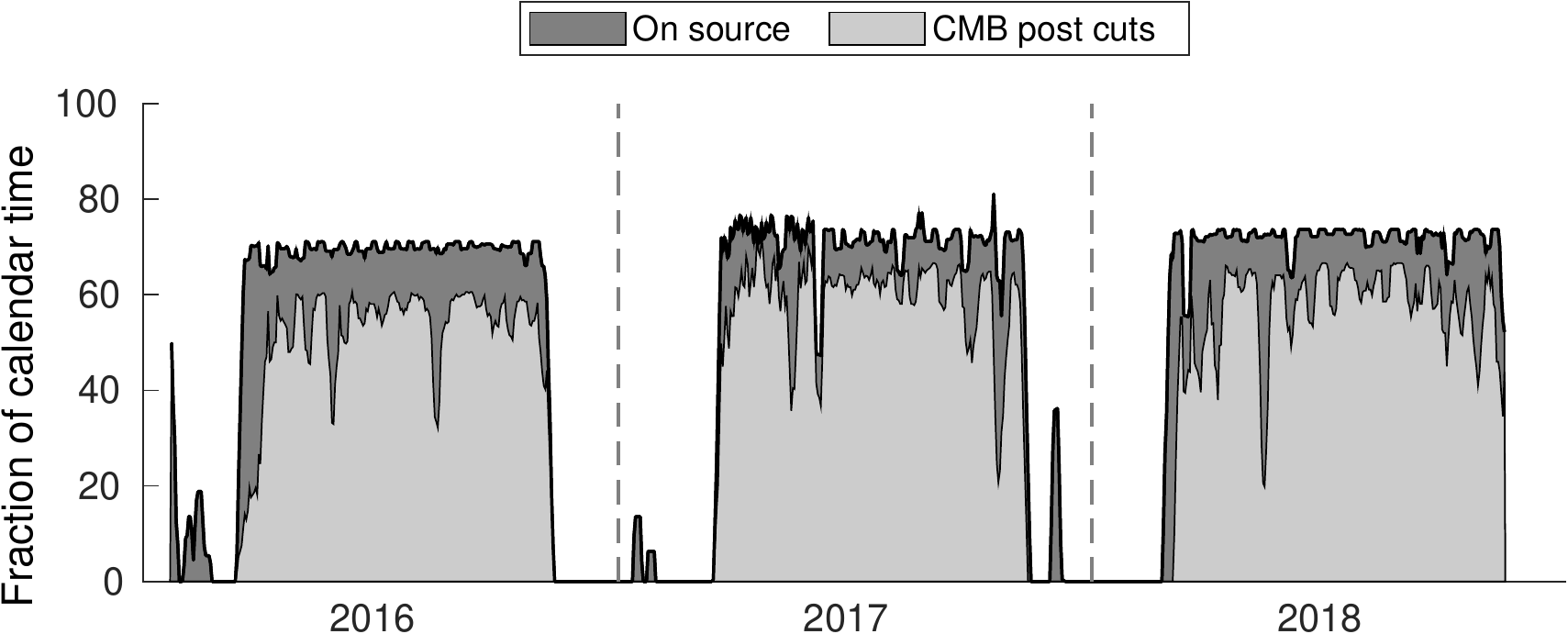}
  \includegraphics[width=0.9\textwidth]{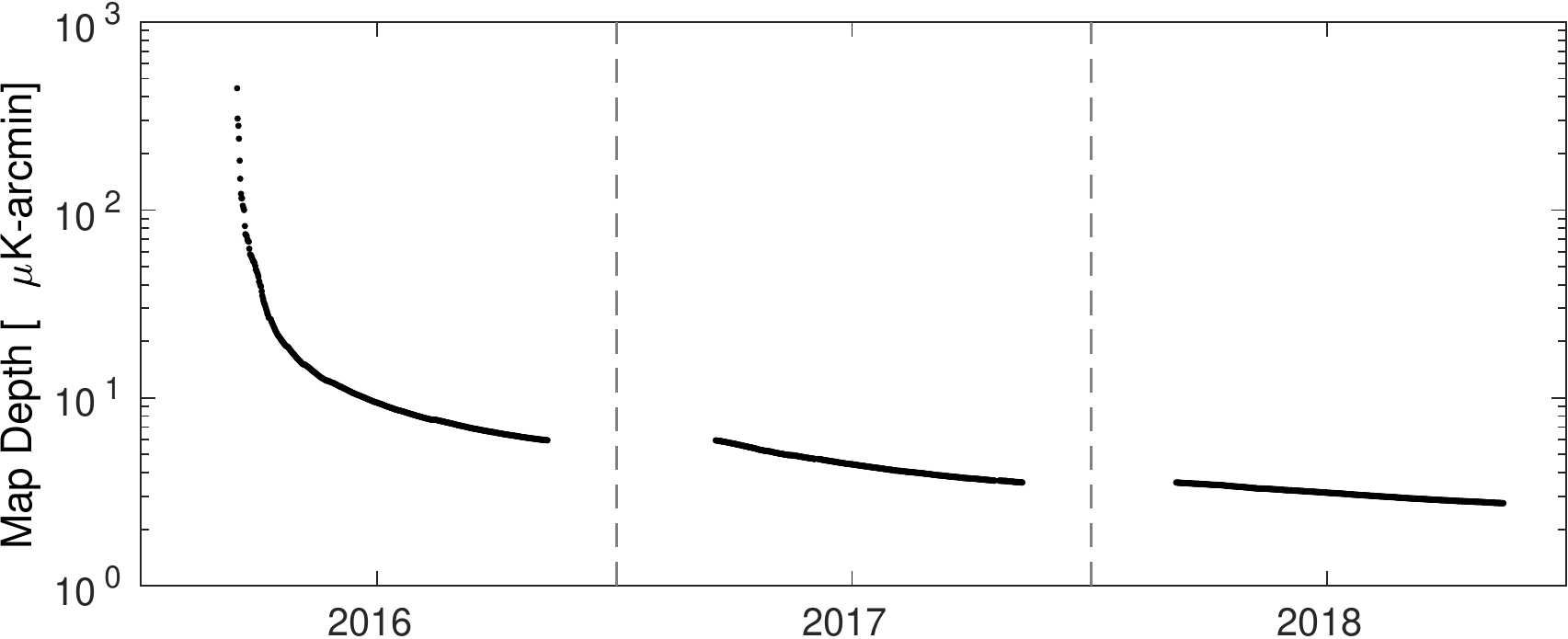}
  \caption{
  Integration of the \bicepthree\ data set from 2016 to 2018, plotting (top) fraction of time per day spent in CMB scans, excluding regular calibrations and refrigerator cycling.
  During austral summers (November to February), the observing schedules were been interspersed with calibration measurements.
  During the austral winter, on-source efficiency is about 70~\%.
  The lower curve includes data quality cuts, but excluding non-functioning channels (Row 2 in Table~\ref{tab:cuts_stat}.)
  The rms map-based sensitivity (bottom) improves over time and reaches 2.8~\ukarcmin\ at the end of 2018.  
}
  \label{fig:b3_livetime}
\end{figure*}

The fully-populated focal plane has 1200 optically active dual-polarized detector pairs.
At optimal TES bias values, there were 930, 992, and 992 responsive detector pairs in 2016, 2017 and 2018, respectively.
Detectors were further down-selected during analysis based on performance in CMB observations and various external calibration measurements.
The achieved efficiency in CMB observations and the progression of map sensitivity are shown in Fig.~\ref{fig:b3_livetime}.
Data selection (`cut') parameters and statistics are shown in Table~\ref{tab:cuts_stat}.
The last round of data down-selection, so-called `channel flags', excludes detectors that are found to be discrepant during map-making and external calibration measurements.
These may include aberrant beams, absolute calibrations, and the exclusion of Tile~1 channel discussed in \S\ref{subsec:FSL}.
These detectors are removed for the entire season, and their data are thus excluded from the final coadded maps.
The rest of the cut parameters probe a wide variety of data quality metrics at per-detector pair, per-scanset, and per-halfscan levels.
Out of all the responsive detector pairs, \bicepthree\ achieved overall pass fractions of 57.0\%, 65.8\%, and 61.8\% in 2016, 2017 and 2018, respectively.

\begin{table*}
  \centering
  \caption{
  Data Cuts as described in Appendix A of \cite{thesis:willmert}.
  These cuts are applied sequentially, in the order listed.
  The first column (\textbf{raw}) shows the fraction of data removed by the cut parameter, if no other cuts were considered.
  For rows after the second, this value is referenced against the total fraction of data from nominally ``functional'' detectors.
  The second column shows the fractional \textbf{remaining} data after each sequentially applied round of cuts.
  Horizontal lines distinguish between (from top to bottom) focal plane yield, cuts applied per-halfscan, cuts applied per-scanset, cuts on cuts,
  and post-hoc per-pair `channel flag' cuts determined at the mapmaking stage.
  }
  \label{tab:cuts_stat}
  \begin{tabular}{l c c| c c| c c}
  & \multicolumn{2}{c|}{2016 Cuts [\%]} & \multicolumn{2}{c|}{2017 Cuts [\%]} & \multicolumn{2}{c}{2018 Cuts [\%]}\\
  Cut Parameter & Raw  & Remaining  & Raw & Remaining  & Raw & Remaining\\
  \midrule
  Before cuts & - & 100.0 & - & 100.0 & - & 100.0 \\ 
  Non-functional detectors & 22.54 & 77.5 & 17.29 & 82.7 & 17.29 & 82.7 \\ 
  \hline
  Timestream glitches/dropped samples & 0.62 & 76.8 & 0.13 & 82.6 & 0.13 & 82.6 \\ 
  Intra-MCE synchronization & 0.00 & 76.8 & 0.06 & 82.6 & 0.00 & 82.6 \\ 
  Inter-MCE synchronization & 0.00 & 76.8 & 0.04 & 82.6 & 0.00 & 82.6 \\ 
  Frac. of passing channels (MUX col.) & 0.10 & 76.8 & 0.08 & 82.6 & 0.02 & 82.6 \\ 
  Frac. of passing channels (full expt.) & 0.00 & 76.8 & 0.07 & 82.6 & 0.02 & 82.6 \\ 
  \hline
  Raw elnod amplitude not negligible & 14.54 & 62.9 & 12.45 & 70.2 & 13.14 & 69.6 \\ 
  Elnod stability over scanset & 11.56 & 60.9 & 9.80 & 68.7 & 11.59 & 66.8 \\ 
  Elnod A/B ratio stability over scanset & 16.25 & 59.3 & 13.14 & 67.8 & 14.03 & 66.0 \\ 
  Glitches in elnod & 2.72 & 57.8 & 1.36 & 66.9 & 1.55 & 65.0 \\
  Pair-sum elnod model goodness-of-fit & 10.65 & 53.4 & 6.29 & 64.5 & 9.53 & 62.1 \\ 
  Pair-difference elnod shows structure & 6.19 & 52.7 & 4.25 & 64.0 & 5.61 & 61.9 \\ 
  TES resistance & 2.14 & 52.1 & 1.27 & 63.7 & 0.94 & 61.7 \\ 
  Median elnod amplitude & 0.40 & 52.1 & 0.48 & 63.7 & 1.46 & 61.7 \\ 
  Pair-difference timestream skewness & 9.23 & 51.3 & 6.54 & 63.3 & 8.85 & 60.4 \\ 
  Change in azimuth-fixed signal during scanset & 2.28 & 51.2 & 1.91 & 63.2 & 2.08 & 60.4 \\
  Excessively high timestream STD & 18.57 & 49.8 & 14.92 & 62.4 & 15.92 & 59.7 \\ 
  Focal plane-correlated noise & 0.00 & 49.8 & 0.00 & 62.4 & 0.00 & 59.7 \\ 
  Noise stationarity (per-detector) & 5.96 & 49.2 & 4.99 & 61.6 & 7.10 & 58.8 \\
  Noise stationarity (per-pair) & 10.64 & 49.1 & 8.00 & 61.6 & 10.51 & 58.7 \\ 
  Mean focal plane temperature & 0.00 & 49.1 & 0.00 & 61.6 & 0.00 & 58.7 \\ 
  Focal plane temperature stability & 0.37 & 49.0 & 0.76 & 61.3 & 0.27 & 58.6 \\ 
  Abnormal azimuth scanning & 0.02 & 49.0 & 0.00 & 61.3 & 0.00 & 58.6 \\
  Too many large steps in timestream & 0.53 & 49.0 & 0.14 & 61.3 & 0.38 & 58.6 \\ 
  Too many de-stepping attempts & 10.04 & 48.1 & 7.85 & 60.2 & 11.91 & 56.6 \\
  Extreme/variable crosstalk & 0.23 & 47.9 & 0.16 & 60.1 & 0.38 & 56.3 \\
  \hline
  Frac. of passing halfscans & 1.22 & 47.9 & 0.49 & 60.1 & 1.35 & 56.3 \\ 
  Frac. of passing data in scanset & 9.02 & 47.4 & 6.52 & 59.9 & 9.59 & 56.1 \\
  \hline
  Per-pair mapmaking cuts (inc. drop tile 1) & - & 44.2 & - & 54.4 & - & 51.1 \\
  \bottomrule
  \end{tabular}
\end{table*}

\subsection{Data reduction}

The data reduction and map making pipeline used in \bicepthree\ is largely the same as the \bicep2/\keck\ procedure detailed in \cite{BKII}, with only minor modifications to accommodate the larger number of detectors.

\subsection{Jackknife Tests}
\label{sec:jackknifes}

As a standard internal consistency check, we split each year's data set into two halves in various ways.
These halves are differenced to cancel out the common signal and leave only noise and the potential contribution of systematic errors.
Following previous experiments, we use 14 different jackknifes listed in \cite{BKIII} to probe temporal, spatial and readout systematics.

For a given jackknife $j$, a vector of bandpowers of $\mathbf{d}_j$ is compared to the bandpowers from the ensemble of 499 signal+noise simulations $\mathbf{s}_j$ by calculating $\chi$ and $\chi^2$ statistics, defined as:

\begin{eqnarray}
 \chi &=&\sum_b\frac{\left[\mathbf{d}_j\right]_b-\left<\mathbf{s}_j\right>_b}{\left[\mathbf{S}_j\right]_{bb}}\nonumber\\
 \chi^2_j &=& \left(\mathbf{d}_j-\left<\mathbf{s}_j\right>\right)^\top \mathbf{S}_j^{-1}\left(\mathbf{d}_j-\left<\mathbf{s}_j\right>\right)
 \label{eq:jacktest}
\end{eqnarray}
where $\mathbf{S}_j=\left<\mathbf{s}_j\mathbf{s}_j^\top\right>$ is the bandpower covariance matrix of the corresponding simulations for a single spectral type.
The $\chi$ and $\chi^2$ values are calculated for real data against the full ensemble, and the real data are judged against the distribution of statistics obtained from the simulations by computing the probability to exceed (PTE).

\begin{figure}
  \centering
  \includegraphics[width=0.45\textwidth]{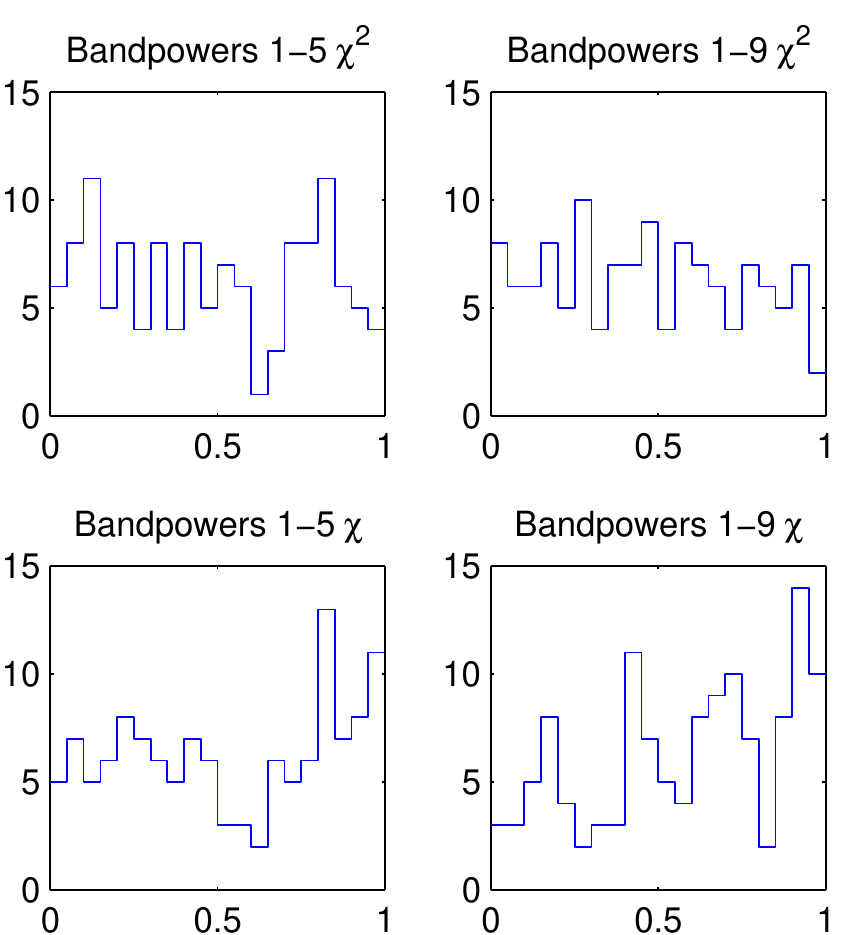}
  \caption{
Distributions of the jackknife $\chi$ and $\chi^2$ PTE values for \bicepthree\ 2016 to 2018 95~GHz data.
They are consistent with a uniform distribution, indicating there is no evidence for systematics at the level of statistical sensitivity.
}
  \label{fig:ptedist_95}
\end{figure}

For the three-year data set, we performed jackknife tests on each year separately, giving a total of $3(\text{EE/BB/EB})\times14(\text{jack})\times2(\chi/\chi^2)\times2(\text{bin})\times3(\text{year})=504$~tests.
A statistical failure led to the discovery and removal of the tile~1 data discussed in \S\ref{subsec:FSL}, and the injection of a flexure term to the pointing model discussed in \S\ref{sec:star_pointing}.
All the jackknife tests were then repeated with the updated data set and the final PTE values from the $\chi$ and $\chi^2$ statistics are listed in Table~\ref{tab:b3_jacknum}.
Fig.~\ref{fig:ptedist_95} shows the distributions of these PTE values, which appear to be uniform, showing no evidence of problems at the level of statistical noise.

\begin{table*}
  \centering
  \caption{Jackknife PTE values from $\chi$ and $\chi^2$ tests.}
  \label{tab:b3_jacknum}
  \scalebox{0.9}{
  \begin{tabular}{l c c| c c| c c}
    \toprule
    & \multicolumn{2}{c|}{2016} & \multicolumn{2}{c|}{2017} & \multicolumn{2}{c}{2018}\\
    & $\chi$ & $\chi^2$ & $\chi$ & $\chi^2$ & $\chi$ & $\chi^2$\\
    Band Power & 1-5/1-9 &  &  & & & \\
    \midrule
    \multicolumn{3}{l}{Deck jackknife} & \multicolumn{2}{|c}{} & \multicolumn{2}{|c}{}\\
    EE & 0.501/0.383 & 0.754/0.719 & 0.982/0.992 & 0.050/0.108 & 0.022/0.172 & 0.122/0.092\\
    BB & 0.936/0.998 & 0.443/0.226 & 0.731/0.419 & 0.848/0.563 & 0.567/0.924 & 0.130/0.152\\
    EB & 0.319/0.283 & 0.840/0.866 & 0.263/0.589 & 0.932/0.838 & 0.265/0.307 & 0.832/0.868\\
    
    \multicolumn{3}{l}{Scan dir jackknife} & \multicolumn{2}{|c}{} & \multicolumn{2}{|c}{}\\
    EE & 0.277/0.112 & 0.956/0.275 & 0.449/0.453 & 0.066/0.070 & 0.198/0.437 & 0.804/0.459\\
    BB & 0.764/0.872 & 0.525/0.196 & 0.283/0.433 & 0.162/0.407 & 0.168/0.172 & 0.012/0.030\\
    EB & 0.904/0.431 & 0.697/0.591 & 0.076/0.228 & 0.517/0.816 & 0.838/0.527 & 0.806/0.798\\
    
    \multicolumn{3}{l}{Temporal split jackknife} & \multicolumn{2}{|c}{} & \multicolumn{2}{|c}{}\\
    EE & 0.998/0.996 & 0.084/0.257 & 0.028/0.068 & 0.277/0.295 & 0.146/0.467 & 0.395/0.152\\
    BB & 0.098/0.200 & 0.261/0.255 & 0.263/0.531 & 0.331/0.317 & 0.826/0.946 & 0.822/0.719\\
    EB & 0.958/0.772 & 0.461/0.713 & 0.956/0.936 & 0.020/0.044 & 0.070/0.034 & 0.497/0.499\\
    
    \multicolumn{3}{l}{Tile jackknife} & \multicolumn{2}{|c}{} & \multicolumn{2}{|c}{}\\
    EE & 0.257/0.150 & 0.429/0.623 & 0.403/0.529 & 0.559/0.248 & 0.002/0.004 & 0.004/0.018\\
    BB & 0.527/0.713 & 0.323/0.495 & 0.952/0.852 & 0.455/0.816 & 0.697/0.862 & 0.705/0.371\\
    EB & 0.707/0.493 & 0.776/0.872 & 0.381/0.633 & 0.517/0.311 & 0.946/0.984 & 0.146/0.257\\
    
    \multicolumn{3}{l}{Azimuth jackknife} & \multicolumn{2}{|c}{} & \multicolumn{2}{|c}{}\\
    EE & 0.575/0.866 & 0.140/0.259 & 0.776/0.727 & 0.916/0.962 & 0.834/0.545 & 0.695/0.687\\
    BB & 0.014/0.126 & 0.082/0.068 & 0.178/0.425 & 0.435/0.667 & 0.487/0.279 & 0.860/0.665\\
    EB & 0.357/0.415 & 0.846/0.212 & 0.487/0.068 & 0.904/0.363 & 0.876/0.998 & 0.164/0.040\\
    
    \multicolumn{3}{l}{Mux col jackknife} & \multicolumn{2}{|c}{} & \multicolumn{2}{|c}{}\\
    EE & 0.309/0.429 & 0.335/0.363 & 0.731/0.745 & 0.232/0.625 & 0.681/0.946 & 0.894/0.778\\
    BB & 0.665/0.182 & 0.960/0.423 & 0.116/0.070 & 0.840/0.950 & 0.210/0.657 & 0.573/0.108\\
    EB & 0.451/0.681 & 0.944/0.992 & 0.335/0.339 & 0.423/0.415 & 0.248/0.353 & 0.988/0.924\\
    
    \multicolumn{3}{l}{Alt deck jackknife} & \multicolumn{2}{|c}{} & \multicolumn{2}{|c}{}\\
    EE & 0.982/0.996 & 0.220/0.166 & 0.972/0.954 & 0.172/0.405 & 0.056/0.182 & 0.102/0.214\\
    BB & 0.062/0.635 & 0.307/0.170 & 0.251/0.236 & 0.050/0.100 & 0.054/0.467 & 0.152/0.042\\
    EB & 0.198/0.192 & 0.477/0.790 & 0.411/0.731 & 0.238/0.118 & 0.667/0.429 & 0.513/0.814\\
    
    \multicolumn{3}{l}{Mux row jackknife} & \multicolumn{2}{|c}{} & \multicolumn{2}{|c}{}\\
    EE & 0.776/0.796 & 0.144/0.068 & 0.914/0.824 & 0.345/0.447 & 0.741/0.359 & 0.707/0.719\\
    BB & 0.822/0.725 & 0.539/0.631 & 0.425/0.631 & 0.561/0.800 & 0.515/0.673 & 0.890/0.583\\
    EB & 0.850/0.471 & 0.060/0.166 & 0.677/0.573 & 0.383/0.677 & 0.367/0.601 & 0.870/0.844\\
    
    \multicolumn{3}{l}{Tile and deck jackknife} & \multicolumn{2}{|c}{} & \multicolumn{2}{|c}{}\\
    EE & 0.631/0.421 & 0.788/0.878 & 0.439/0.427 & 0.888/0.920 & 0.886/0.926 & 0.715/0.902\\
    BB & 0.902/0.904 & 0.531/0.477 & 0.601/0.786 & 0.441/0.407 & 0.411/0.567 & 0.349/0.695\\
    EB & 0.311/0.461 & 0.429/0.569 & 0.842/0.709 & 0.204/0.377 & 0.896/0.944 & 0.733/0.485\\
    
    \multicolumn{3}{l}{Focal plane inner or outer jackknife} & \multicolumn{2}{|c}{} & \multicolumn{2}{|c}{}\\
    EE & 0.355/0.635 & 0.822/0.311 & 0.204/0.224 & 0.579/0.633 & 0.174/0.120 & 0.208/0.327\\
    BB & 0.800/0.922 & 0.711/0.555 & 0.663/0.928 & 0.617/0.295 & 0.148/0.194 & 0.204/0.283\\
    EB & 0.483/0.760 & 0.303/0.373 & 0.836/0.974 & 0.711/0.549 & 0.132/0.130 & 0.880/0.635\\
    
    \multicolumn{3}{l}{Tile top or bottom jackknife} & \multicolumn{2}{|c}{} & \multicolumn{2}{|c}{}\\
    EE & 0.942/0.641 & 0.064/0.010 & 0.768/0.960 & 0.505/0.397 & 0.910/0.679 & 0.204/0.463\\
    BB & 0.974/0.764 & 0.124/0.012 & 0.224/0.703 & 0.046/0.090 & 0.226/0.705 & 0.774/0.503\\
    EB & 0.353/0.717 & 0.675/0.593 & 0.786/0.932 & 0.411/0.451 & 0.136/0.345 & 0.148/0.174\\
    
    \multicolumn{3}{l}{Tile inner or outer jackknife} & \multicolumn{2}{|c}{} & \multicolumn{2}{|c}{}\\
    EE & 0.745/0.665 & 0.397/0.798 & 0.828/0.870 & 0.756/0.930 & 0.002/0.012 & 0.014/0.124\\
    BB & 0.337/0.667 & 0.224/0.421 & 0.196/0.667 & 0.956/0.818 & 0.810/0.924 & 0.076/0.138\\
    EB & 0.820/0.900 & 0.840/0.922 & 0.216/0.405 & 0.583/0.756 & 0.321/0.545 & 0.321/0.635\\
    
    \multicolumn{3}{l}{Moon jackknife} & \multicolumn{2}{|c}{} & \multicolumn{2}{|c}{}\\
    EE & 0.218/0.709 & 0.485/0.487 & 0.860/0.882 & 0.780/0.878 & 0.904/0.683 & 0.104/0.160\\
    BB & 0.976/0.824 & 0.255/0.607 & 0.996/0.946 & 0.108/0.246 & 0.206/0.164 & 0.142/0.385\\
    EB & 0.487/0.900 & 0.778/0.693 & 0.088/0.128 & 0.583/0.463 & 0.840/0.912 & 0.701/0.064\\
    
    \multicolumn{3}{l}{A and B offset best and worst jackknife} & \multicolumn{2}{|c}{} & \multicolumn{2}{|c}{}\\
    EE & 0.860/0.794 & 0.723/0.924 & 0.571/0.661 & 0.315/0.537 & 0.860/0.625 & 0.908/0.565\\
    BB & 0.453/0.561 & 0.022/0.044 & 0.970/0.972 & 0.194/0.293 & 0.860/0.942 & 0.814/0.780\\
    EB & 0.435/0.455 & 0.259/0.549 & 0.806/0.760 & 0.421/0.285 & 0.806/0.623 & 0.776/0.551\\

    \bottomrule
    \vspace{1 mm}
  \end{tabular}
  }
\end{table*}

\begin{figure*}
  \centering
  \includegraphics{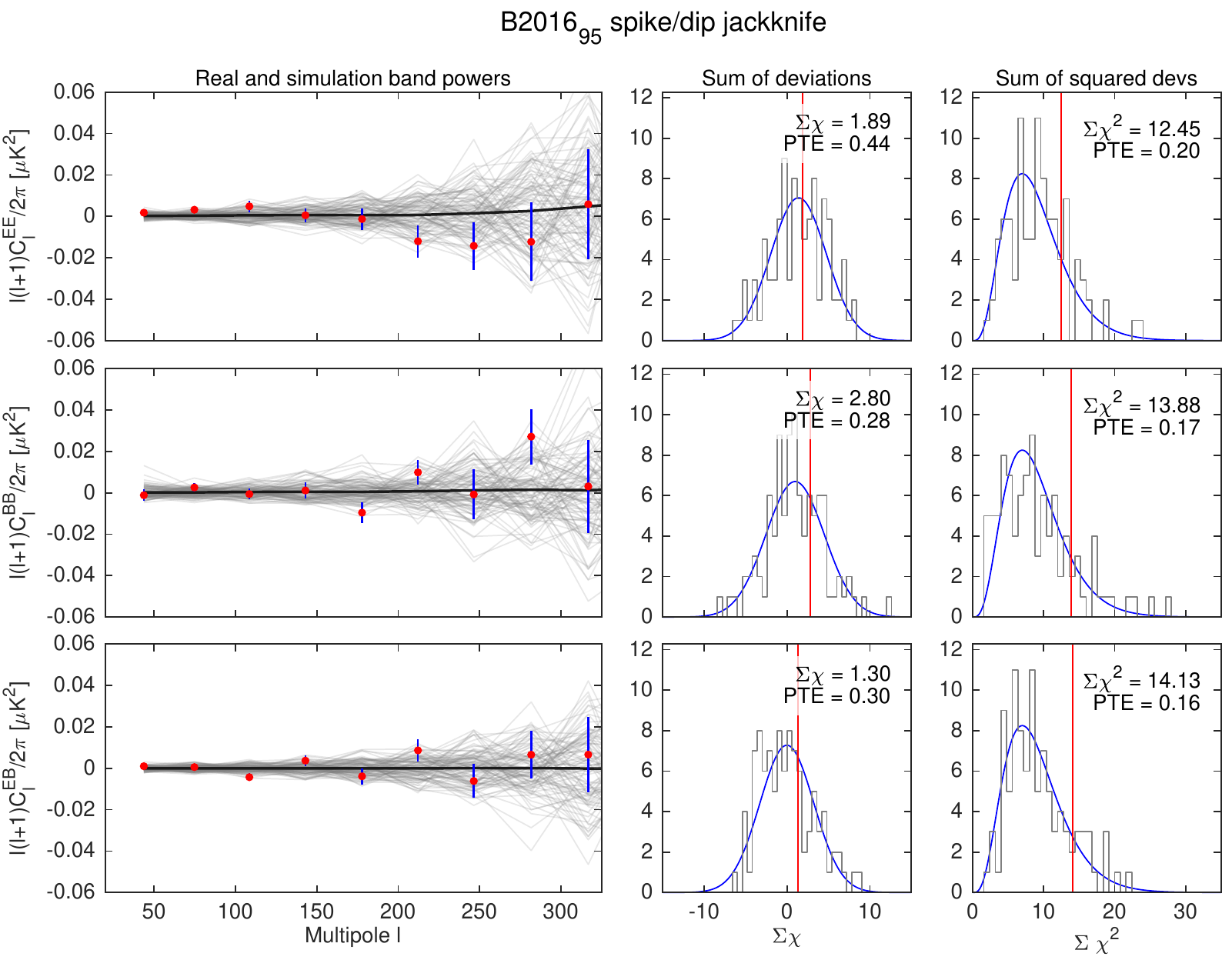}
  \caption{
  \textit{Left:} Bandpowers for the \bicepthree\ 2016 spike-dip jackknife $EE, BB$, and $EB$ spectra.
  The welter of light gray lines are the bandpowers of the ensemble of 99 lensed-$\Lambda$CDM+dust+noise simulations, with the mean of these simulations given by the thick black line.
  The real map bandpowers are given by the red circles, where the error bars are for the standard deviation of the simulations.
  \textit{Right:} Histograms of the $\chi$ and $\chi^2$ statistics for each simulation, with the expected Gaussian and $\chi^2$ distribution over-plotted in blue.
  The real data value is marked by the vertical red line, with the value and corresponding probability to exceed (PTE) annotated.
}
  \label{fig:spike_dip_jack}
\end{figure*}

In addition to the standard 14 jackknife tests, we created an additional spectral jackknife to test the spectral anomaly in the 2016 data due to the metal mesh low-pass edge filter delamination described in \S\ref{sec:ade_filters}.
Using the definition in Eq.~\ref{eq:spec_estimator}, we selected the worst 50\% (the most spikes and dips) of the data, and compared it to the coadded 2017/18 data which has an uniform spectral response. 
With 99 simulations, we found the difference is consistent with a null result (Fig.~\ref{fig:spike_dip_jack}).

\subsection{Instantaneous sensitivity}
\label{sec:NET}

\begin{figure}
  \centering
  \includegraphics{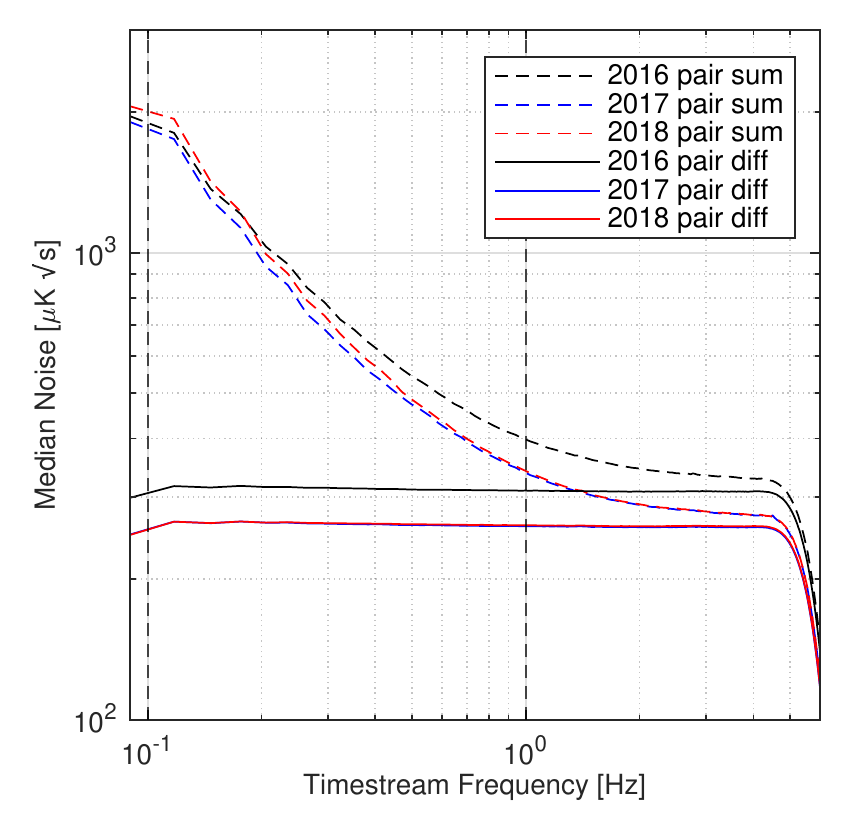}
  \caption{
Median pair-sum and pair-diff noise spectra, evaluated from minimally-processed timestreams from the \bicepthree\ 2016-2018 seasons.
The median per-detector NET is the average between the science band from 0.1-1~Hz.
  }
  \label{fig:net_year_time}
\end{figure}

During normal CMB observation, the detector timestreams are multiplexed, downsampled and filtered to 30~Hz, which is sufficient for the 0.1-1~Hz science band.
1/$f$ noise dominated by atmospheric noise, is greatly reduced after pairs of co-located polarization sensitive detectors are differenced.
Contamination from the atmosphere and ground signals is further suppressed by filtering the timestreams with a third-order polynomial and subtracting a ground-averaged template from the scanning data (Fig.~\ref{fig:net_year_time}).

\begin{figure}
  \centering
  \includegraphics{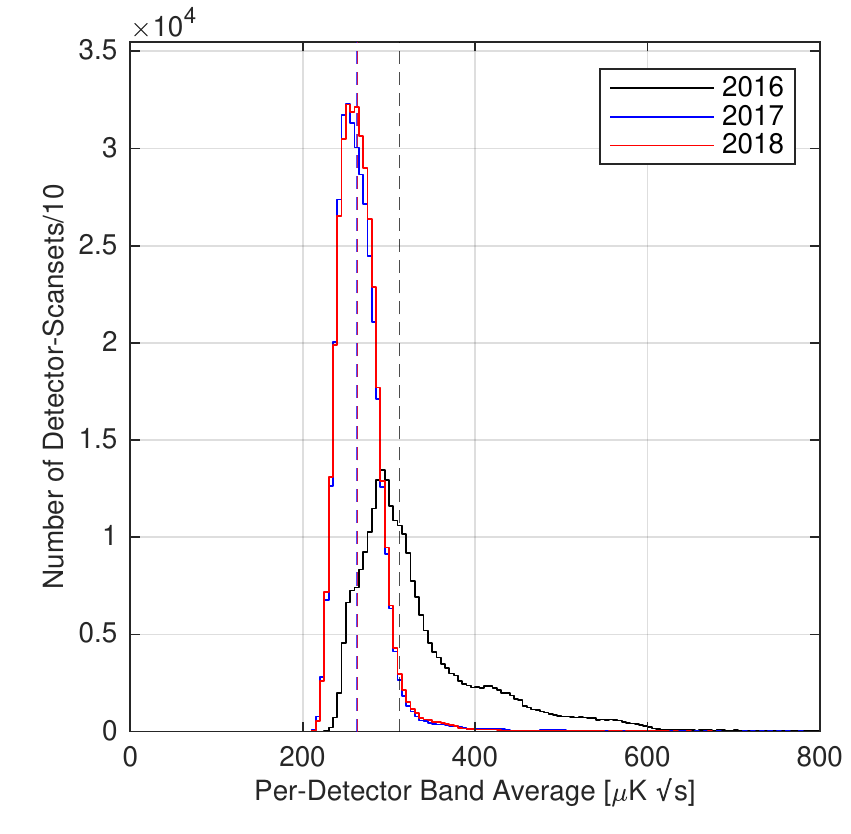}
  \caption{
Histogram of per-detector, per-scanset noise for every 10th scanset from 2016 to 2018, after applying a 3rd-order polynomial filter and averaging across the 0.1-1 Hz science band.
Median values of 312~\ukrts and 263~\ukrts are marked by vertical dashed lines for the 2016 and 2017/18 data, respectively.
The reduced internal receiver loading after switching from the metal-mesh infrared filters to the Zotefoam filters, as well as swapping four detector modules, leads to the improved noise performance shown here.
  }
  \label{fig:net_year}
\end{figure}

The per-detector, per-scanset noise distribution is shown in Fig.~\ref{fig:net_year}.
The median pair-difference, per-detector sensitivity of 312~\ukrts and 263~\ukrts is achieved in 2016 and 2017/18 data, respectively.
The sensitivity includes the weighted combination of every individual detector in every scanset during the three-year observation, capturing both the telescope performance and Antarctic seasonal variation.

We estimate the noise performance of the full \bicepthree\ instrument by calculating the inverse variance weight used in the final maps, giving the average \bicepthree\ noise equivalent temperature (NET) as 9.15, 6.82 and 7.14~\ukrts in 2016, 2017 and 2018, respectively, after all cuts are applied.

\subsection{Map depth}

The map depth is a measure of the noise level in the polarization maps.
Together with the area of the maps, the map depth~$D$, which we define as the deepest, central part of the map~\citep{BKII}, sets the final statistical sensitivity of the experiment.
This calculation accounts for the nonuniform coverage of the field, weighting each map pixel by its contribution relative to the deepest part of the map.
The effective area~$A_{\text{eff}}$ is calculated using the apodization mask to ensure the maps smoothly fall to zero, thus accounting for the higher variance and lower weight at the edges of the map.
The total sensitivity, $T = D/\sqrt{A_{\text{eff}}}$, gives a single number in temperature units indicating the total $B$-mode statistical sensitivity.

\begin{figure*}
  \centering
  \includegraphics[width=0.9\textwidth]{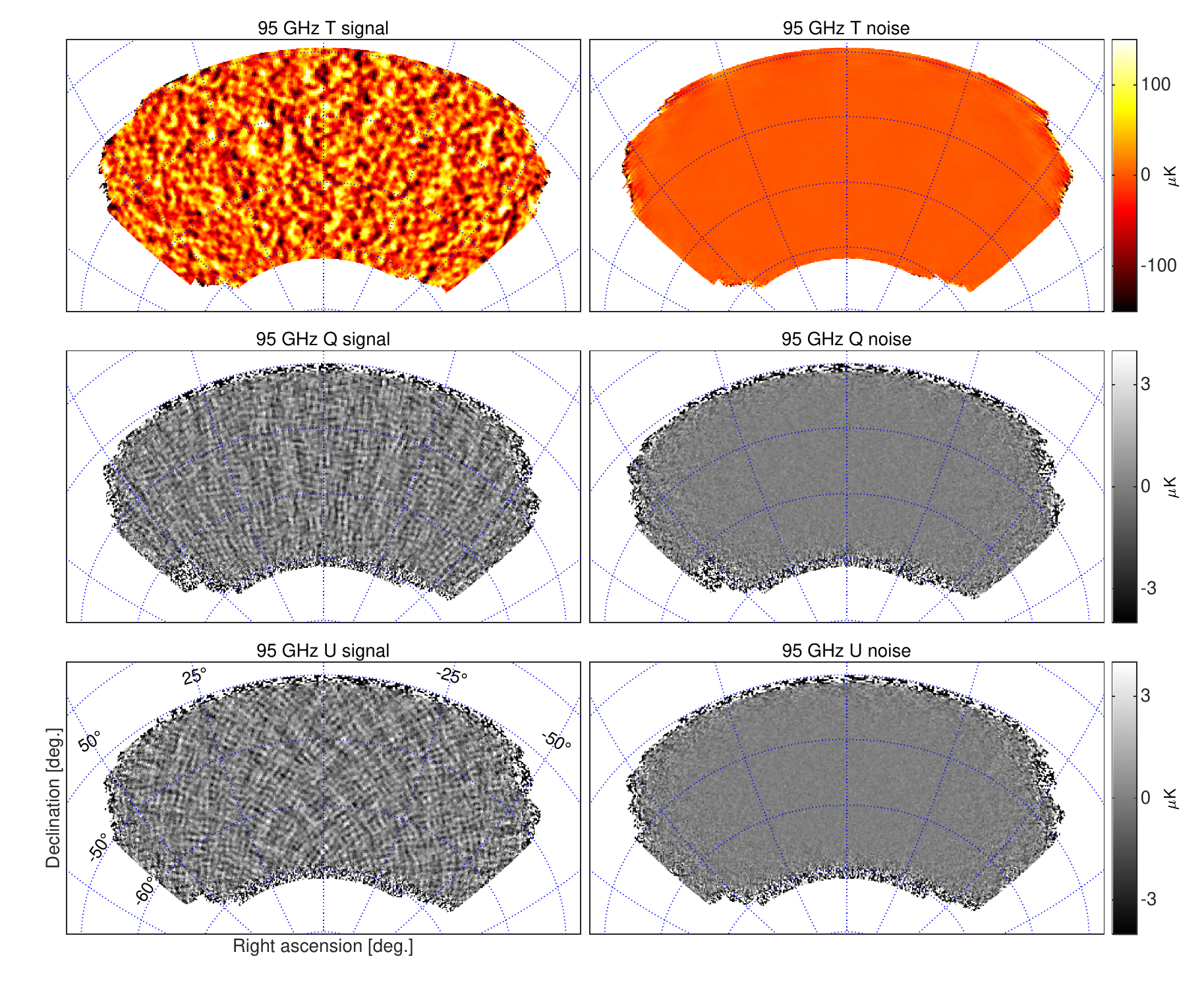}
  \caption {
T, Q, U map and its jackknife using the 3-years dataset from \bicepthree.
E-mode polarization is measured to high S/N, evidenced by visually seeing the $+$ and $\times$ patterns in the Q and U maps.
}
  \label{fig:netspec}
\end{figure*}

Table~\ref{tab:b3_sensitivity} lists the performance for \bicepthree.
We achieved a map depth of 2.8~$\mu$K-arcmin over an effective area of 584.9~$\text{deg}^2$, for a total sensitivity of  $T=1.3$~nK from the three-year \bicepthree\ data set.
The final T, Q, U maps and their corresponding noise are shown in Fig.~\ref{fig:netspec}.

\begin{table}
  \centering
  \caption{\bicepthree\ map based sensitivity by season}
  \label{tab:b3_sensitivity}
  \begin{tabular}{c c c c c} 
    \toprule
    Season & Map Depth ($D$) & Eff. Area ($A_{\text{eff}}$) & Total Sen. ($T$) \\
    \midrule
    2016 & 5.9~$\mu\text{K-arcmin}$ & 569.2~$\text{deg}^2$ & 2.9~nK \\
    2017 & 4.4~$\mu\text{K-arcmin}$ & 588.1~$\text{deg}^2$ & 2.2~nK \\
    2018 & 4.4~$\mu\text{K-arcmin}$ & 584.7~$\text{deg}^2$ & 2.1~nK \\
    Total & 2.8~$\mu\text{K-arcmin}$ & 584.9~$\text{deg}^2$ & 1.3~nK \\
    \bottomrule
    \vspace{1 mm}
  \end{tabular}
\end{table}

\section{Conclusions}
\label{sec:conclusion}

We presented the design and performance of \bicepthree, which has been observing CMB polarization from the South Pole since 2016.
The three-year data set, from 2016-2018, reached a map depth of 2.8~\ukarcmin\ (46~nK-deg) over an effective area of 584.9~square degrees, corresponding to a total sensitivity of $T=1.3$~nK.
A suite of jackknife tests show possible sources of systematic false polarization are controlled below the level of statistical sensitivity in each test.

In \cite{BKXIII} we present a cosmological analysis using this three-year \bicepthree\ data set with the combination of data from \keck.
The combination improve the sensitivity on $r$ to $\sigma(r)=0.009$.
During the austral summers at 2018 and 2019, we observed the CMB cold spot anomaly, which is detailed in \cite{spie:Jang_2020}.
With its achieved sensitivity, \bicepthree\ can also be utilized for a number of additional science targets, including primordial magnetic fields and axion-like particles through anisotropic and time-variable cosmic birefringence~\citep{BKIX,BKXII}, the properties of interstellar dust grains~\citep{ClarkHensley2019}, and patchy reionization and gravitational lensing through the use of distortion metrics~\citep{Yadav2010}.

Additional \bicepthree\ data taken from 2019 to 2021 are expected to further reduce noise in the 95~GHz power spectrum by a factor greater than $\sqrt{2}$.
\biceparray, which is based on the design of \bicepthree, deployed the first of four receivers at 30/40~GHz to the South Pole in 2020.
In coming years, additional receivers will be deployed at 95, 150 and 220/270 GHz with 32,000+ total detectors \citep{spie:LM}.
We project the \bicep/\keck\ experiment to reach $\sigma(r)$ between 0.002 and 0.004 at the end of \biceparray, depending on foreground complexity and degree of removal of $B$-modes due to gravitational lensing \citep{wu2021}.

\acknowledgements

\bicep/\keck\ has been made possible through a series of grants from the National Science Foundation including 0742818, 0742592, 1044978, 1110087, 1145172, 1145143, 1145248, 1639040, 1638957, 1638978, 1638970, 1836010 and by the Keck Foundation.
The construction of the \bicepthree\ receiver was supported by the Department of Energy, Laboratory Directed Research and Development program at SLAC National Accelerator Laboratory, under contract DE-AC02-76SF00515.
The development of antenna-coupled detector technology was supported by the JPL Research and Technology Development Fund and Grants No. 06-ARPA206-0040 and 10-SAT10-0017 from the NASA APRA and SAT programs.
The development and testing of focal planes were supported by the Gordon and Betty Moore Foundation at Caltech.
Readout electronics were supported by a Canada Foundation for Innovation grant to UBC.
The computations in this paper were run on the Cannon cluster supported by the FAS Science Division Research Computing Group at Harvard University.
We thank the staff of the U.S. Antarctic Program and in particular the South Pole Station without whose help this research would not have been possible.
Most special thanks go to our heroic winter-overs Sam Harrison, Hans Boenish, Grantland Hall, Ta-Lee Shue, Paula Crock, and Calvin Tsai.
We thank all those who have contributed past efforts to the \bicep/\keck\ series of experiments, including the \bicep1 team.
We would also like to thank Jonathon Hunacek at JPL/Caltech for exchange of information regarding detector readout and laboratory housekeeping design, and Aritoki Suzuki, Oliver Jeong, Yuki Inoue, Tomotake Matsumura at Berkeley and in Japan for discussions regarding alumina optics and AR coatings.

\bibliographystyle{aasjournal}

\bibliography{main}{}

\end{document}